\documentclass[10pt,a4paper]{article}
\usepackage[utf8]{inputenc}
\usepackage{graphicx}
\usepackage{authblk}
\usepackage{amsmath}
\usepackage{caption}
\usepackage{subcaption}
\usepackage{subcaption}
\title{Control of the vortex lattice
formation in coupled atom-molecular Bose-Einstein condensate in a
double well potential: Role of atom-molecule coupling, trap
rotation frequency and detuning parameter}
\author[1]{Moumita Gupta}
\author[2,*]{Krishna Rai Dastidar}
\affil[1]{Charuchandra College, Kolkata-700029, INDIA} \affil[2]
{Indian Association for the Cultivation of Science,
Kolkata-700032, INDIA}

\date{}

\begin{document}

\maketitle
* spkrd@iacs.res.in

\begin{abstract}
We study the vortex formation in coupled atomic and molecular
condensates in a rotating double well trap by numerically solving
the coupled Gross-Pitaevskii like equations. Starting with the
atomic condensate in the double well potential we considered
two-photon Raman photo-association for coherent conversion of
atoms to molecules. It is shown that the competition between
atom-molecule coupling strength and repulsive atom-molecule
interaction controls the spacings between atomic and molecular
vortices and the rotation frequency of the trap is the key player
for controlling the number of visible atomic and molecular
vortices. Whereas the Raman detuning controls the spacing between
atomic and molecular vortices as well as the number of atomic and
molecular vortices in the trap. We have shown by considering the molecular 
lattices  the distance between two molecular vortices can be controlled by 
varying the Raman detuning. In addition we have found that the
Feynman rule relating the total number of vortices and average
angular momentum both for atoms and molecules can be satisfied by
considering the atomic and molecular vortices those are hidden in
density distribution and seen as singularities in phase
distribution of the coupled system except for the lattice
structure where molecular vortices are overlapped with each other.
It is found that although the number of visible/core vortices in
atomic and molecular vortex lattices depends significantly on the
system parameters the number of atomic and molecular hidden
vortices remains constant in most of the cases.

\end{abstract}

\section{Introduction}

Realization of molecular Bose-Einstein condensate (BEC) via
two-photon Raman photoassociation \cite{Wynar, Winkler, Danzl} and
magnetic Feshbach resonance \cite{Thompson} has led to a plethora
of theoretical and experimental studies in the coupled
atomic-molecular system \cite{Timmermans_1, Timmermans_2, Heinzen,
Olsen, M3, M4}. Coherent conversion of an atomic BEC to a
molecular BEC via two-photon Raman photoassociation is important
to study the quantum superchemistry of the system \cite{Heinzen,
Olsen, Hope}. This coupled atomic-molecular BEC is also an
interesting system to study the vortex lattices which has not been
much explored. Formation of quantized vortices as well as highly
ordered vortex lattices by rapidly rotating the trapping potential
is one of the distinct characteristics of BEC which is a signature
of superfluidity \cite{ Madison, Shaeer}. For higher values of
rotation frequencies the vortex lattice structure can be observed.
The detailed dynamics of vortex lattice formation of an atomic BEC
in a rotating harmonic potential has been studied \cite{Tsubota,
Kasamatsu_1}. The influence of dipole-dipole interaction on the
formation and stability of atomic vortex lattice structure has
been also demonstrated \cite{Kumar_1, Kumar_2}. Ground state
vortex properties have been investigated in two-component BEC
\cite{Wang} and vortex lattice pattern has been emerged in
two-component mixture of $^{85}Rb$-$^{133}Cs$ \cite {Kumar_3}.
Formation of quantized ring vortices in coupled atom-molecular BEC
of $^{87}Rb$ atoms trapped in a rotating three dimensional
anisotropic cylindrical trap has been investigated and the
stability of these atomic and molecular vortices has been examined
\cite{Dutta}. A vortex molecule has been predicted in rotating
two-component BEC whose internal states are coherently coupled
\cite{Kasamatsu_3} and stable vortex structure of molecules have
been also studied in multicomponent BECs \cite{Nitta}. Vortex
lattices has been found in a rotating atomic-molecular BEC by Woo
et al and the coherent coupling between the atoms and molecules
gives a pairing between atomic and molecular vortices like a
carbon-di-oxide molecule \cite{Woo}. Controlling atom-molecule
interaction and Raman detuning parameter formation of vortex
lattice structure in rotating atomic molecular BECs could be
changed from overlapped atomic-molecular to carbon-di-oxide type
\cite{Liu}.

The theoretical and experimental studies of vortices and vortex
lattice structure in atomic BECs, two-component BECs or coherently
coupled atomic-molecular BECs described so far are for the system
which is trapped inside single harmonic potential well. With
technological advancements ultracold atoms can now be obtained in
double well (DW) potential as well \cite{Albiez}. DW potential
helps us to explore some wonderful properties of cold atoms like
Josephson effects \cite{Levy} and matter wave interference
\cite{Andrews}. In a rotating trap the vortices and vortex
lattices are formed depending on the speed of rotation. The
Feynman rule says that the total number of vortices $N_v$
increases linearly with the angular frequency of rotation $\Omega$
\cite{Feynman}. For an atomic BEC trapped in DW potential Feynman
rule has been verified considering the hidden vortices along the
central barrier of the trap together with the core vortices
\cite{Wen_1, Mithun}. Though these hidden vortices do not exist in
density distribution of the atomic condensate give rise to phase
singularities along the central barrier and appear in the density
distribution after free expansion of the condensate \cite{Wen_1}.
Hidden vortices in dipolar atomic BECs in rotating DW trap has
also been explored \cite{Sabari}. Hidden vorticity has also been
studied considering binary BECs described by 2D Gross-Pitaevskii
equations \cite{Brtka}. Enhancement in the vortex number by
introducing artificial guage potential in a condensate confined in
DW potential has been examined \cite {Bai}.

The motivation of our present work is to study the dependence of
the structure of vortex lattices formed in the coupled
atomic-molecular BECs in a rotating DW trap on the variation of
different system parameters e.g. atom-molecular coupling strength,
Raman detuning and the frequency of rotation of the trap, which
has not been explored in detail to the best of our knowledge. We
have considered two-photon Raman photo-association method for the
coherent atomic to molecular formation in BEC. To investigate the
effect of variation of different system parameters as mentioned
above on the vortex lattice formation we carry out a numerical
analysis of the coupled Gross-Pitaevskii equations considering
rotating DW trap following Crank-Nicholson method. We analyzed how
the variation of atom-molecular coupling strength, rotation
frequency of the trap and the Raman detuning affects the structure
of atomic and molecular vortex lattices, in particular the number
of atomic and molecular vortices and the overlapping of atomic and
molecular vortices in the lattice. The presence of atomic as well
as molecular hidden vortices along the central barrier of rotating
DW trap are clearly visible in the phase distribution and this
hidden vortices have also been revealed in the density
distributions when the DW trap is released and condensates are
freely expanded. The Feynman rule of vortices is found to be
satisfied well only after the inclusion of hidden atomic and
molecular vortices along with the visible ones except in the cases
where adjacent molecular vortices overlap with each other. It is
found that although the number of visible vortices changes
significantly with the variation of system parameters the number
of hidden vortices remains unaffected in most of the cases.

In the next section 2, our theoretical model is formulated which
describes the coupled atomic-molecular condensates in a rotating
DW trap. Results of the numerical analysis have been presented in
sections 3.1, 3.2 and 3.3 to discuss the effects of variation of
(i) atom-molecule coupling strength, (ii) rotation frequency and
the (iii) Raman detuning, respectively. Finally we conclude in
section 4.

\section{Theoretical model}

\begin{figure}\begin{minipage}{4in}
\includegraphics[width=4in,height=3.5in,angle=0]{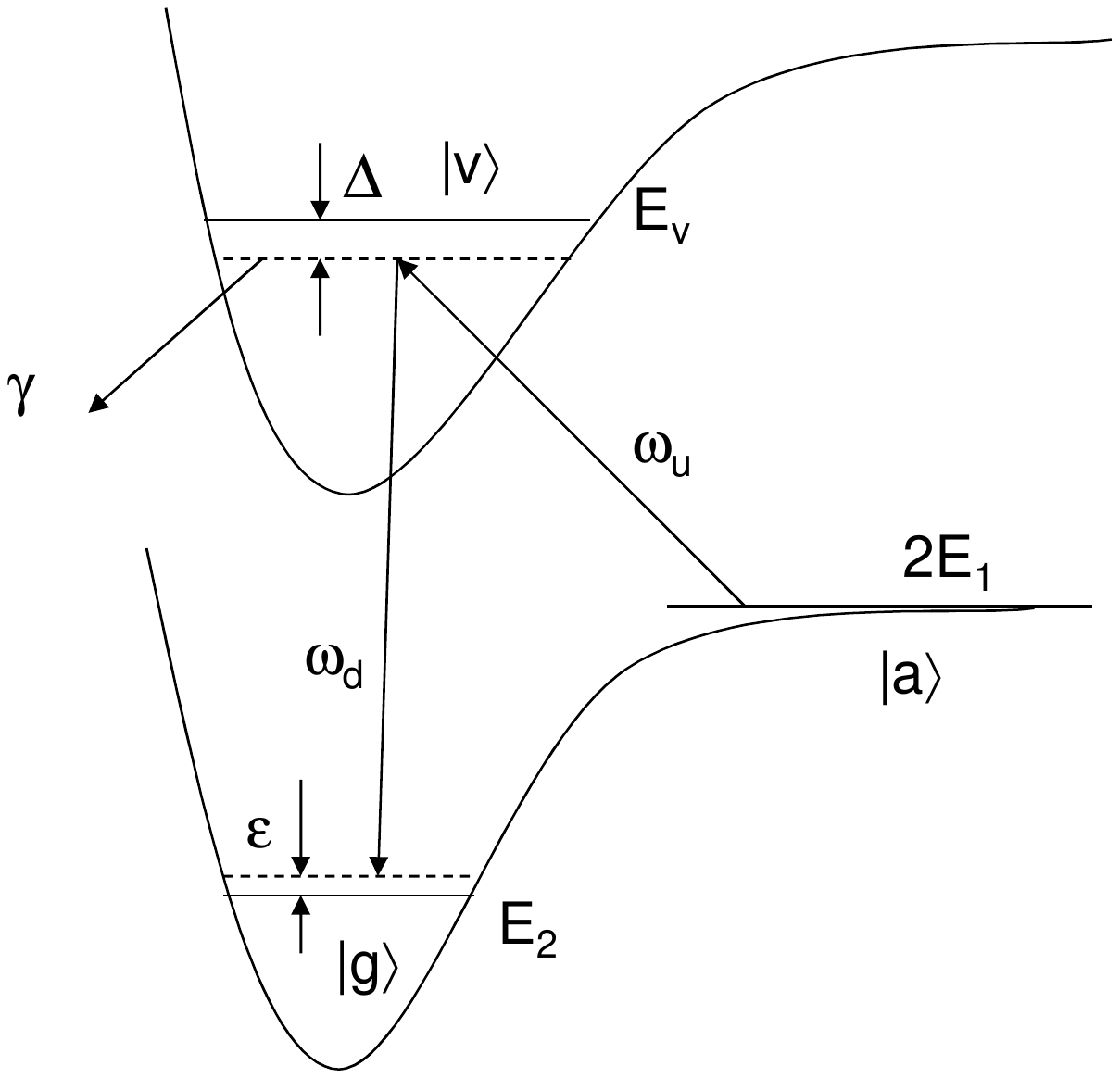}\end{minipage}
 \begin{minipage}{5in}\caption{Schematic representation of Raman
photoassociation}\end{minipage}
\end{figure}

The  model system considered here is shown schematically in Fig.
1 which is also described in Ref. \cite{Heinzen, M3}. The two atoms trapped in a DW potential 
in the atomic BEC state ($|a\rangle$) with total
energy $2E_1$ form a molecular ground state $|g\rangle$
of energy $E_2$ via a rovibrational level $|v\rangle$ of energy
$E_v$ of an excited electronic state by two-photon Raman photoassociation.
Levels $|a\rangle$ and $|v\rangle$ are coupled by a laser field
with frequency $\omega_u$ and the levels $|v\rangle$ and
$|g\rangle$ are coupled by a laser field with frequency
$\omega_d$. The process of Raman coupling becomes resonant when
the two-photon Raman detuning
$\epsilon=(2E_1-E_2)/\hbar-(\omega_d-\omega_u)$ is zero. Both the atomic and the molecular condensate
wave functions in the DW trap satisfy the coupled Gross-Pitaevskii (GP) equations with a
Raman coupling constant $\chi$.

Atomic and molecular BECs trapped in an external potential is
described by the macroscopic wave functions $\psi_a(\bf r, t)$ and
$\psi_m(\bf r, t)$ and the corresponding mean field energy density
for the system in the rotating frame with frequency of rotation
$\Omega$ may be given as

$$E[\psi_a,\psi_m]=\psi_a^*[-{\hbar^2\nabla^2\over 2m}-\Omega
L_z+V_a({\bf r})+{\lambda_{aa}\over 2}\psi_a^*\psi_a]\psi_a
+$$$$\psi_m^*[-{\hbar^2\nabla^2\over 4m}-\Omega L_z+V_m({\bf
r})+\epsilon+{\lambda_{mm}\over
2}\psi_m^*\psi_m]\psi_m+$$\begin{eqnarray}
\lambda_{am}\psi_a^*\psi_a\psi_m^*\psi_m+{\chi\over 2}
[\psi_m^*\psi_a\psi_a+\psi_m\psi_a^*\psi_a^*]\end{eqnarray} where
$\lambda_{aa}$, $\lambda_{mm}$ and $\lambda_{am}$ represent the
atom-atom, molecule-molecule and atom-molecule interactions,
respectively. $\chi$ represents the strength of coupling between
atoms and molecules due to stimulated Raman transitions and
$\epsilon$ characterizes Raman detuning.
$L_z=i\hbar(y{\partial\over
\partial x}-x{\partial\over
\partial y})$ is the z-component of angular momentum. Here the
system of coupled condensates is considered to be in an axially
symmetric trap $(\omega_x=\omega_y=\omega_\perp)$ which is very
strongly axial so that $\omega_x,\omega_y<<\omega_z$ i.e. the trap
is effectively two dimensional. The DW potential traps for atoms
and molecules are given by \begin{eqnarray}V_a({\bf
r})=V_a(x,y)={1\over 2}m\omega_\perp ^2 (x^2+\lambda^2
y^2)+V_0e^{-{x^2\over 2\sigma^2}}\end{eqnarray} and
\begin{eqnarray}V_m({\bf r})=V_m(x,y)=m\omega_\perp ^2 (x^2+\lambda^2 y^2)+V_0e^{{-x^2\over 2\sigma^2}}\end{eqnarray}
where $V_0$ and $\sigma$ denote the depth and width of the DW
potential barrier, respectively. $\lambda={\omega_y\over
\omega_x}$ is the anisotropy parameter of the trap.

At very low temperature the ground state of the system is
described by the coupled Gross-Pitaevskii (GP) equations given by
\begin{eqnarray}i\hbar{\partial\psi_a\over \partial t}={\delta E\over \delta \psi_a^*}\end{eqnarray}
and
\begin{eqnarray}i\hbar{\partial\psi_m\over \partial t}={\delta E\over \delta \psi_m^*}\end{eqnarray}
Using the energy density given in Eq.(1) the coupled GP equations
(4) and (5) take the following form:
\begin{eqnarray}i\hbar{\partial\psi_a\over
\partial t}=[-{\hbar^2\nabla^2\over 2m}-\Omega
L_z+V_a({x,y})+{\lambda_{aa}}|\psi_a|^2+\lambda_{am}|\psi_m|^2]\psi_a
+\chi\psi_m\psi_a^*\end{eqnarray}
\begin{eqnarray}i\hbar{\partial\psi_m\over
\partial t}=[-{\hbar^2\nabla^2\over 4m}-\Omega
L_z+V_m({x,y})+\epsilon+{\lambda_{mm}}|\psi_m|^2+\lambda_{am}|\psi_a|^2]\psi_m
+{\chi\over 2}\psi_a^2\end{eqnarray}

As the trap is strongly axial ($\omega_x,\omega_y<<\omega_z$) the
system is reduced to effectively two-dimensional x-y space by
taking the atomic and molecular wave functions as $\psi_a({\bf
r})=\psi_a(z)\psi_a(x,y)$ and $\psi_m({\bf
r})=\psi_m(z)\psi_m(x,y)$. $\psi_a(z)$ and $\psi_m(z)$ are taken
as normalized Gaussian functions as: $\psi_{a,m}(z)={1\over {(\pi
d_z^2)}^{1/4}}e^{{-z^2\over 2 {d_z}^2}}$ where
$d_z=\sqrt{{\omega_\perp\over \omega_z}}$.

In order to reduce the equations (6) and (7) in dimensionless form
the length and time are scaled as $x=a_hx_1$, $y=a_hy_1$ and
$t=t_1/w_\perp$ with $a_h=\sqrt{{\hbar\over 2mw_\perp}}$. To
include the effect of dissipation in vortex dynamics which may be
due to the loss of atoms and molecules from the trap resulting
from inelastic collisions the term
$-\gamma{\partial\psi_{a,m}\over \partial t}$ is added in
equations (6) and (7) where $\gamma$ is a dimensionless parameter
characterizing the dissipation. Then the coupled equations (6) and
(7) are reduced to the dimensionless form
$$(i-\gamma){\partial\psi_a\over
\partial t}=[-({d^2\over dx_1^2}+{d^2\over dy_1^2})-i\Omega_1(y_1{\partial\over \partial x_1}-x_1{\partial \over \partial y_1})
+{1\over 4}(x_1^2+y_1^2)+V_{01}e^{-x_1^2\over 2
\sigma_1^2}+$$
\begin{eqnarray}{\lambda_{aa_1}}\eta_1|\psi_a|^2+\lambda_{am_1}\eta_2|\psi_m|^2]\psi_a
+\chi_1\eta_3\psi_m\psi_a^*\end{eqnarray} and

$$(i-\gamma){\partial\psi_m\over
\partial t}=[-{1\over 2}({d^2\over dx_1^2}+{d^2\over dy_1^2})-i\Omega_1(y_1{\partial\over \partial x_1}-x_1{\partial \over \partial y_1})
+{1\over 8}(x_1^2+y_1^2)+V_{01}e^{-x_1^2\over 2 \sigma_1^2}+$$
\begin{eqnarray}{\lambda_{mm_1}}\eta_{1m}\psi_m|^2+\lambda_{am_1}\eta_{2m}|\psi_a|^2+\epsilon_1]\psi_m
+{\chi_1\over 2}\eta_{3m}\psi_a^2\end{eqnarray} where
$\lambda_{aa_1}={\lambda_{aa}\over \hbar \omega_\perp}$,
$\lambda_{mm_1}={\lambda_{mm}\over \hbar \omega_\perp}$,
$\lambda_{am_1}={\lambda_{am}\over \hbar \omega_\perp}$,
$\chi_1={\chi\over \hbar\omega_\perp}$, $\epsilon_1={\epsilon\over
\hbar\omega_\perp}$, $\Omega_1={\Omega\over \hbar \omega_\perp}$,
and
\begin{eqnarray}
\eta_1={\int{|\psi_a(z)|^4dz\over |\psi_a(z)|^2dz}}
\end{eqnarray}
\begin{eqnarray}
\eta_2={\int{|\psi_m(z)|^2 |\psi_a(z)|^2dz\over |\psi_a(z)|^2dz}}
\end{eqnarray}
\begin{eqnarray}
\eta_3={\int{\psi_m(z) \{\psi_a^*(z)\}^2dz\over |\psi_a(z)|^2dz}}
\end{eqnarray}
\begin{eqnarray}
\eta_{1m}={\int{|\psi_m(z)|^4dz\over |\psi_m(z)|^2dz}}
\end{eqnarray}
\begin{eqnarray}
\eta_{2m}={\int{|\psi_a(z)|^2 |\psi_m(z)|^2dz\over
|\psi_m(z)|^2dz}}
\end{eqnarray}
\begin{eqnarray}
\eta_{3m}={\int{\psi_m^*(z) \{\psi_a(z)\}^2dz\over
|\psi_m(z)|^2dz}}
\end{eqnarray}

The normalization of the atomic and molecular wave functions can
be given by the condition
\begin{eqnarray}
|\psi_a(x,y)|^2+2|\psi_m(x,y)|^2=1
\end{eqnarray}

In the following section we study the vortices of a coupled
atomic-molecular BEC by solving the equations (8) and (9) with the
help of Crank-Nicolson scheme \cite{M3, M4}. The iteration in time
has been started with the initial atomic wave function taken as
the normalized ground state solution of GP equation for atomic BEC
in DW potential for $\Omega_1$=0 which can be written as
\begin{eqnarray}[-({d^2\over dx_1^2}+{d^2\over dy_1^2})
+{1\over 4}(x_1^2+y_1^2)+V_{01}e^{-x_1^2\over 2 \sigma_1^2}+
{\lambda_{a_1}}\eta_1|\psi_a|^2]\psi_a =\mu_1\psi_a\end{eqnarray}
where $\mu_1$ is the chemical potential in the units of
$\hbar\omega_\perp$.

The energies corresponding to atomic-molecular scattering
\begin{eqnarray}E_{am}=\lambda_{am}\int{d{\bf r}|\psi_a|^2|\psi_m|^2}\end{eqnarray}
The energies corresponding to atomic-molecular coupling
\begin{eqnarray}E_{c}={\chi\over 2}\int{d{\bf r}[\psi_a^*\psi_a^*\psi_m+\psi_m^*\psi_m^*\psi_a}]\end{eqnarray}
The energies corresponding to rotation
\begin{eqnarray}E_{rot}=(-\Omega)\int{d{\bf r}[\psi_a^*L_z\psi_a+\psi_m^*L_z\psi_m]}\end{eqnarray}

\begin{figure}\begin{subfigure}[t]{0.3\textwidth}
\includegraphics[width=1.5in,height=1.8in,angle=0]{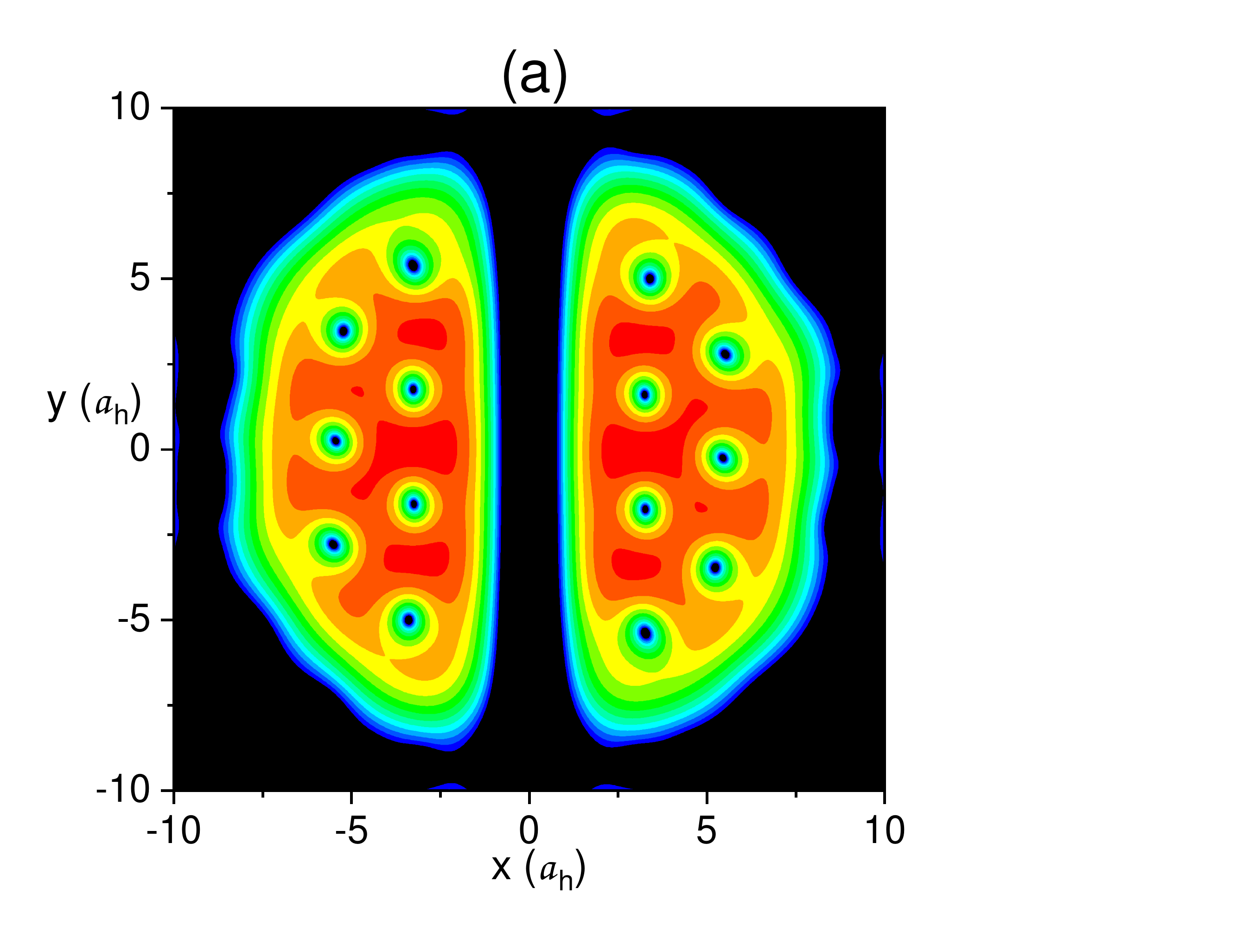}
\end{subfigure}\hfill
\begin{subfigure}[t]{0.3\textwidth}
\includegraphics[width=1.5in,height=1.8in,angle=0]{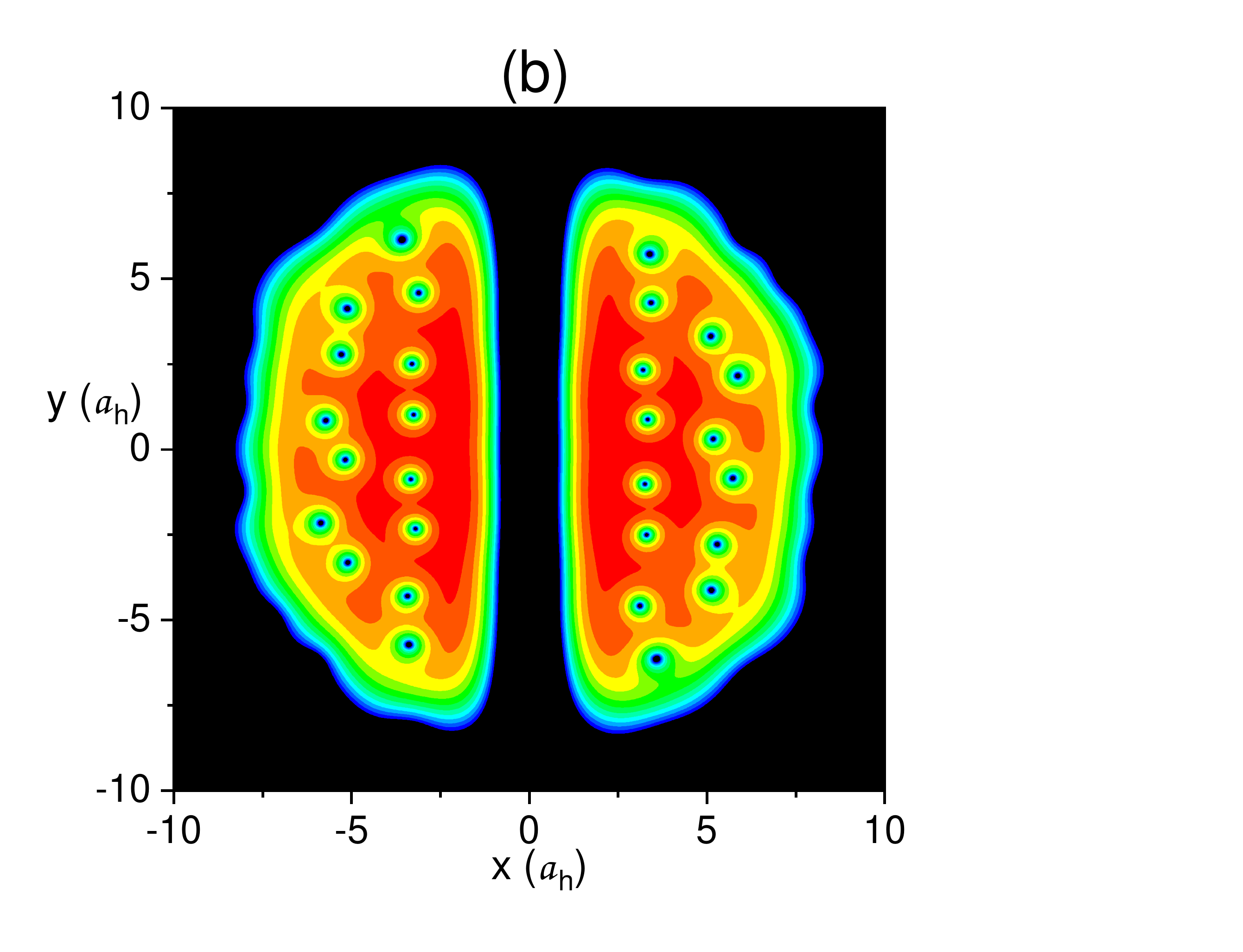}
\end{subfigure}\hfill
\begin{subfigure}[t]{0.3\textwidth}
\includegraphics[width=1.5in,height=1.8in,angle=0]{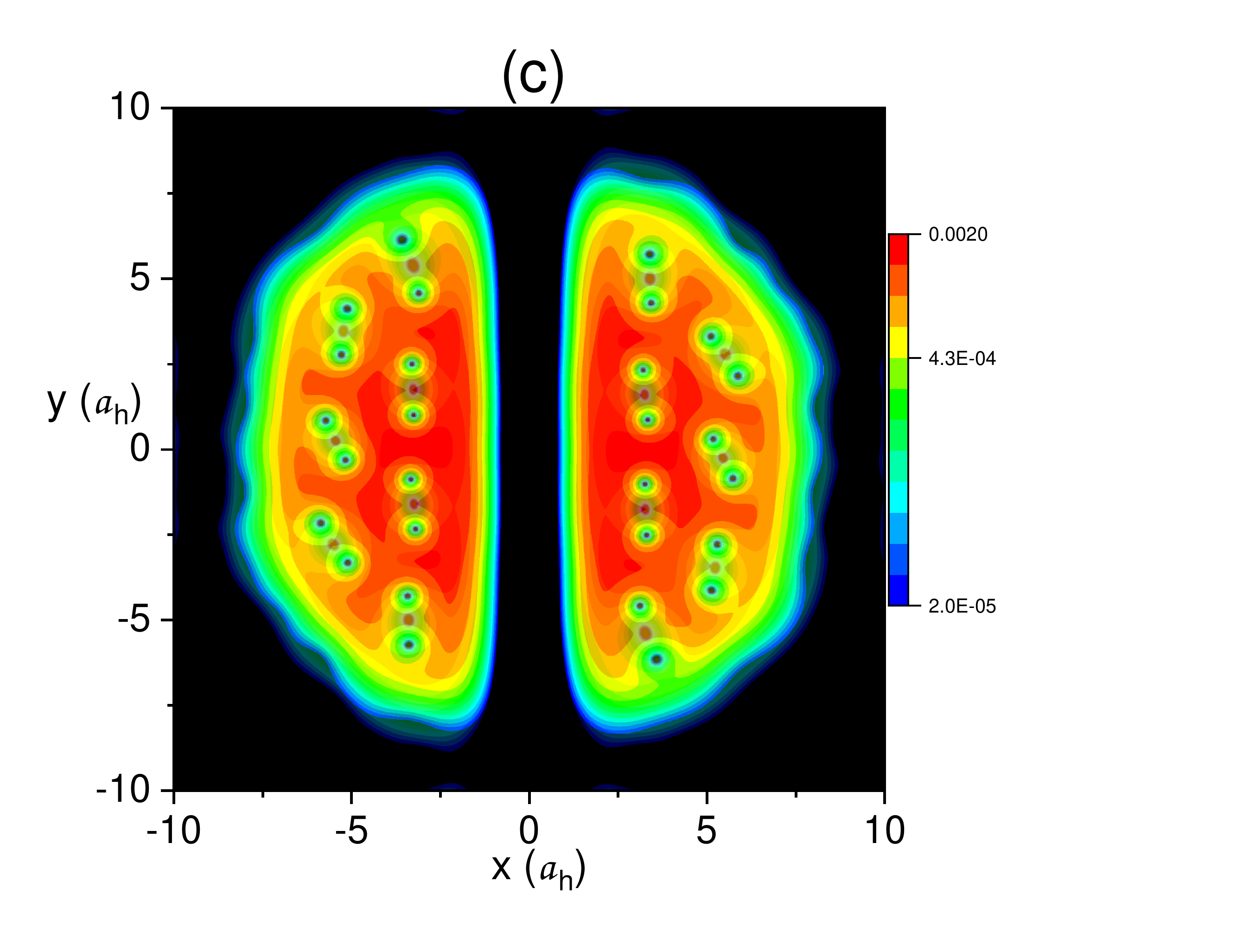}
\end{subfigure}

\begin{subfigure}[t]{0.3\textwidth}
\includegraphics[width=1.5in,height=1.8in,angle=0]{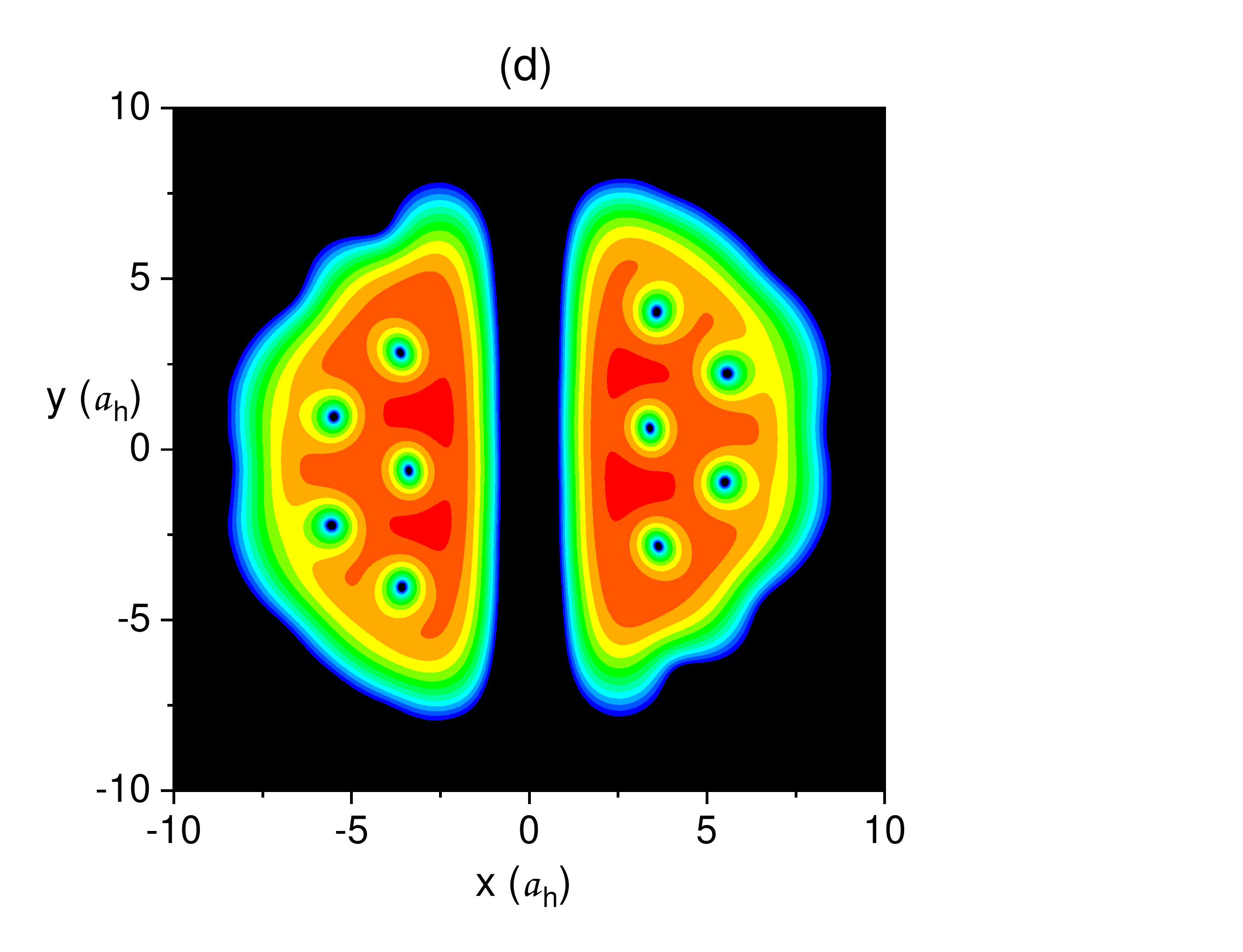}
\end{subfigure}\hfill
\begin{subfigure}[t]{0.3\textwidth}
\includegraphics[width=1.5in,height=1.8in,angle=0]{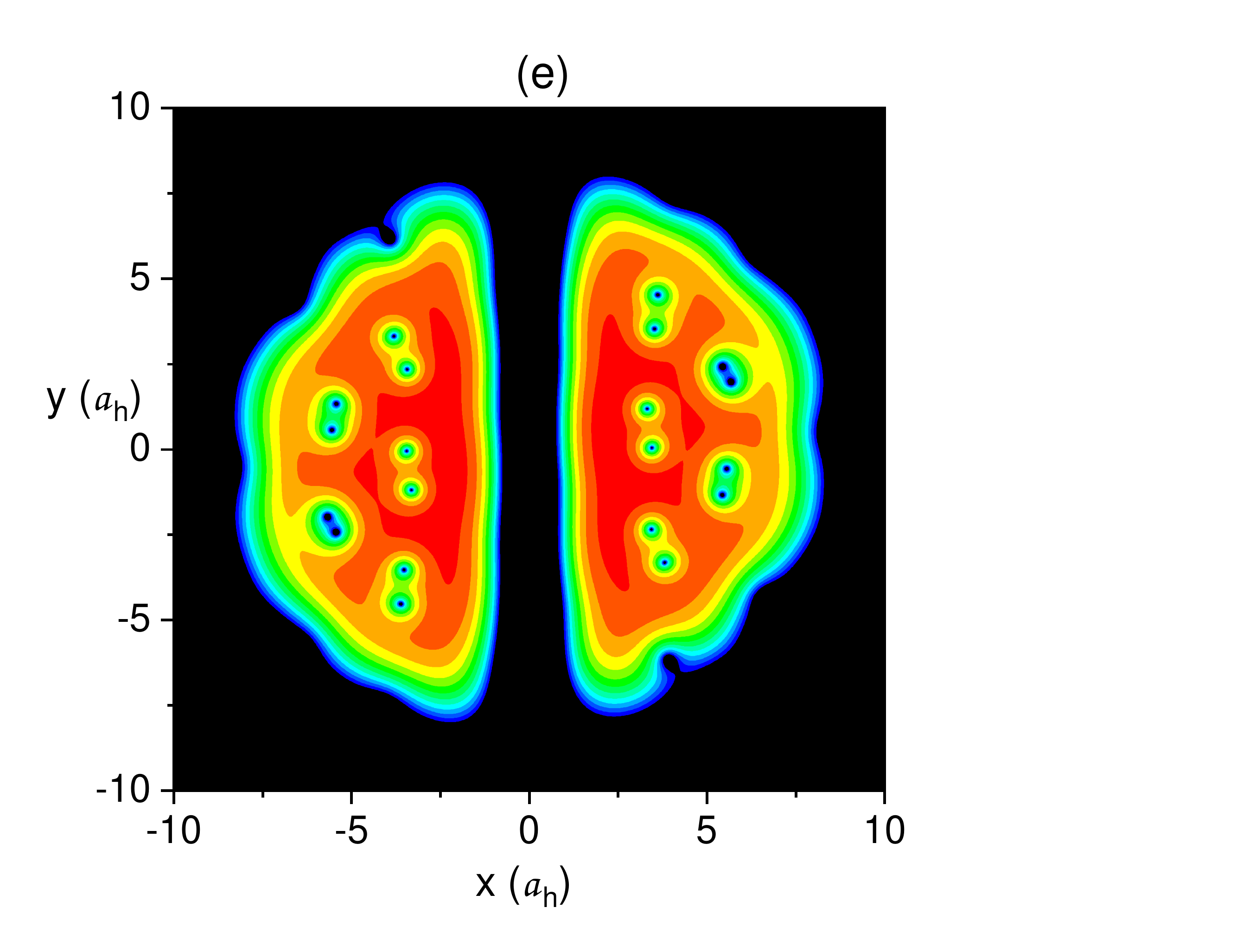}
\end{subfigure}\hfill
\begin{subfigure}[t]{0.3\textwidth}
\includegraphics[width=1.5in,height=1.8in,angle=0]{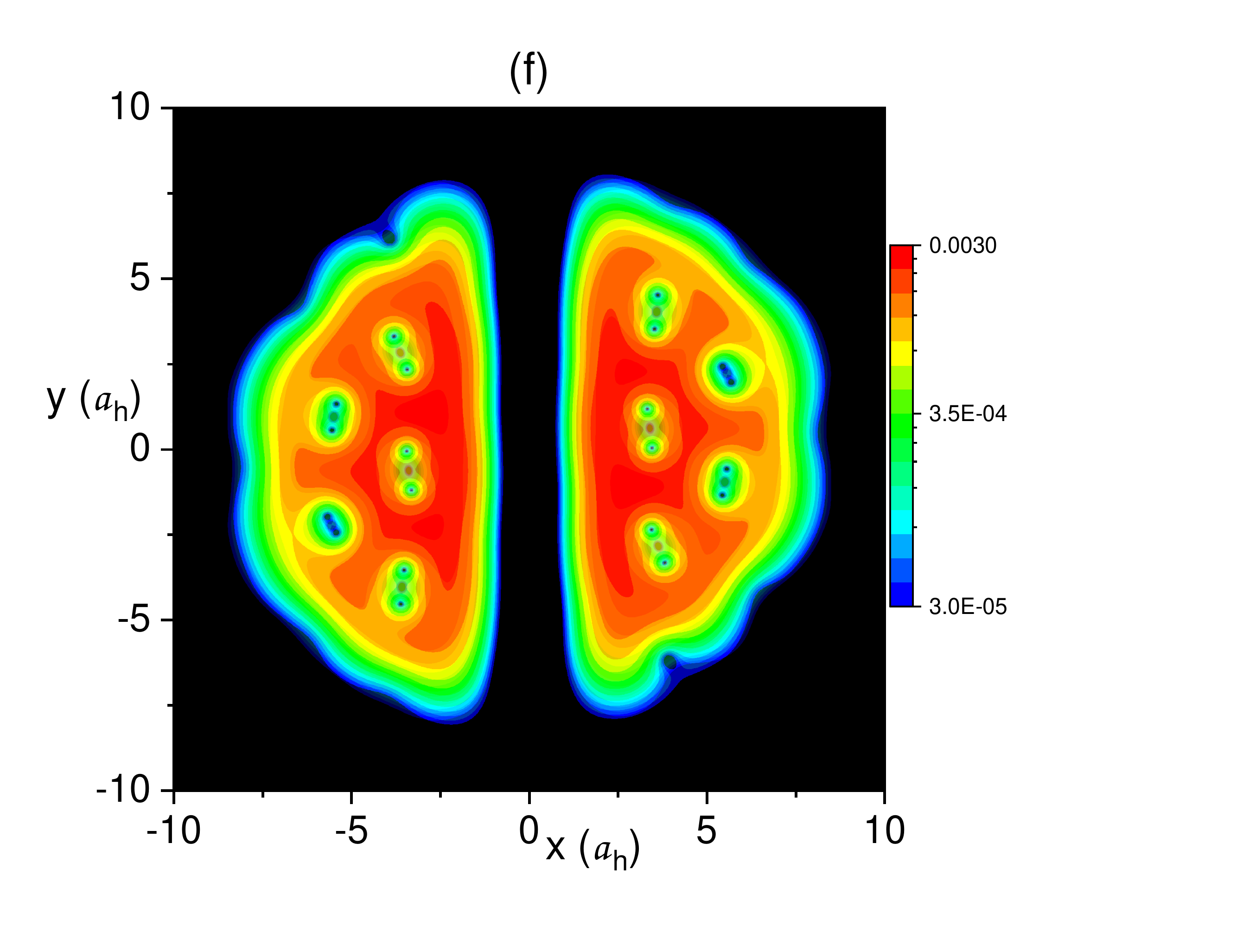}
\end{subfigure}

\begin{subfigure}[t]{0.3\textwidth}
\includegraphics[width=1.5in,height=1.8in,angle=0]{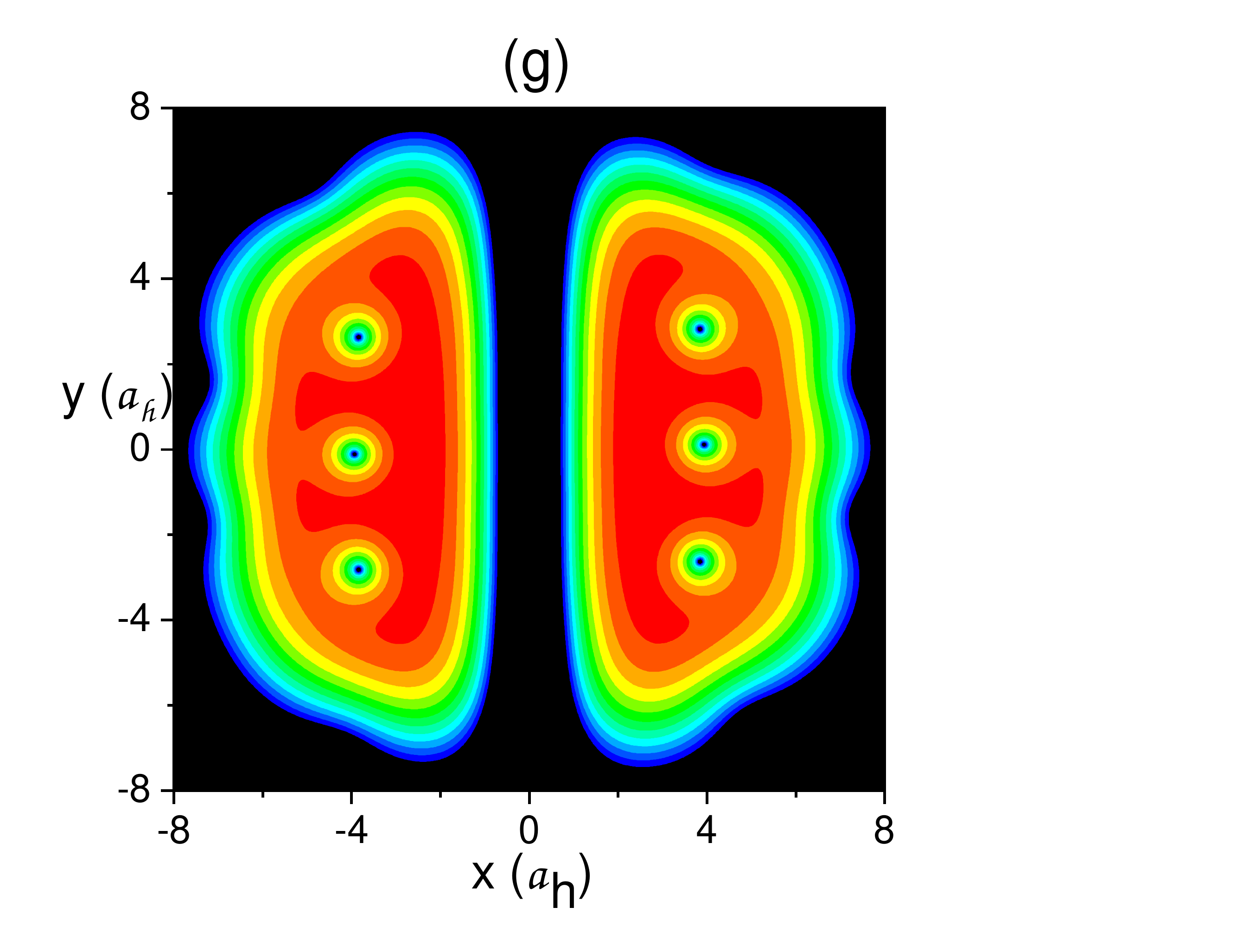}
\end{subfigure}\hfill
\begin{subfigure}[t]{0.3\textwidth}
\includegraphics[width=1.5in,height=1.8in,angle=0]{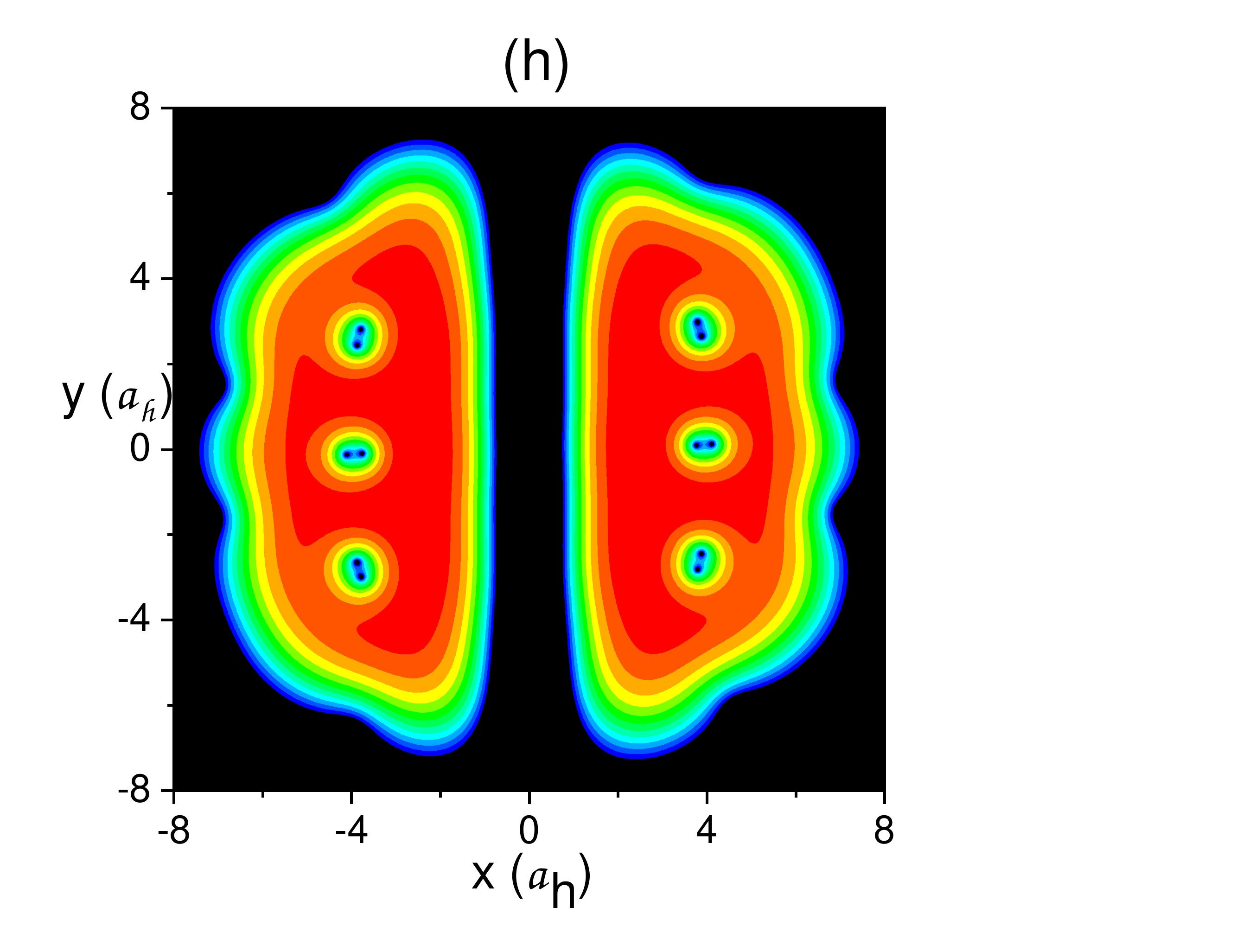}
\end{subfigure}\hfill
\begin{subfigure}[t]{0.3\textwidth}
\includegraphics[width=1.5in,height=1.8in,angle=0]{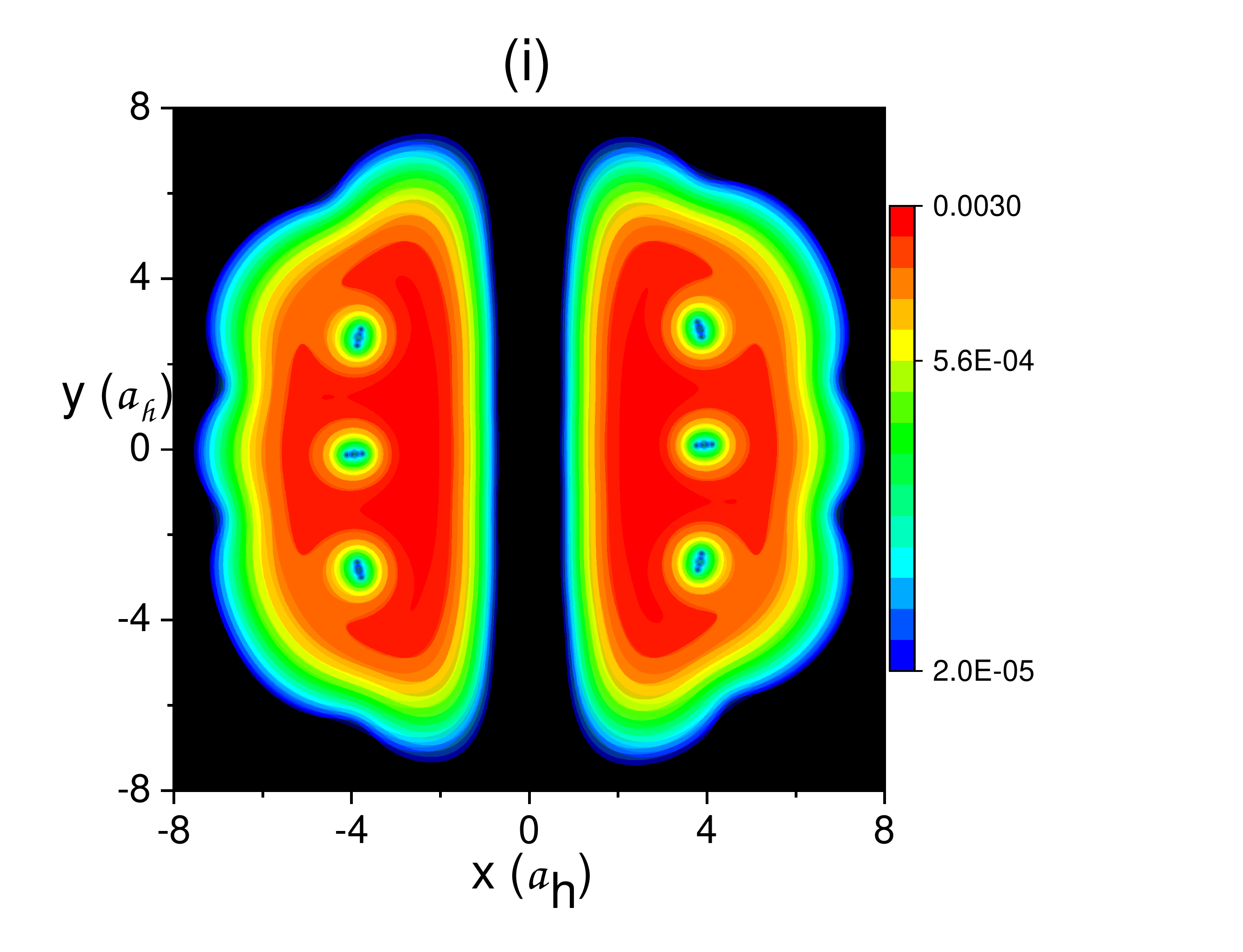}
\end{subfigure}

\begin{minipage}{5in}
\caption{(Color online) Density distributions for atomic [Figs. (a), (d), (g)], molecular  [Figs. (b), (e), (h)] and 
atomic with molecular  [Figs. (c), (f), (i)] vortex lattice configuration for different coupling strengths 
$\chi_1$= 30 [Figs. (a), (b), (c)], 50 [Figs. (d), (e), (f)] and 70 [Figs. (g), (h), (i)] at $t_1$= 300. $\Omega_1$= 0.95 and
$\epsilon_1$= 0. Red color corresponds to higher
densities and blue color corresponds to lower densities. Darker
color corresponds to lower density. x and y are in the units of
$a_h$=$\sqrt{\hbar\over{2m\omega_\perp}}$}
\end{minipage}
\end{figure}

\begin{figure}\begin{subfigure}[t]{0.3\textwidth}
\includegraphics[width=1.5in,height=1.8in,angle=0]{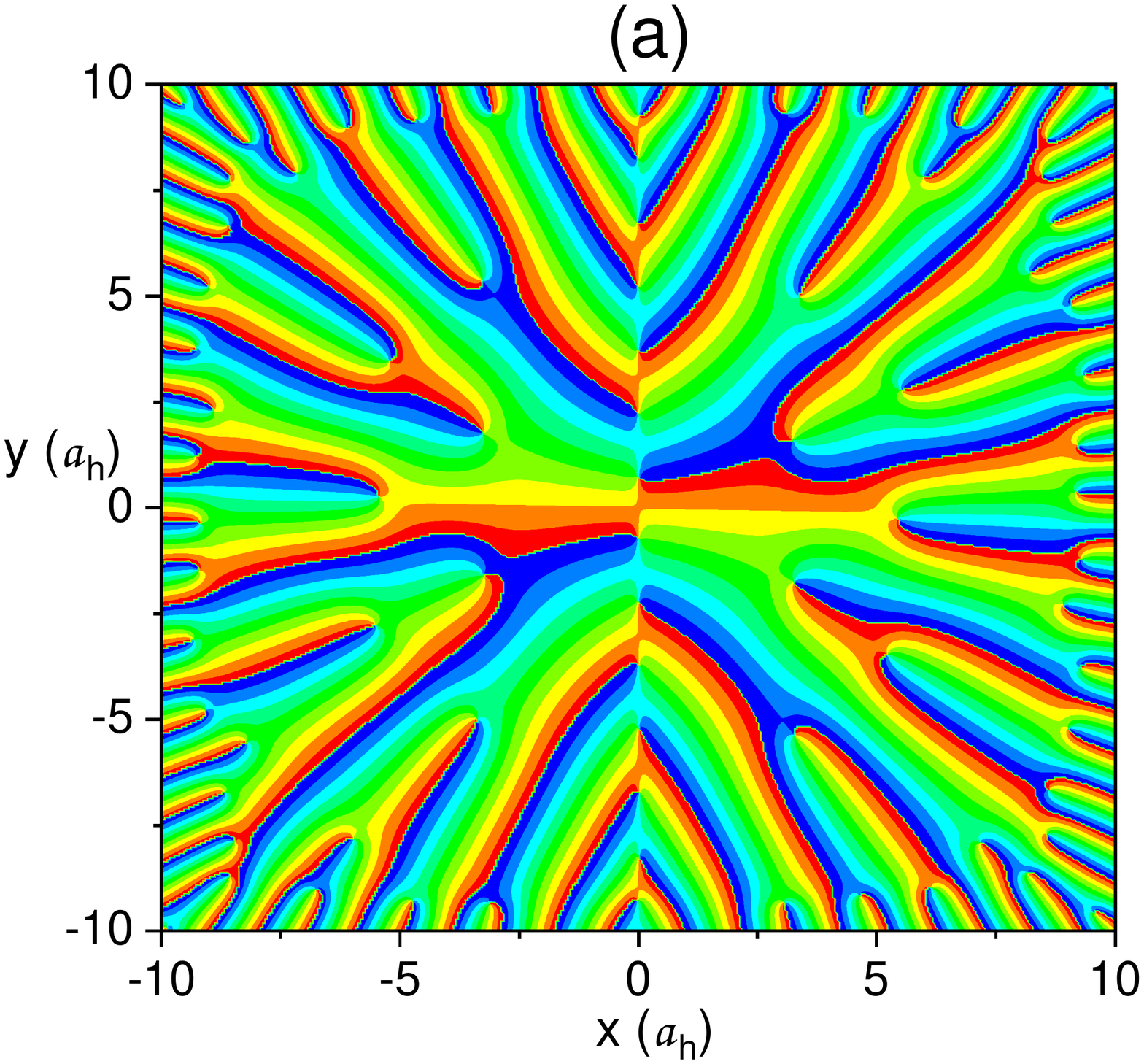}
\end{subfigure}\hfill
\begin{subfigure}[t]{0.3\textwidth}
\includegraphics[width=1.5in,height=1.8in,angle=0]{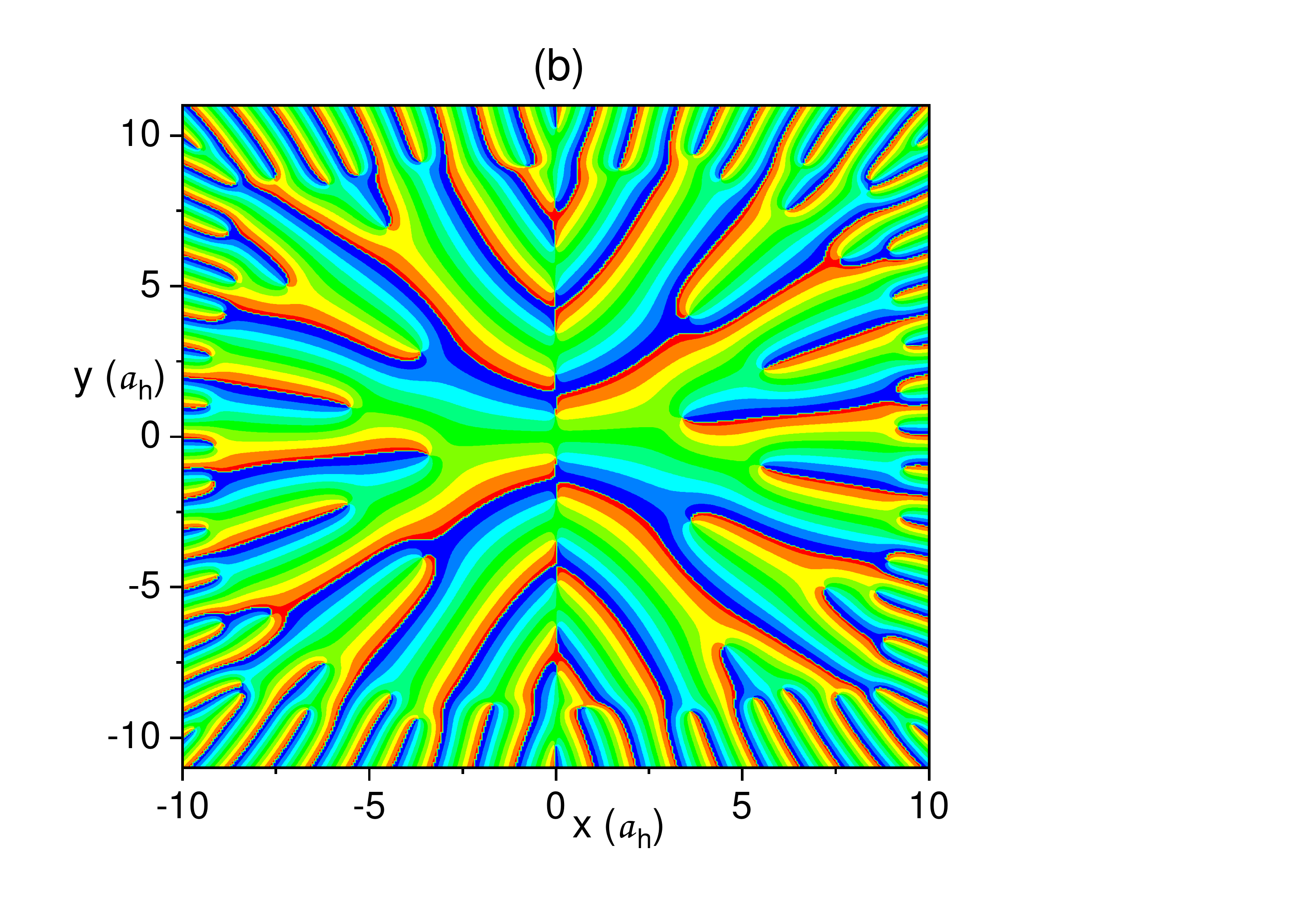}
\end{subfigure}\hfill
\begin{subfigure}[t]{0.3\textwidth}
\includegraphics[width=1.5in,height=1.8in,angle=0]{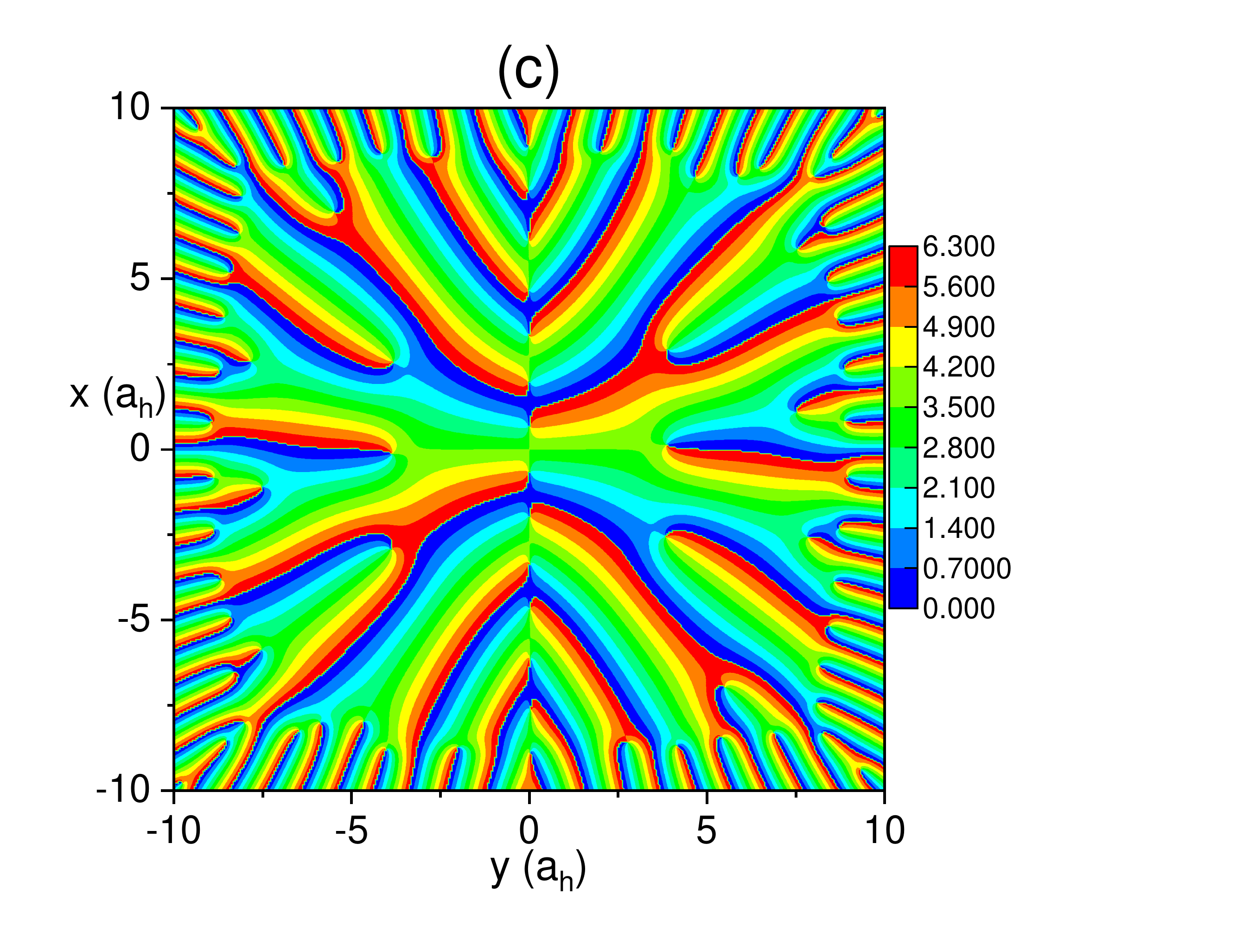}
\end{subfigure}

\begin{subfigure}[t]{0.3\textwidth}
\includegraphics[width=1.5in,height=1.8in,angle=0]{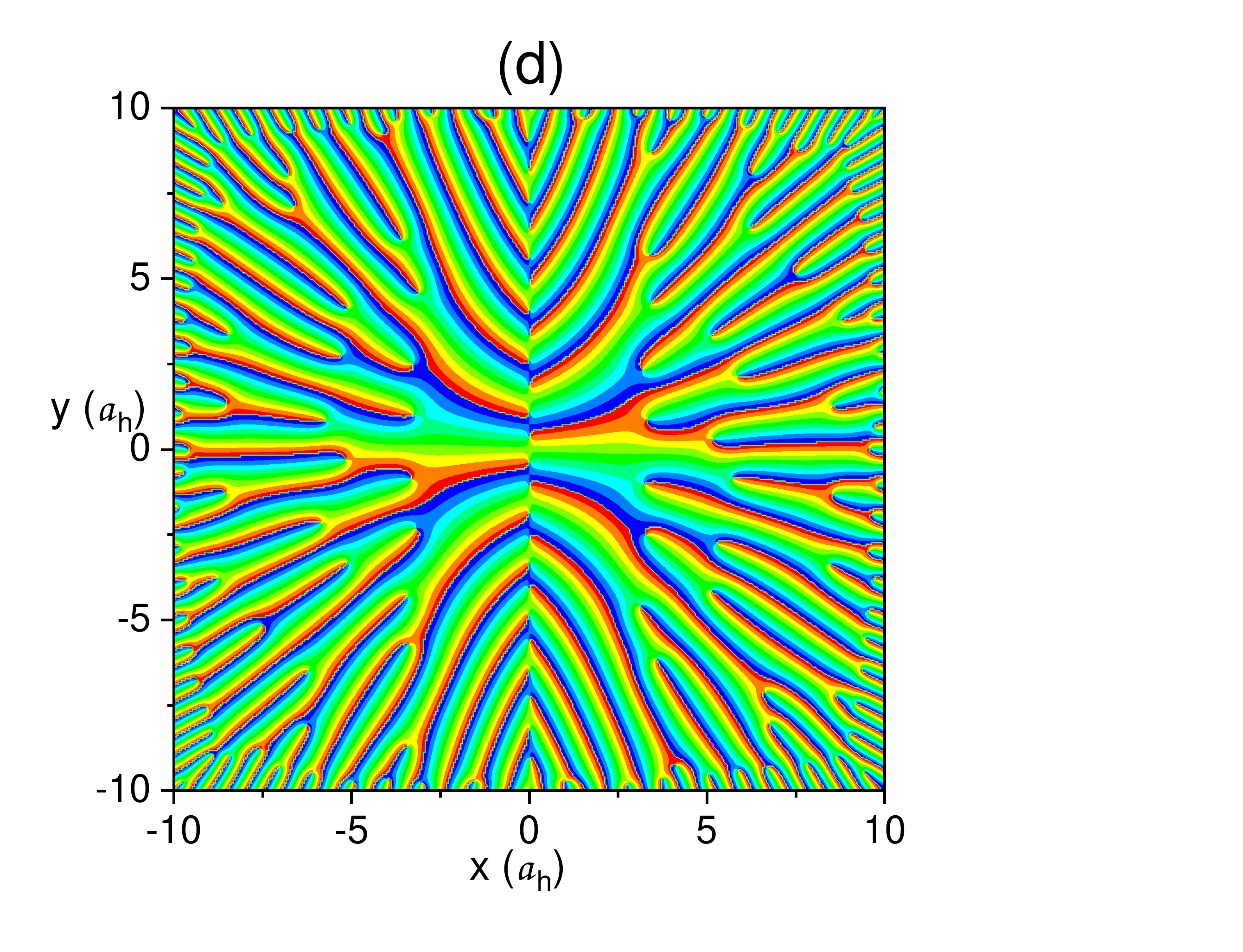}
\end{subfigure}\hfill
\begin{subfigure}[t]{0.3\textwidth}
\includegraphics[width=1.5in,height=1.8in,angle=0]{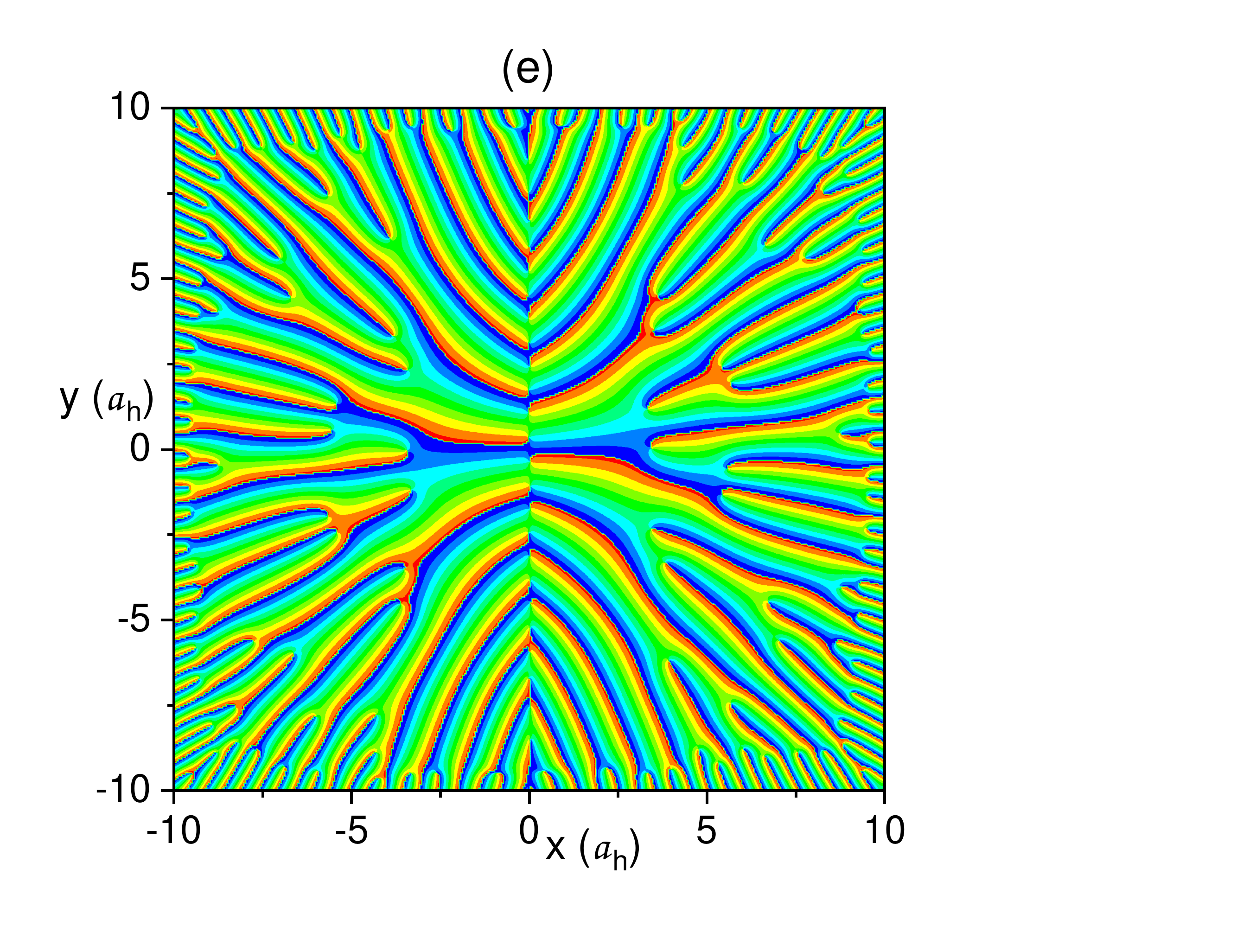}
\end{subfigure}\hfill
\begin{subfigure}[t]{0.3\textwidth}
\includegraphics[width=1.5in,height=1.8in,angle=0]{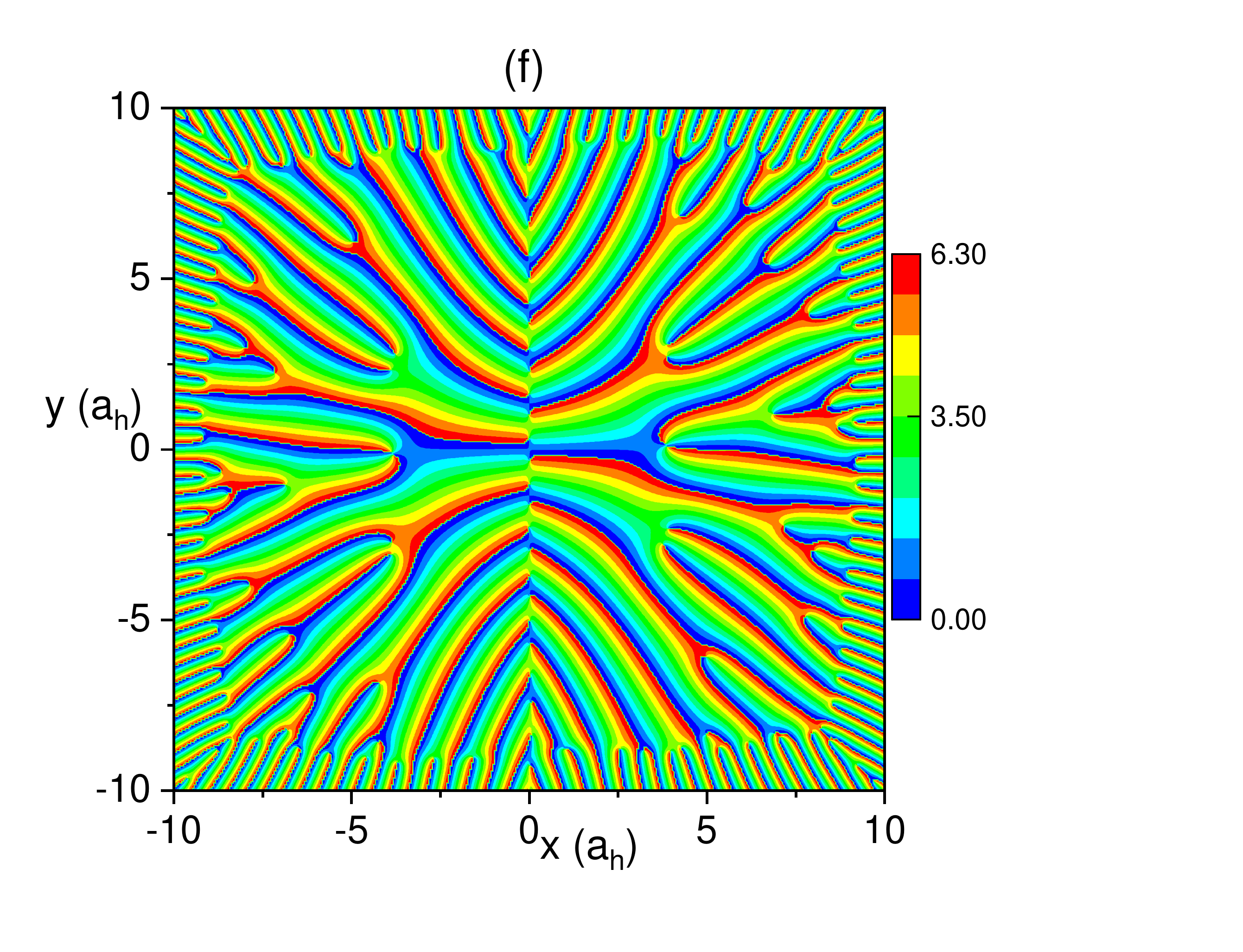}
\end{subfigure}

\begin{minipage}{5in}
\caption{(Color online) Phase profile of atoms $\psi_a$ for different
coupling strengths $\chi_1$= 30 [Fig. (a)], 50 [Fig. (b)] and 70
[Fig. (c)]  and  phase profile of molecules $\psi_m$ for different
coupling strengths $\chi_1$= 30 [Fig. (d)], 50 [Fig. (e)] and 70
[Fig. (f)]  at $t_1$= 300. $\Omega_1$= 0.95 and $\epsilon_1$= 0.
Phase varies from 0 to 2$\pi$. Red color corresponds to higher
values and blue color corresponds to lower values. Darker color
corresponds to lower phase. x and y are in the units of $a_h$=
$\sqrt{\hbar\over {2m\omega_\perp}}$.}
\end{minipage}
\end{figure}

\section{Results and Discussion}

We have carried out a detailed numerical analysis of the effect of
relative strength of atom-molecular coherent coupling and the
interspecies interaction, Raman detuning and the rotation
frequency of the trap on the vortex lattice formation in the
coupled atom-molecular BECs in DW potential. To start with we
assume the presence of atomic BEC only (with no molecules) inside
the DW trap by making $\lambda_{am1}$= $\lambda_{mm1}$= $\chi_1$=
0 in absence of rotation and the ground state solution of the 2D
GP equation [Eq. (17)] for atomic BEC of $^{87}Rb$ atoms trapped
in DW potential has been found out considering $\Omega_1$= 0 using
the imaginary time propagation method [34]. This ground state wave
functions were used as initial wave functions to solve the 2D
coupled atomic molecular GP equations [Eqs. (8) and (9)] with non
zero values of $\Omega_1$ by using Crank-Nicolson scheme \cite{M3,
M4}. The space step and the time step are taken as $\Delta x$ =
0.05 and $\Delta t$ = 0.0005 respectively. The trap parameters are
chosen as $\omega_x$= $\omega_y$= $\omega_\perp$= 2$\pi \times$ 40
Hz, $\omega_z$= 2$\pi \times$ 800 Hz, $V_0$= 40 and $\sigma$=0.5 as taken by Wen et al. \cite{Wen_1}.
We have taken $\lambda_{aa1}$= 600, $\lambda_{am1}$= 0.85 $\times
\lambda_{aa1}$ (as considered by Woo et al. \cite{Woo}), $\lambda_{mm1}$= 2 $\times \lambda_{aa1}$. Here we
consider atom-atom, molecule-molecule and atom-molecule
interactions to be repulsive. $\Omega_1$ ranges from 0.85 to 0.95,
$\chi_1$ ranges from 30 to 70 and $\epsilon_1$ ranges from -0.5 to 5.
The initial value of $\chi_1$ is so chosen that carbon-di-oxide
like atomic-molecular vortex lattice structure could be obtained.
The atom-atom, molecule-molecule and atom-molecule interactions
have been kept fixed. The unit of length and time are
$\sqrt{\hbar\over {2m\omega_\perp}}$ and $\omega_\perp^{-1}$
respectively. The effect of dissipation is included by adding the
term $-\gamma {\partial \psi \over \partial t}$ in Eqs. (8) and
(9). The value of $\gamma$ is taken as 0.03 \cite{Wen_1}. In this
system three competing forces e.g. (i) atom-molecule repulsive
interaction, (ii) atom-molecule coherent coupling strength which
is attractive in this system and (iii) the centrifugal force from
the rotation of the trap control the structure and the number of
vortices in the vortex lattice of both atoms and molecules. The
repulsive atom-molecule interaction energy tends to increase the
distance between the atomic and molecular vortices whereas
attractive atom-molecule coupling tends to reduce the distance
between the atomic and molecular vortices. On the other hand
increase in centrifugal forces due to the increase in rotation
frequency leads to spread in density distribution effectively
resulting in increase in the number of vortices. Conversely the
decrease in rotation frequency tends to squeeze the
atomic-molecular density distribution by restricting its spread
which causes decrease in the number of vortices. Moreover it is
found that the variation in the detuning $\epsilon_1$ can also
affect the nature of vortex lattice formation. It is found that
the variation of any one of these parameters leads to variation in
the overlap of the wave functions and hence leading to change in
the energies. To explore these features of vortex lattice
formation we have compared the three components of energies with
variation of $\chi_1$, $\Omega_1$ and $\epsilon_1$ as presented in
tables 1, 2 and 3, respectively.

\subsection{Effect of variation of atom-molecule coupling strength on atom-molecular vortex lattices}
 In this section we study the effect of the change in atom-molecule
coupling on the vortex lattice formation for a fixed value of
rotation frequency ($\Omega_1$= 0.95) and detuning ($\epsilon_1$=
0). The vortex lattice structure of atoms, molecules and
atoms with molecules for three different values of atom molecular
coupling $\chi_1$ =30, 50 and 70, respectively is plotted in Fig.2.
Density distributions for atoms ($|\psi_a|^2$), for molecules
($|\psi_m|^2$) and for atoms with molecules
($|\psi_a|^2$ and $|\psi_m|^2$) have been plotted in Figs.
2(a), 2(b) and 2(c), respectively for $\chi_1$= 30. Similarly those for
$\chi_1$=50 have been plotted in Figs. 2(d), 2(e) and 2(f) and for
$\chi_1$=70 in Figs. 2(g), 2(h) and 2(i), respectively. These figures
have been plotted as functions of $x$ and $y$ at $t_1$=300 as this
is the time at which the wave functions become stable. By comparing atomic vortex lattices (Fig. 2(a), Fig. 2(d)
and Fig. 2(g)) for $\chi_1$= 30, 50, 70 it is found that the number of
atomic vortices decreases with increase in the strength of
atom-molecular coupling and the number of visible atomic vortices
are 14, 10 and 6, respectively. However with the increase in
atom-molecular coupling strength both the distance between pair of
molecular vortices as well as the number of molecular vortices
decreases (Fig. 2(b), Fig. 2(e) and Fig. 2(h)) and the total number
of visible molecular vortices are 28, 20 and 12 for $\chi_1$ = 30,
50 and 70, respectively. This indicates that the density of molecular vortices is
twice the density of atomic vortices which satisfies the relation
that density of vortices is proportional to the mass of the
constituent particles as given by Feynman \cite{Feynman}. By comparing the vortex 
lattices for atoms and molecules combined (Figs. 2(c), 2(f) and 2(i)) it is found that 
initially for $\chi_1$=30 (Fig. 2(c)) vortices (atoms and molecules) are arranged as $CO_2$ 
like structure, two molecular vortices are on the two opposite sides of atomic vortices. 
However with increase in  the coupling strength ($\chi_1$) the distance between atomic and 
molecular vortices decreases  (Fig. 2(f)) and for the highest value of $\chi_1$= 70, both 
the atomic and molecular vortices almost overlap  (Fig. 2(i)). Hence the variation
of atom-molecular coupling strength controls the vortex lattice
structure significantly by changing the distance between atomic
and molecular vortices as well as the total number of vortices. To
explore the physics behind the effect of variation of
atom-molecular coupling strength on the vortex lattice structure
we have analyzed the dependence of vortex lattice formation on the
energy components of the coupled system, (i) energies
corresponding to atom-molecule coupling $E _c$, (ii)
atom-molecular interaction energy $E _{am}$ and (iii) the rotational
energy $E_{rot}$ and the values of these three energies have been
given in Table-1 for three different values of $\chi_1$ (30, 50
and 70). Atom-atom and molecule-molecule interaction energies are
not tabulated as these are lower than $E _c$, $E _{am}$ and $E
_{rot}$ by few orders of magnitude. From Table-1 it is found that
as $\chi_1$ increases, the attractive coupling energy $E _c$
dominates over the atom-molecular repulsive energy $E_{am}$ and hence
the ratio $|E_c/E_{am}|$ increases. In this system the attractive
interaction $E _c$ leads to decrease in the distance between
atomic and molecular vortices effectively squeezing the molecular
vortices towards the atomic vortices, whereas repulsive atom-molecular interaction leads to increase in the distance between
atomic molecular vortices effectively pushing the molecular
vortices away from the atomic vortices. Hence the interplay
between these two forces determine the distance between atomic and
molecular vortices. Therefore as the coupling energy $E _c$
increases and dominates over $E_{am}$, the distance between atomic
and molecular vortices as a result that between molecular vortices
decreases and finally overlaps as shown in Fig 2. In Table- 1, the
rotational energy has  been found to decrease with the
increase in the value of $\chi_1$. Although the rotational
frequency remains the same the variation in the parameter $\chi_1$
leads to variation in the wave functions of the system leading to
the decrease in the rotational energy. Decrease in rotational
energy restricts the spread of the lattices and hence leads to
decrease in the number of vortices as it is evident from Fig. 2.

Furthermore this result is supported by the nature of variation of
atomic and molecular angular momentum $l_z$ per atom (or molecule)
averaged over the whole condensate which is given as follows.
\begin{eqnarray} l_z={\int \psi^* L_z\psi dx dy \over \int |\psi
|^2 dx dy }
\end{eqnarray} It is found that with increase
in $\chi_1$, $l_z$ for both atoms and molecules decreases. Hence
the number of vortices in the lattice decreases. However in each
case, $l_z$ is found to be much greater than the half of visible
number of atomic and molecular vortices. Here values of atomic
(molecular) $l_z$ for $\chi_1$= 30, 50 and 70 are 14 (26), 12 (23)
and 10 (19), respectively. However from vortex lattices shown in
Figs. 2(a), 2(b), 2(d), 2(e), 2(g) and 2(h) for $\chi_1$ =30, 50
and 70 the number of visible atomic and molecular vortices are
$N_{v,a}$ ($N_{v,m}$)= 14 (28), 10 (20) and 6 (12), respectively.
According to Feynman’s rule $l_z$ should be . $N_v/2$ where $N_v$
is the total number of vortices. Hence the Feynman’s rule is
apparently violated here. To investigate this disparity we have
calculated the corresponding atomic and molecular phase
distributions from the final stable wave functions $\psi_a(x,y)$
and $\psi_m(x,y)$ and the phase varies from 0 to $2\pi$. The
corresponding atomic and molecular phase profiles are plotted in
Fig. 3. It is evident from the phase distribution that there are
some phase singularities in the central barrier region of the trap
other than the position of visible vortices. If $N_{h,a}$ and $N_{h,m}$
are the total number of phase singularities corresponding to the 
atomic and molecular hidden vortices then from Figs. 2(a), 2(b), 3(a) and 3(b) 
for $\chi_1$= 30: $l_z$ (atomic) = ($N_{v,a}$+$N_{h,a}$)/2= (14+14)/2= 14 and 
$l_z$ (molecular) = ($N_{v,m}$+$N_{h,m}$)/2= (28+24)/2 =26. Similarly from Figs. 2(d), 
2(e), 3(c) and 3(d) for $\chi_1$= 50: $l_z$ (atomic) = ($N_{v,a}$+$N_{h,a}$)/2=(10+14)/2=12 
and $l_z$ (molecular) = ($N_{v,m}$+$N_{h,m}$)/2 = (20+26)/2 =23. And from Figs. 2(g), 
2(h), 3(e) and 3(f) for $\chi_1$= 70: $l_z$ (atomic) =
($N_{v,a}$+$N_{h,a}$)/2=(6+14)/2=10 and $l_z$ (molecular) =
($N_{v,m}$+$N_{h,m}$)/2 = (12+26)/2 =19.

This indicates that the hidden vortices carry the angular momentum just like the visible vortices. 
Fig. 3 shows that the number of atomic and molecular hidden vortices remain nearly constant with 
the variation in $\chi_1$ unlike the visible vortices for which
the number diminishes with the increase in $\chi_1$. This may be
due to the fact that the magnitude of the atomic and molecular
wave functions in the forbidden region (the central barrier) is
much less than that within the trap. Hence the change in the
overlap of the wave functions in the forbidden region becomes less
effective to cause prominent change in the energy components
mentioned above. Atomic and molecular condensates are expanded
freely for  $t_1$= 1 and the density and the phase distributions
for $\chi_1= 30$ and 50, for $\Omega_1= 0.95$ and $\epsilon_1= 0$ are plotted
in Fig. 4. These figures reveal that free expansion make a couple
of new vortices visible along the symmetry axis of the trap which
are otherwise invisible. These new visible vortices must have been
originated from hidden vortices which are evident in phase
singularities.

\begin{table}
\caption {Comparison between different components of energies for
three different values of $\chi_1$, $\Omega_1$ =0.95 and
$\epsilon_1$ =0. Energies are in the units of
$\hbar\omega_\perp$.}
\begin{tabular}{lllllllll}
\hline\hline
&& $E_c$ && $E_{am}$ && $|E_c/ E_{am}|$ && $|E_{rot}|$\\
\hline
 $\chi_1$=30 && -0.142 && 0.111 && 1.280 && 5.050\\\\
\hline
 $\chi_1$=50 && -0.303 && 0.133 && 2.278 && 4.149\\\\
\hline
 $\chi_1$=70 && -0.382 && 0.116 && 3.293 && 2.772\\\\
 \hline\hline
\end{tabular}
\end{table}

\begin{table}
\caption {Comparison between different components of energies for
two different values of $\Omega_1$, $\chi_1$ =30 and $\epsilon_1$
=0, Energies are in the units of $\hbar\omega_\perp$.}
\begin{tabular}{lllllllll}
\hline\hline
&& $E_c$ && $E_{a,m}$ && $|E_c/ E_{a,m}|$ && $|E_{rot}|$\\
\hline
$\Omega_1$=0.85 && -0.278 && 0.304 && 0.914 && 2.663\\\\
\hline
 $\Omega_1$=0.9 && -0.211 && 0.209 && 1.009 && 4.127\\\\
\hline
 $\Omega_1$=0.94 && -0.161 && 0.134 && 1.201 && 4.893\\\\
\hline
 $\Omega_1$=0.95 && -0.142 && 0.111 && 1.279 && 5.050\\\\
 \hline\hline
\end{tabular}
\end{table}

\begin{table}
\caption {Comparison between different components of energies for
three different values of $\epsilon_1$, $\Omega_1$ =0.95 and
$\chi_1$ =50. Energies are in the units of $\hbar\omega_\perp$.}
\begin{tabular}{lllllllll}
\hline\hline
&& $E_c$ && $E_{a,m}$ && $|E_c/ E_{a,m}|$ && $|E_{rot}|$\\
\hline
 $\epsilon_1$= -0.5 && -0.298 && 0.148 && 2.013 && 3.194\\\\
\hline
 $\epsilon_1$= 0 && -0.303 && 0.133 && 2.278 && 4.149\\\\
\hline
 $\epsilon_1$= 2 && -0.306 && 0.101 && 3.029 && 4.882\\\\
\hline
 $\epsilon_1$= 5 && -0.309 && 0.082 && 3.768 && 5.134\\\\
 \hline\hline
\end{tabular}
\end{table}

\begin{figure}\begin{subfigure}[t]{0.25\textwidth}
\includegraphics[width=1.5in,height=1.8in,angle=0]{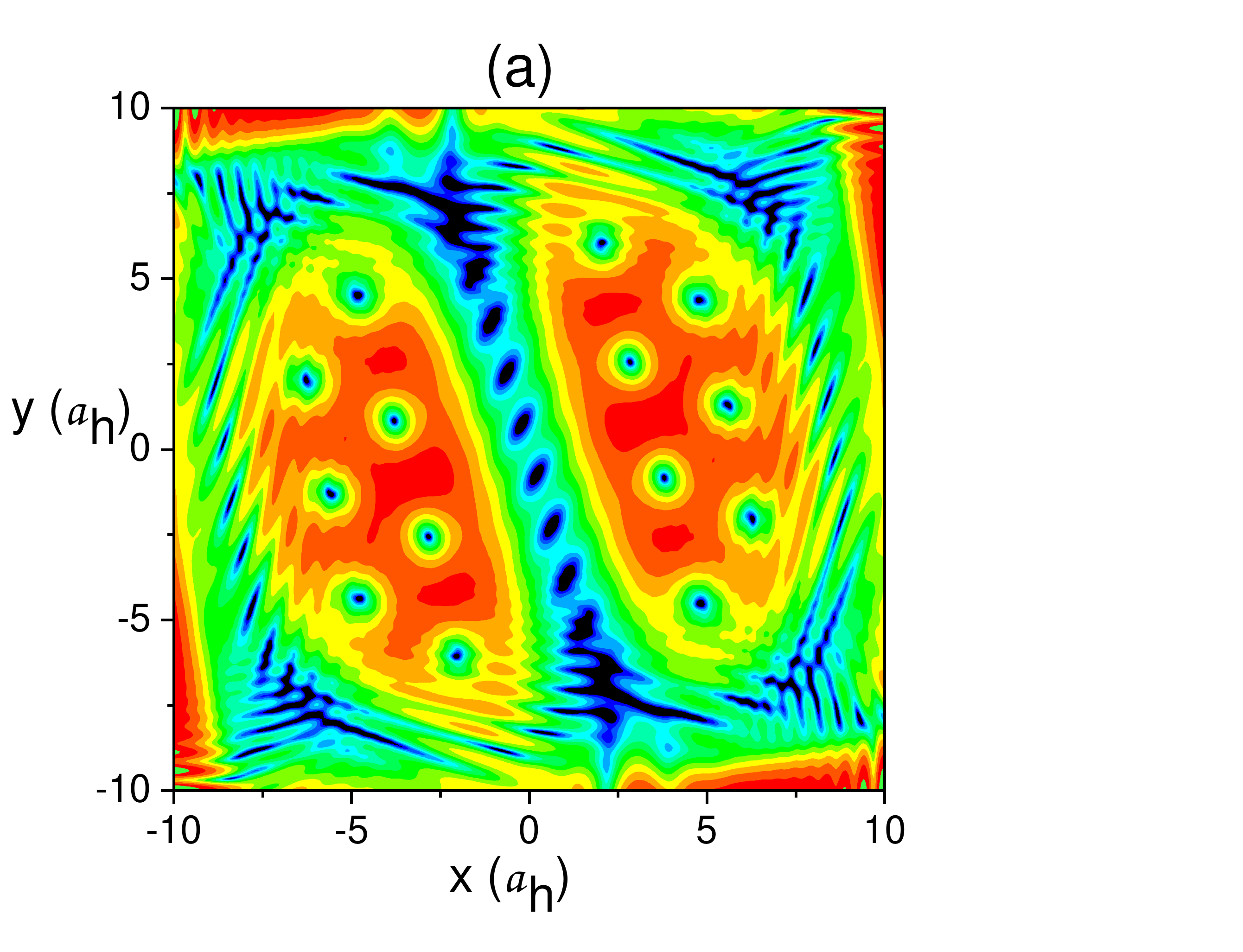}
\end{subfigure}\hfill
\begin{subfigure}[t]{0.25\textwidth}
\includegraphics[width=1.5in,height=1.8in,angle=0]{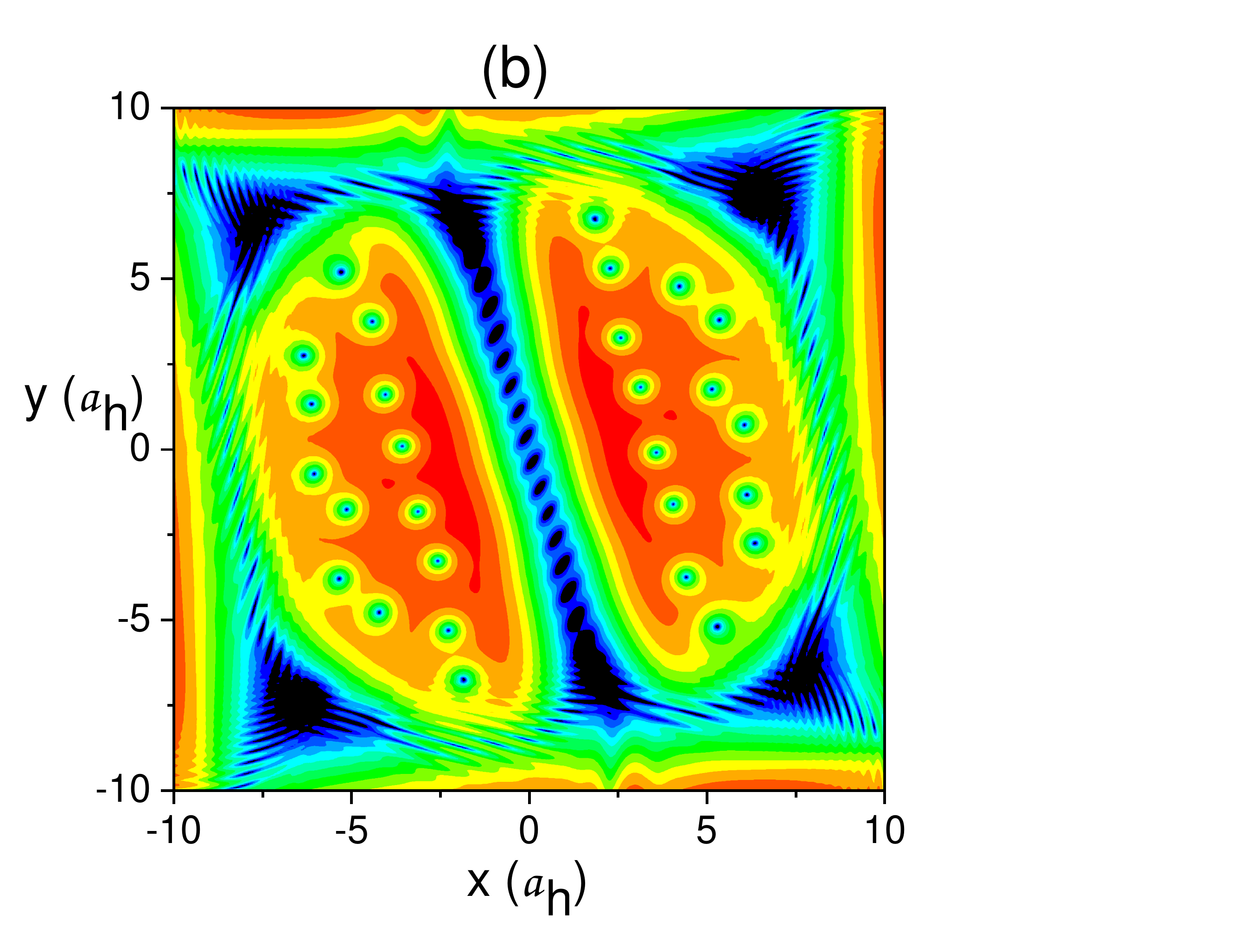}
\end{subfigure}\hfill
\begin{subfigure}[t]{0.25\textwidth}
\includegraphics[width=1.5in,height=1.8in,angle=0]{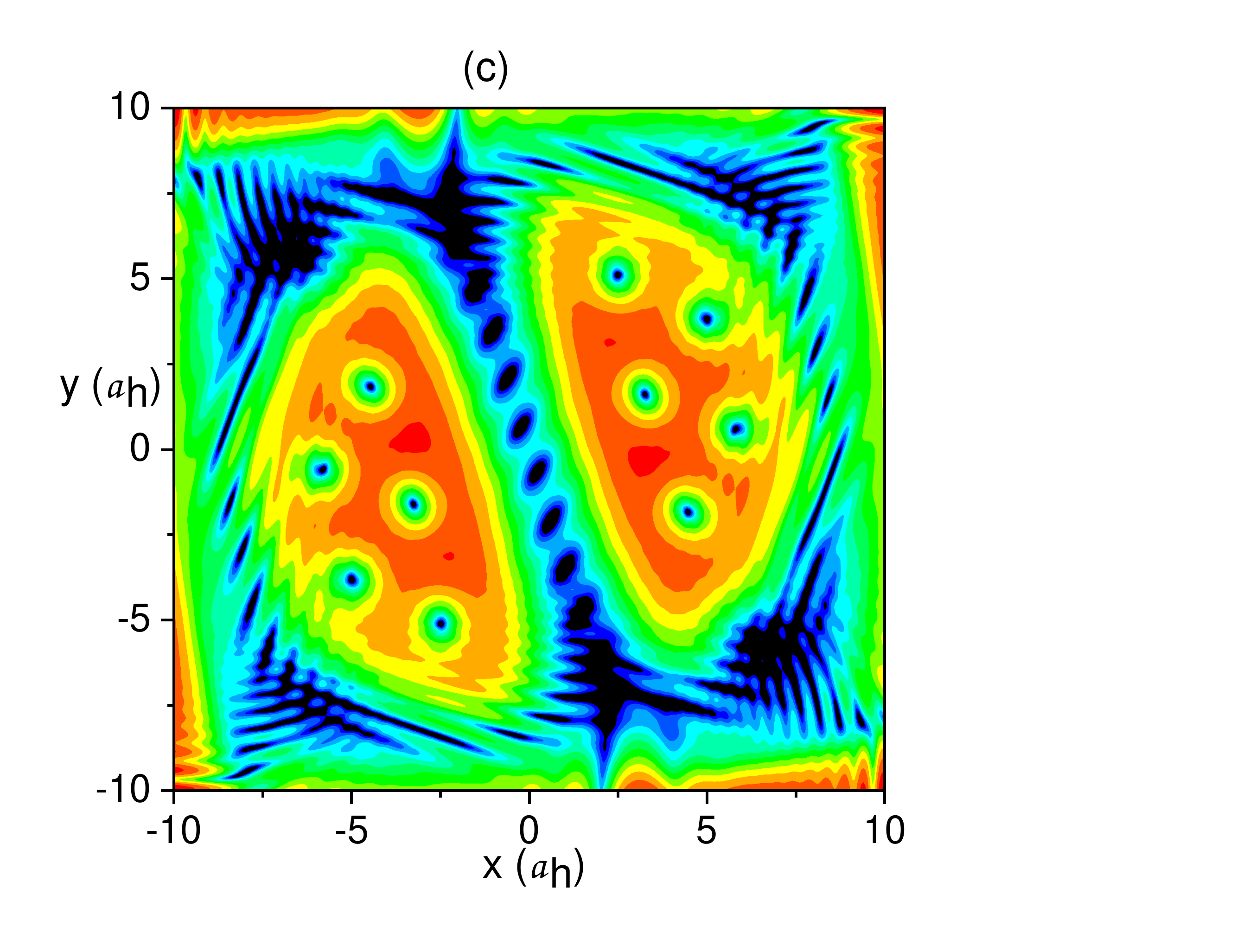}
\end{subfigure}\hfill
\begin{subfigure}[t]{0.25\textwidth}
\includegraphics[width=1.5in,height=1.8in,angle=0]{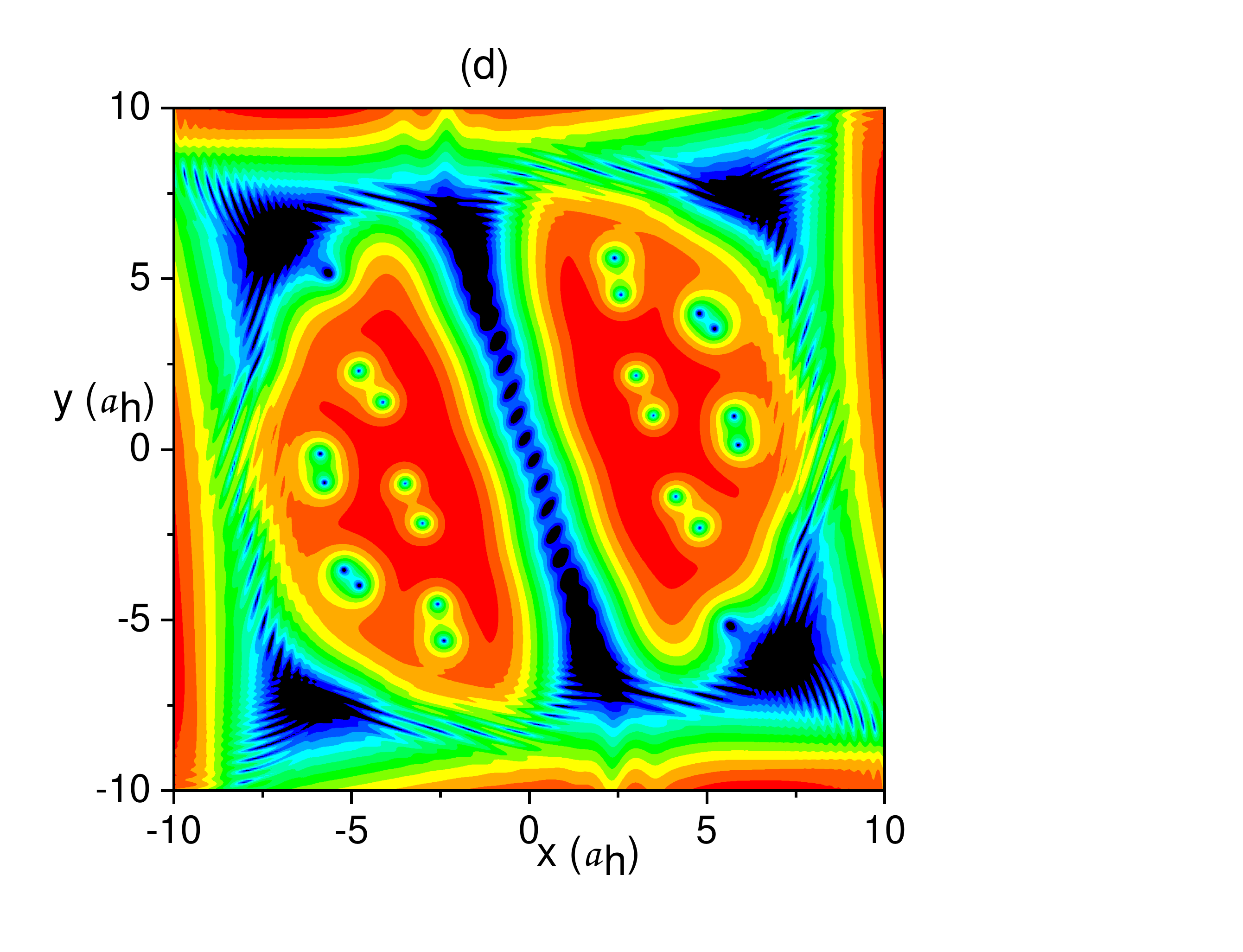}
\end{subfigure}\hfill
\begin{subfigure}[t]{0.25\textwidth}
\includegraphics[width=1.5in,height=1.8in,angle=0]{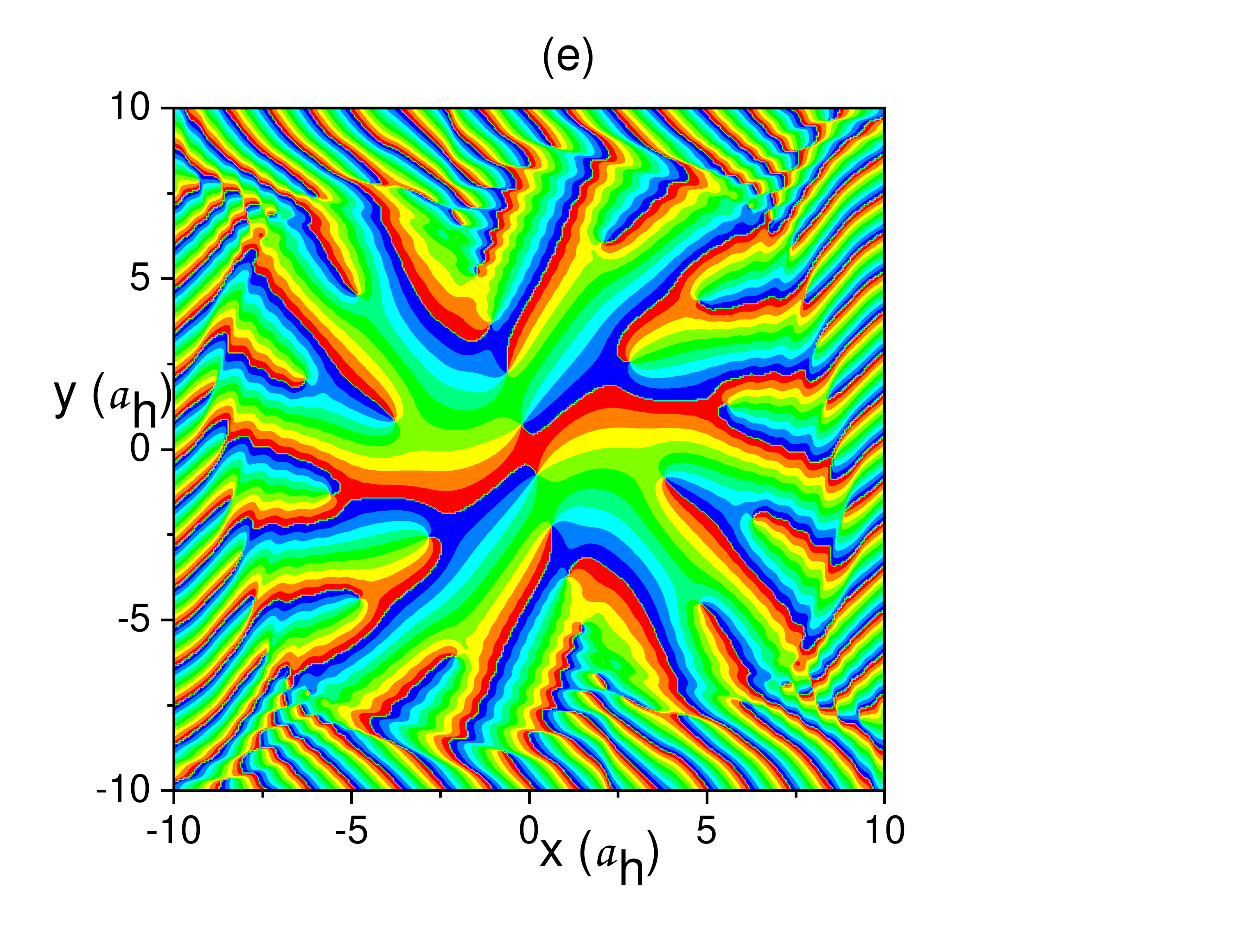}
\end{subfigure}\hfill
\begin{subfigure}[t]{0.25\textwidth}
\includegraphics[width=1.5in,height=1.8in,angle=0]{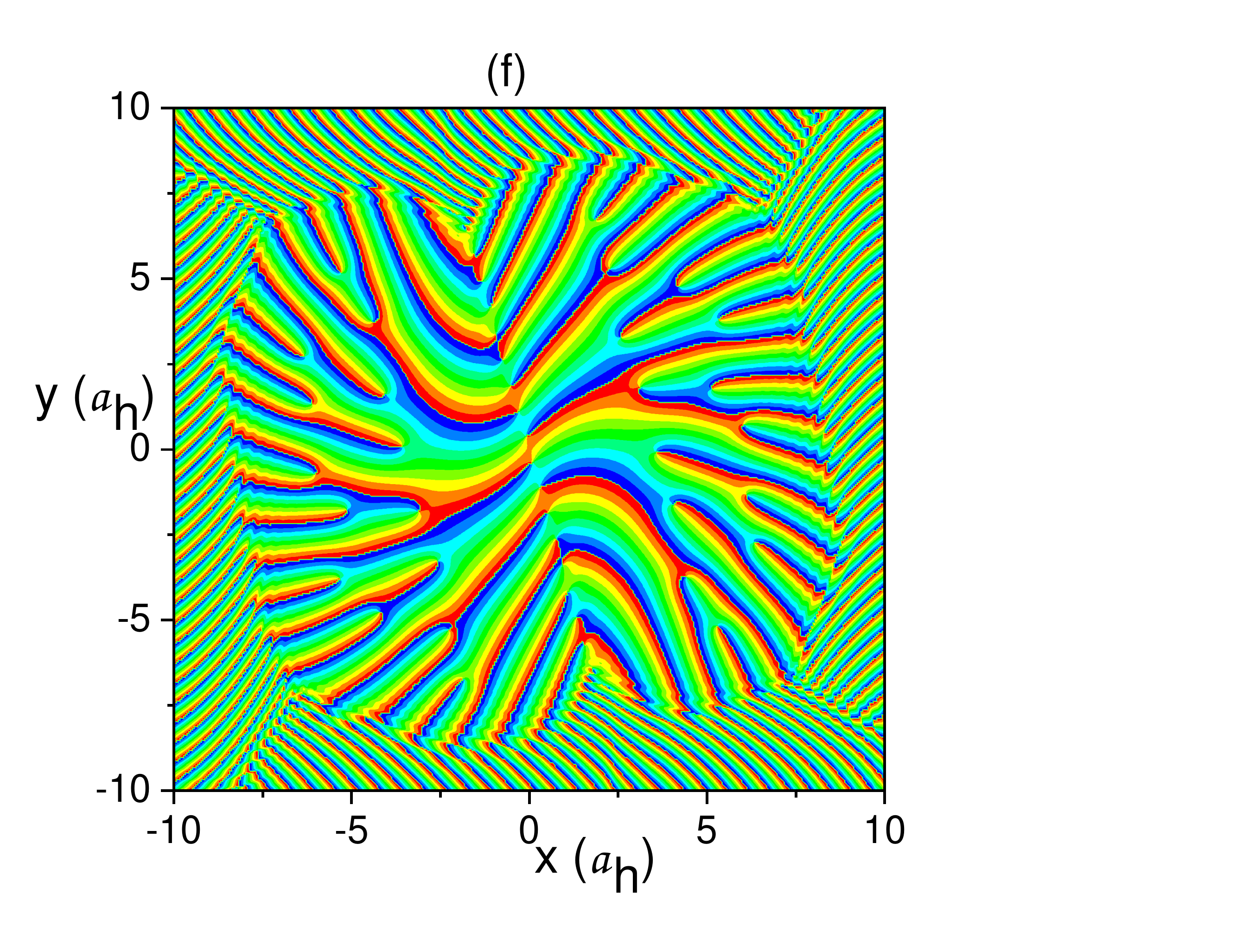}
\end{subfigure}\hfil
\begin{subfigure}[t]{0.25\textwidth}
\includegraphics[width=1.5in,height=1.8in,angle=0]{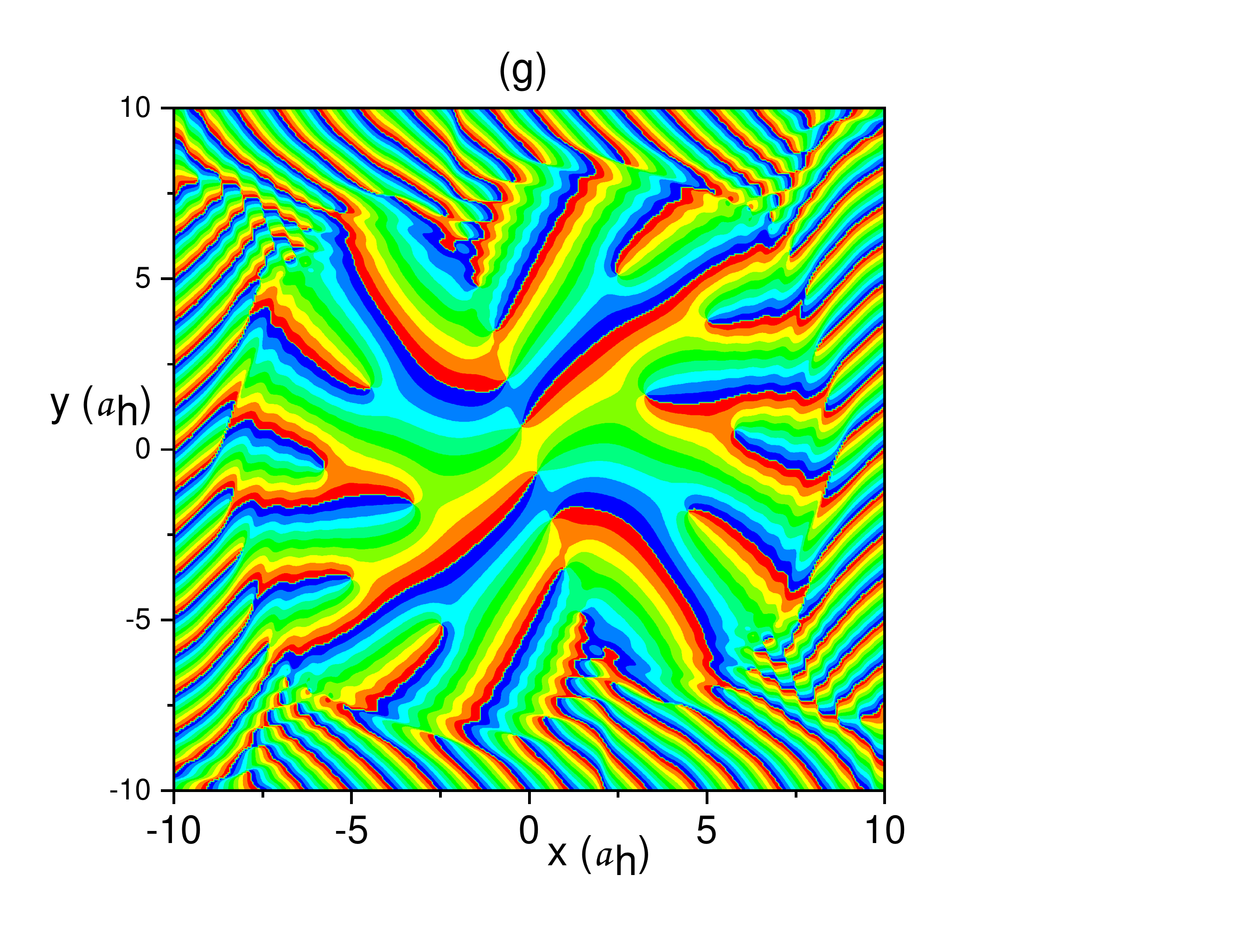}
\end{subfigure}\hfill
\begin{subfigure}[t]{0.25\textwidth}
\includegraphics[width=1.5in,height=1.8in,angle=0]{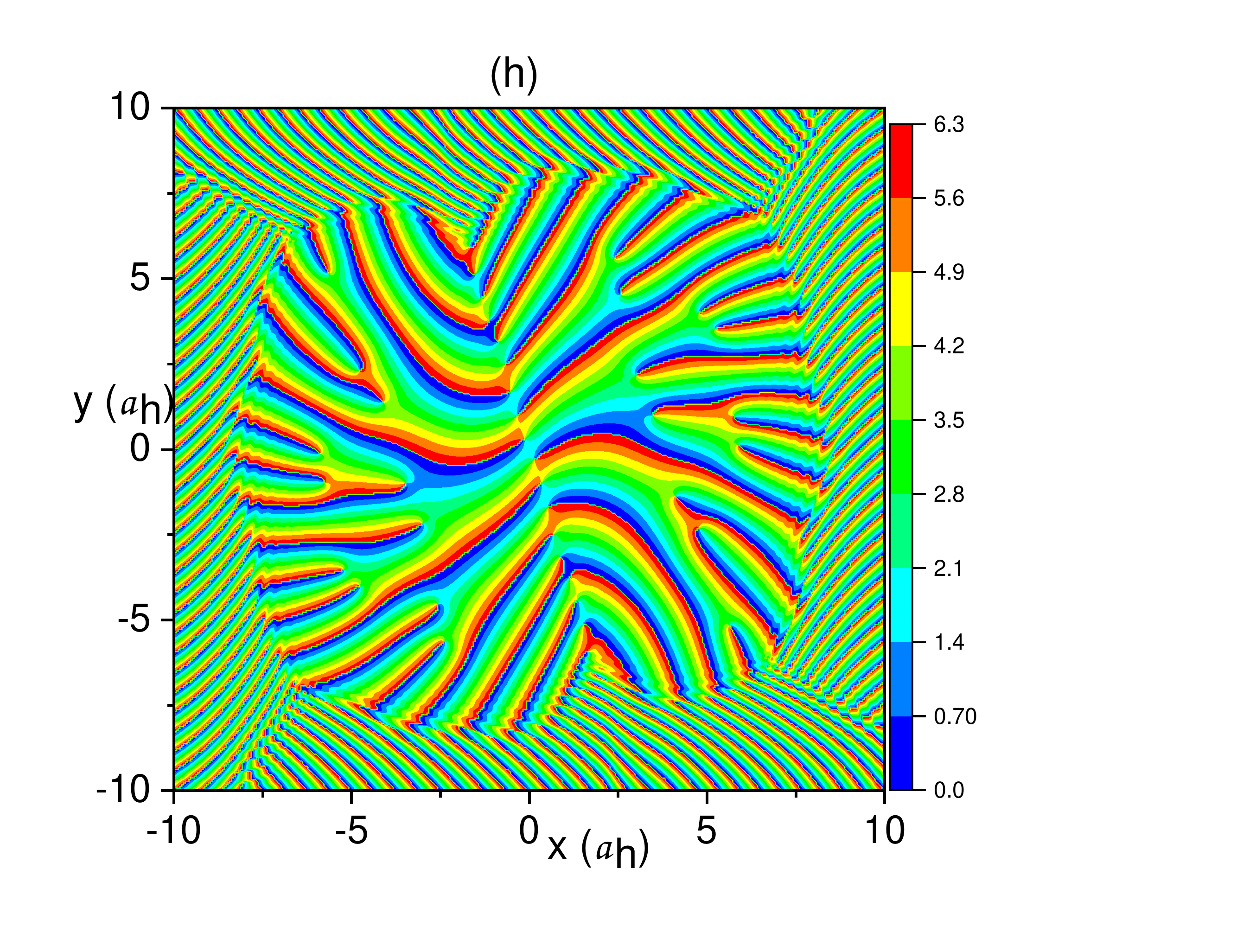}
\end{subfigure}
\begin{minipage}{5in}
\caption{(Color online) Density distribution of atomic [Figs. (a), (c)] and molecular [Figs. (b), (d)]
condensate along with phase profile of $\psi_a$ [Figs. (e), (g)] and $\psi_m$ [Figs.
(f), (h)] after free expansion for
$t_1$= 1. The corresponding parameters are $\Omega_1$= 0.95, $\epsilon_1$= 0
and $\chi_1$= 30 [ Figs. (a), (b), (e), (f)] and 50 [ Figs. (c), (d), (g), (h)]. Red color corresponds to
higher values and blue color corresponds to lower values. Darker
color corresponds to lower density. x and y are in the units of
$a_h$=$\sqrt{\hbar\over{2m\omega_\perp}}$.}
\end{minipage}
\end{figure}

\begin{figure}\begin{subfigure}[t]{0.3\textwidth}
\includegraphics[width=1.5in,height=1.8in,angle=0]{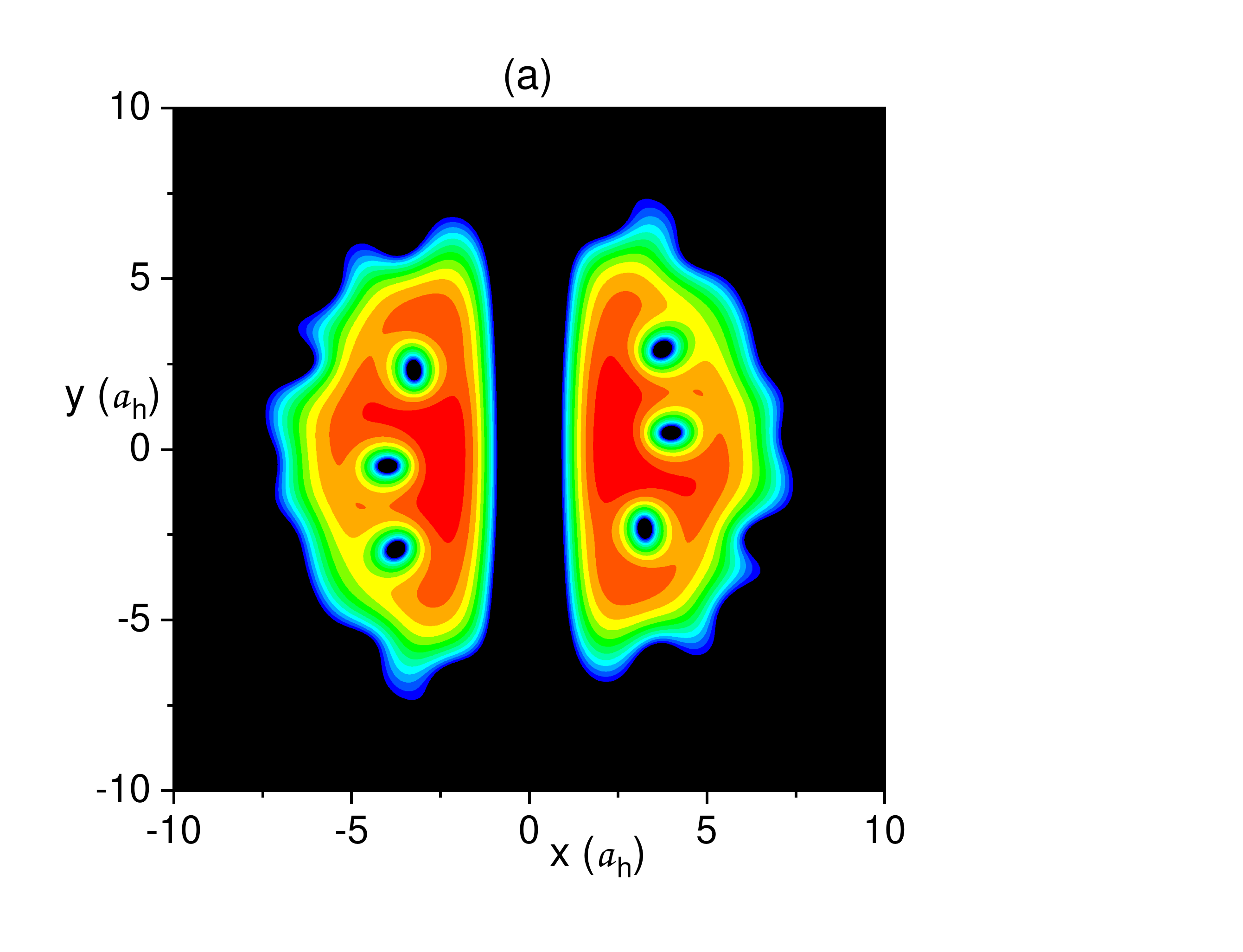}
\end{subfigure}\hfill
\begin{subfigure}[t]{0.3\textwidth}
\includegraphics[width=1.5in,height=1.8in,angle=0]{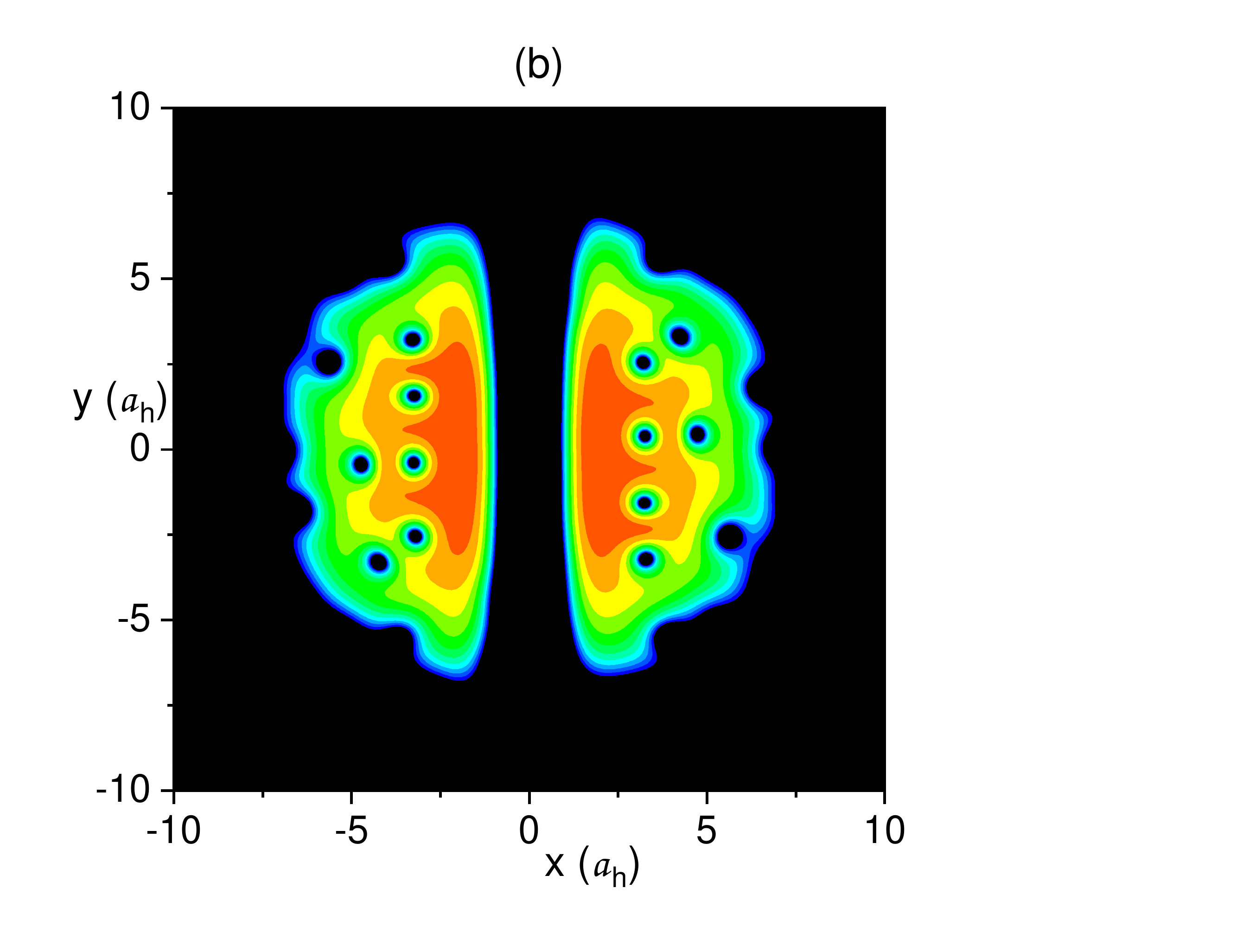}
\end{subfigure}\hfill
\begin{subfigure}[t]{0.3\textwidth}
\includegraphics[width=1.5in,height=1.8in,angle=0]{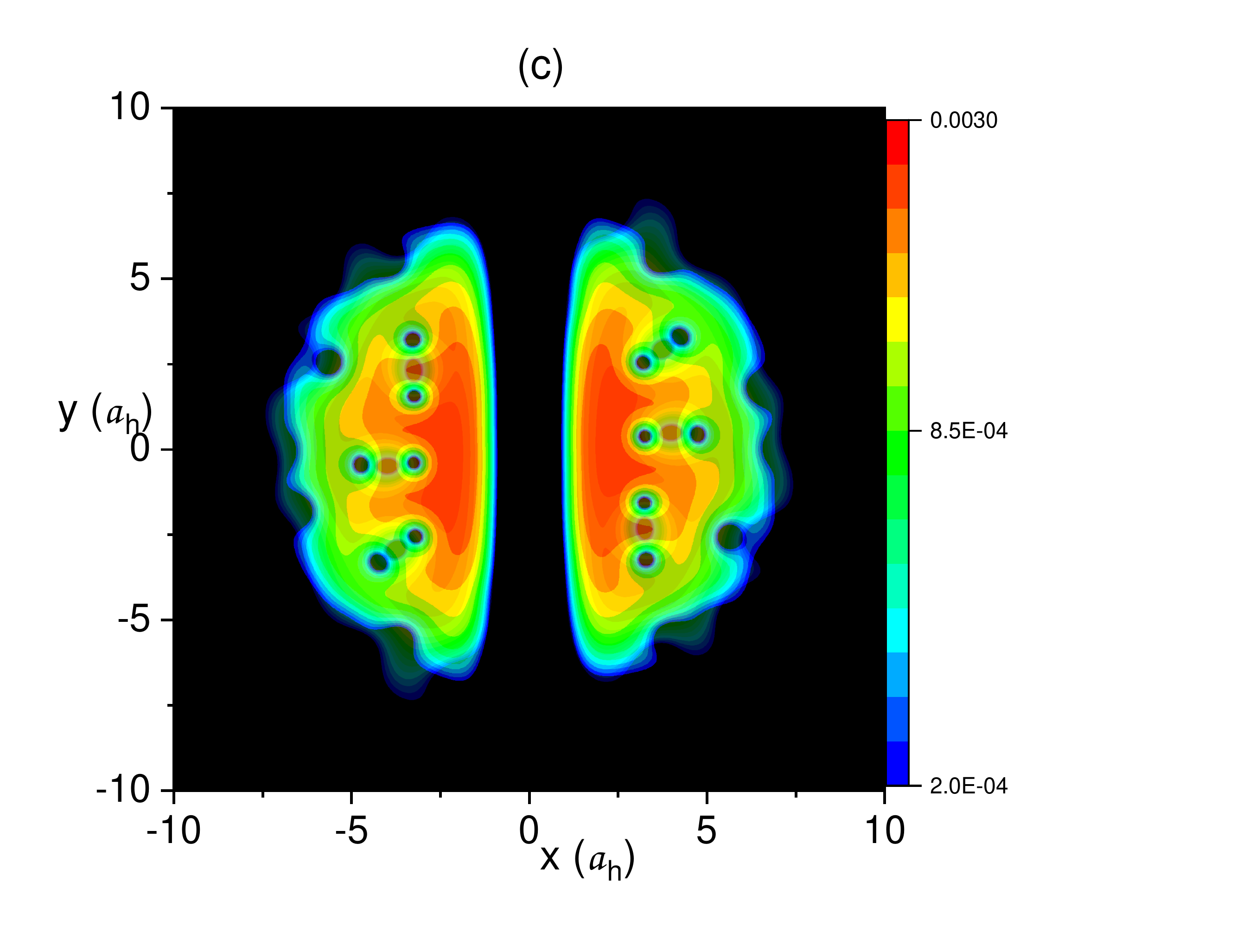}
\end{subfigure}

\begin{subfigure}[t]{0.3\textwidth}
\includegraphics[width=1.5in,height=1.8in,angle=0]{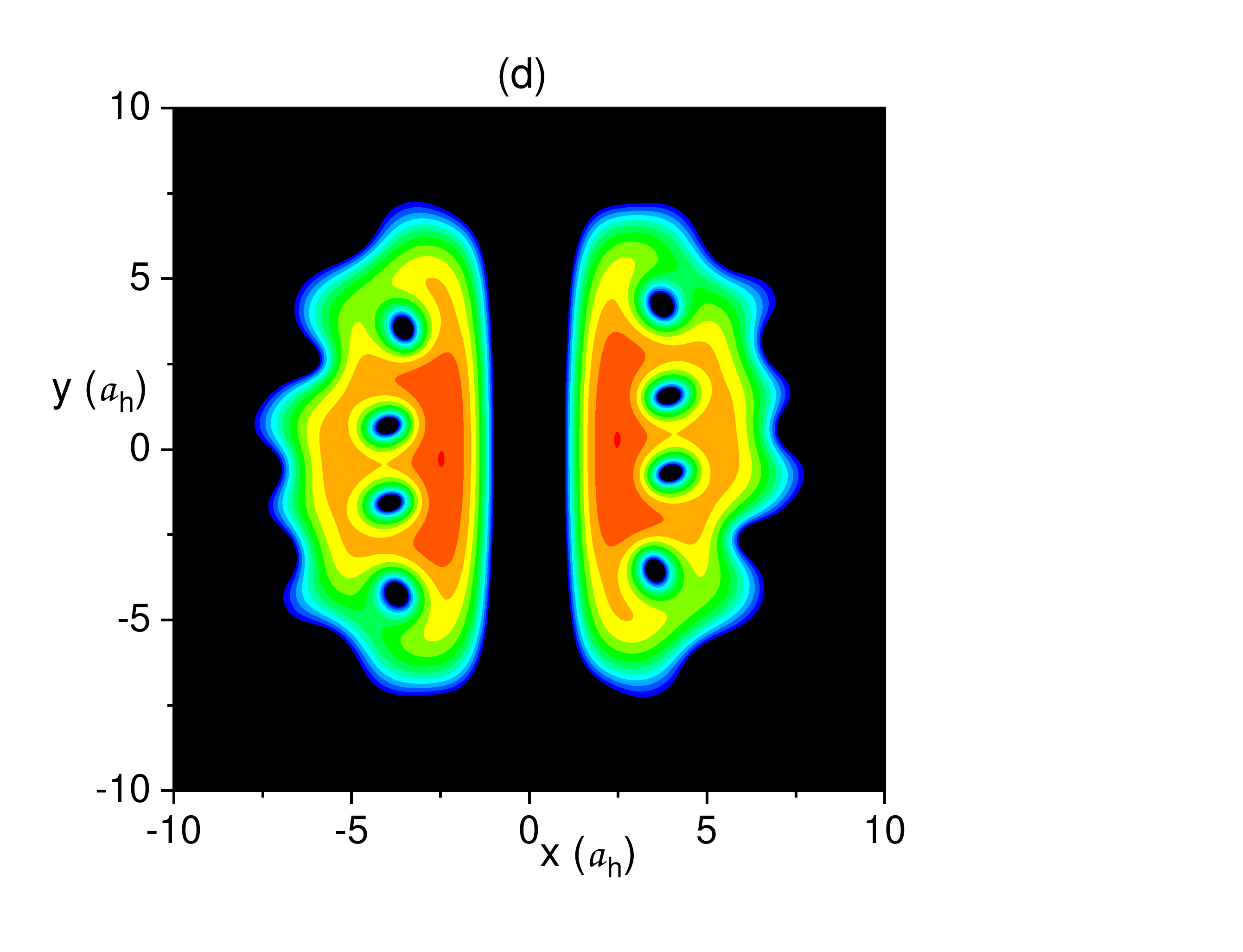}
\end{subfigure}\hfill
\begin{subfigure}[t]{0.3\textwidth}
\includegraphics[width=1.5in,height=1.8in,angle=0]{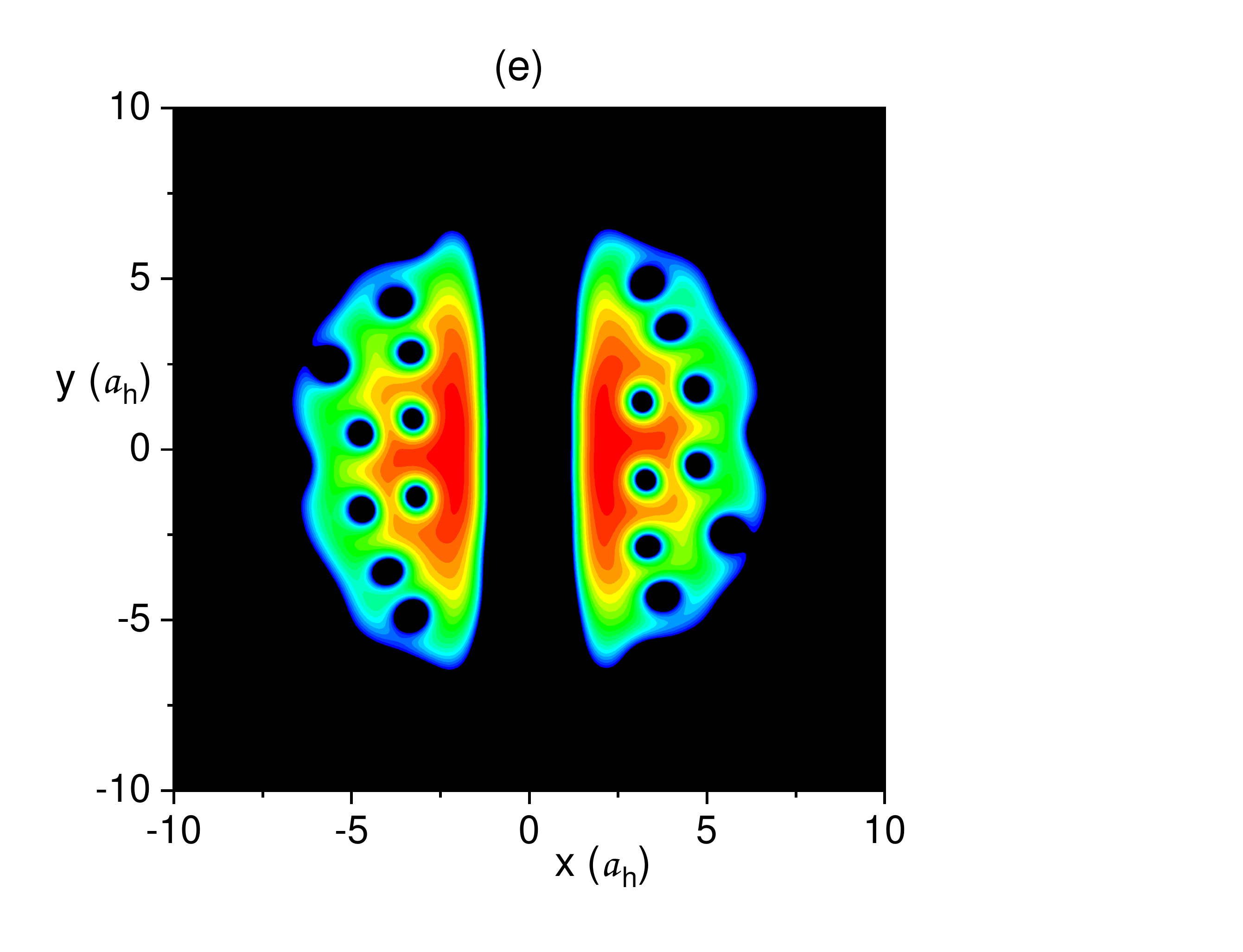}
\end{subfigure}\hfill
\begin{subfigure}[t]{0.3\textwidth}
\includegraphics[width=1.5in,height=1.8in,angle=0]{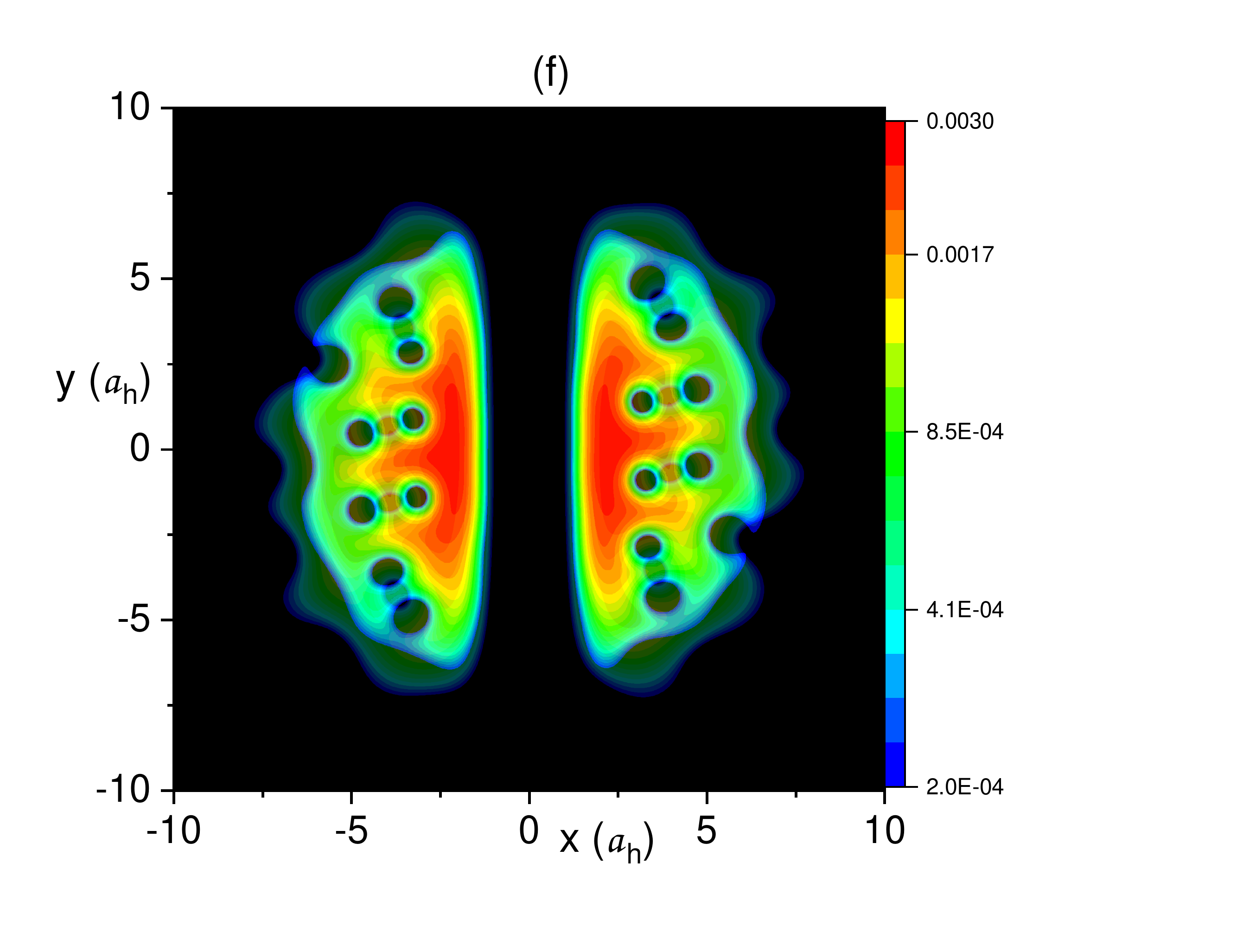}
\end{subfigure}

\begin{subfigure}[t]{0.3\textwidth}
\includegraphics[width=1.5in,height=1.8in,angle=0]{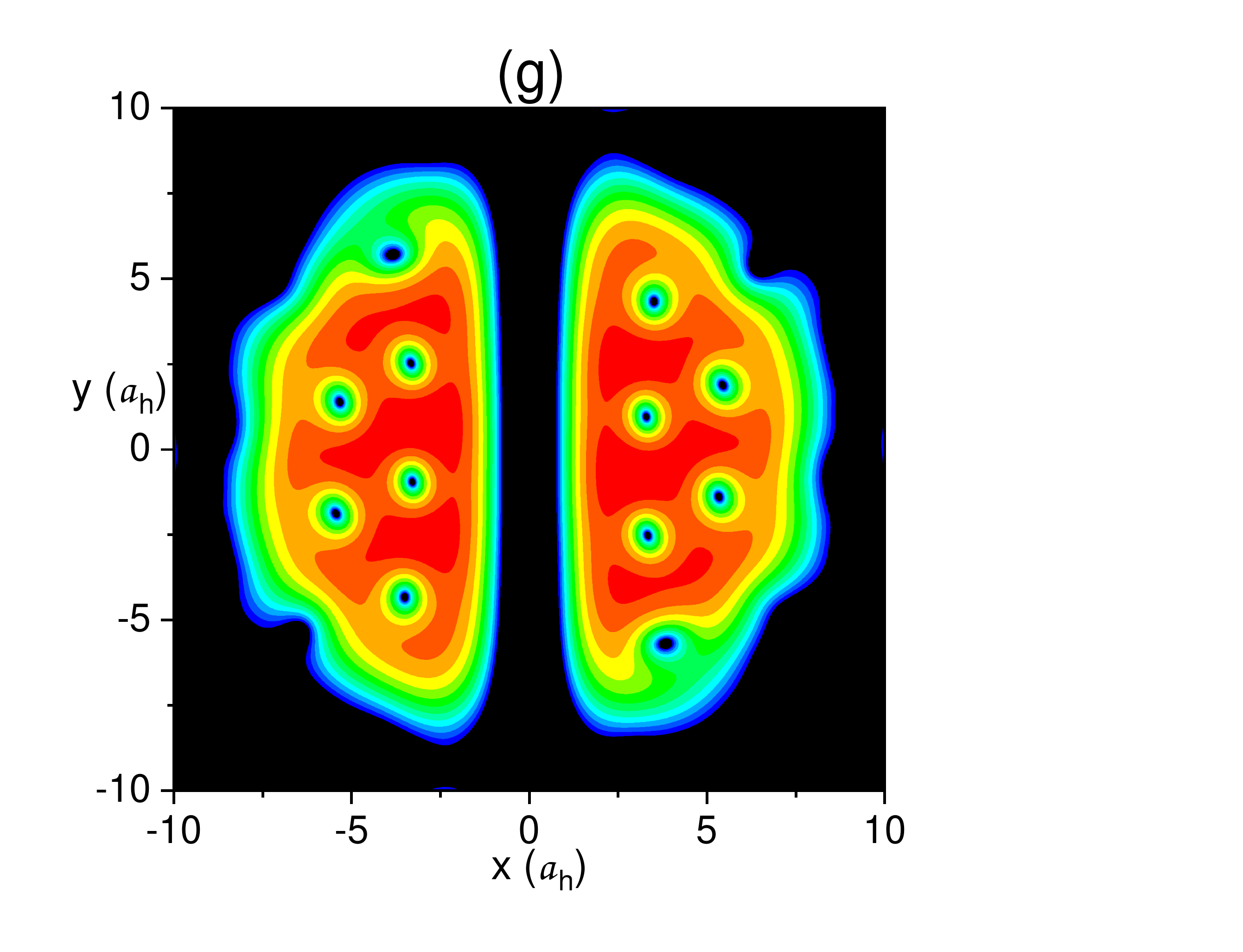}
\end{subfigure}\hfill
\begin{subfigure}[t]{0.3\textwidth}
\includegraphics[width=1.5in,height=1.8in,angle=0]{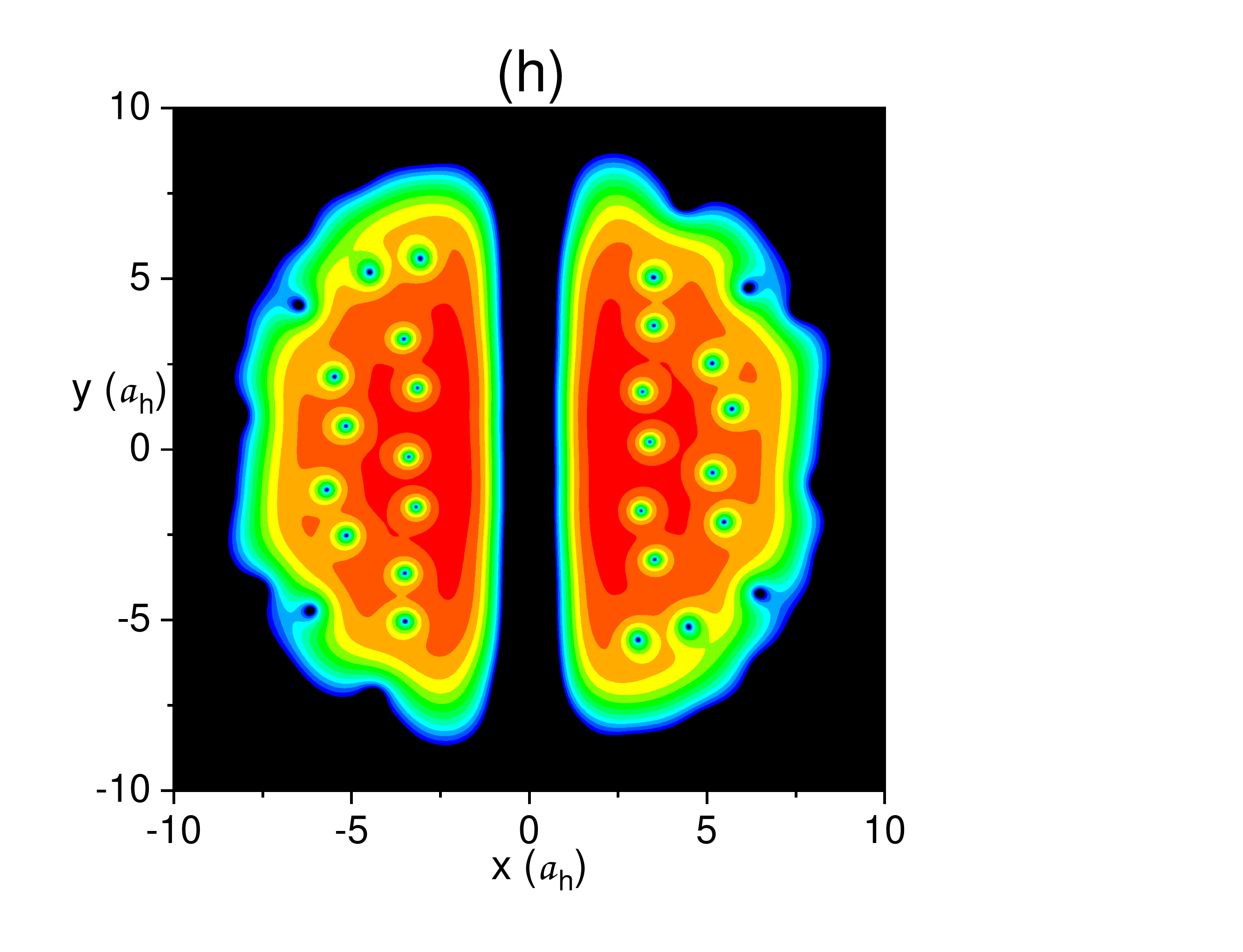}
\end{subfigure}\hfill
\begin{subfigure}[t]{0.3\textwidth}
\includegraphics[width=1.5in,height=1.8in,angle=0]{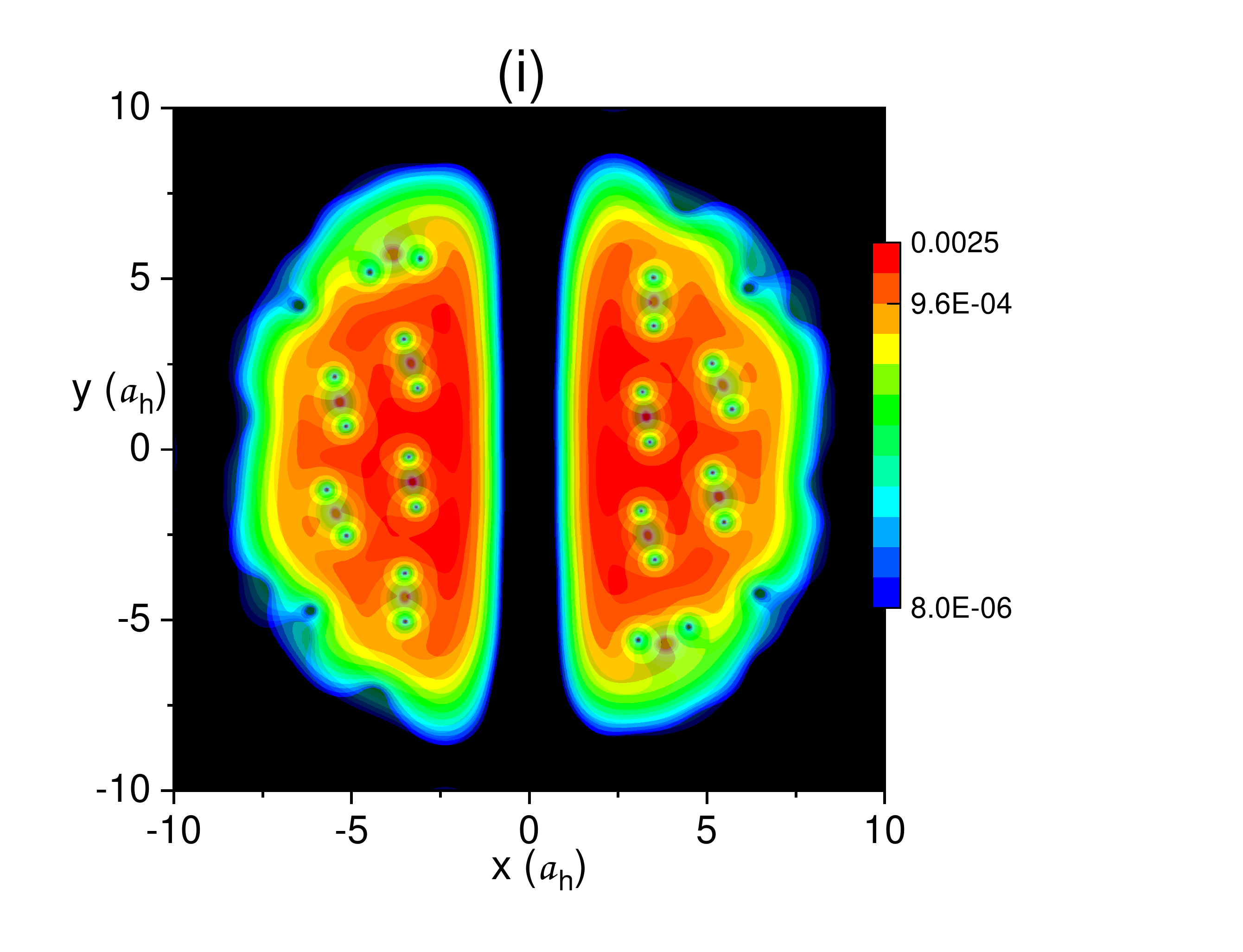}
\end{subfigure}

\begin{minipage}{5in}
\caption{(Color online) Density distributions for atomic [Figs. (a), (d), (g)], molecular [Figs. (b),
(e), (h)] and atomic with molecular [Figs. (c), (f), (i)] vortex lattice
configurations for
different rotation frequencies $\Omega_1$= 0.85 [Figs. (a), (b), (c)] , 0.9 [Figs.
(d), (e), (f)] and 0.94
[Figs. (g), (h), (i)] at $t_1$= 300 . $\chi_1$= 30 and $\epsilon_1$= 0.
Red color corresponds to higher densities and blue color
corresponds to lower densities. Darker color corresponds to lower density. $x$
and $y$ are in the units of $a_h$= $\sqrt{\hbar\over{2m\omega_\perp}}$.}
\end{minipage}
\end{figure}

\subsection{Effect of variation of rotational frequency on atom-molecular vortex lattices}

To study the effect of variation of rotational frequency
($\Omega_1$) on the vortex lattice structure we consider the fixed
values of atom-molecule coupling strength ($\chi_1$ =30) and
detuning ($\epsilon_1$=0). We varied the rotational frequency
($\Omega_1$) from values 0.85 to 0.95 to examine its effect on the
atomic-molecular vortex lattice structure and plotted the atomic
and molecular densities as functions of $x$ and $y$ in Fig. 5.
Previously in Figs. 2(a) and 2(b) we have shown the vortex lattice
formed for $\Omega_1$= 0.95, $\chi_1$= 30 and $\epsilon_1$= 0. In
Fig. 5 we have plotted the vortex lattices for atoms and molecules
for the values of $\Omega_1$= 0.85, 0.90 and 0.94. Figs. 5(a), 5(d)
and 5(g) show the atomic vortex lattices and Figs. 5(b), 5(e) and
5(h) show the molecular vortex lattices for $\Omega_1$= 0.85, 0.90
and 0.94, respectively. Combined vortex lattices for atoms with
molecules are shown in Figs. 5(c), 5(f) and 5(i) for values of
$\Omega_1$ = 0.85, 0.90 and 0.94, respectively. Corresponding phase
profiles for atoms and molecules are plotted in Fig. 6 for $\Omega_1$= 0.9. 
It is found that the number of
atomic and molecular vortices increases with the increase in
rotational frequency. With the increase in rotational frequency the
centrifugal force increases and it forces the condensate to spread
more towards the edges of the trap. The corresponding energy
components are given in Table-2. From Table- 2 it is found that
with the increase in rotational frequency the rotational energy
increases leading to increase in the number of atomic and
molecular vortices. However the values of $|E_c/E_{a,m}|$ increase
by very small amount on the increase in $\Omega_1$ and hence there
is no significant change in the relative distance between atomic
and molecular vortices. Numerically calculated values of $l_z$
(atomic) and  $l_z$ (molecular) are 11 and 20 respectively for
$\Omega_1$= 0.9. Whereas the number of real atomic ($N_{v,a}$) and
molecular ($N_{v,m}$) vortices are 8 and 16 [from Figs. 5(d) and
5(e)]. From the corresponding phase distribution of $\psi_a$ and $\psi_m$ plotted in Figs.
6(a) and 6(b), the number of hidden atomic ($N_{h,a}$) and
molecular ($N_{h,m}$) vortices are 14 and 24. This satisfies the
Feynman rule $l_z$ (atomic)= ($N_{v,a}$+$N_{h,a}$)/2= (8+14)/2=11
and $l_z$ (molecular)= ($N_{v,m}$+$N_{h,m}$)/2 = (16+24)/2 =20.  
Previously in section 3.1 we have shown that Feynman rule is satisfied for 
$\Omega_1$= 0.95 by considering both the visible and hidden vortices. We have also found 
that Feynman rule is well satisfied (results are not shown here) for other values of 
$\Omega_1$ considered here.

\subsection{Effect of variation of detuning on atom-molecular vortex lattices}

In order to study the effect of variation of Raman detuning
parameter $\epsilon_1$ on the formation of atomic and molecular
vortex lattices the density profiles for atoms ($|\psi_a|^2$),
for molecules ($|\psi_m|^2$) and for atom and molecule combined 
($|\psi_a|^2$ with $|\psi_m|^2$) are plotted as functions of $x$
and $y$ in Fig. 7 for three different values of Raman detuning
$\epsilon_1$= -0.5, 2 and 5, keeping $\chi_1$ and $\Omega_1$ fixed
at 50 and 0.95. Figs. 7(a), 7(d) and 7(g) show atomic lattices,
Figs. 7(b), 7(e) and 7(h) show molecular lattices and Figs.
7(c), 7(f) and 7(i) show lattices for atomic-molecular vortices combined for
three values of detuning $\epsilon_1$= -0.5, 2 and 5, respectively.
It is found that the number of visible atomic (Figs. 7(a), 7(d)
and 7(g)) and molecular (Figs. 7(b), 7(e) and 7(h)) vortices are 6
and 12 for $\epsilon_1$= -0.5, those are 14 and 28 for
$\epsilon_1$= 2, whereas those for $\epsilon_1$= 5 are 18 and 36
respectively. From Figs. 2(d) and 2(e) we find the number of
visible atomic and molecular vortices are 10 and 20 respectively
for $\epsilon_1$= 0. By comparing results from Fig. 7 and Fig. 2
it is found that the number of atomic and molecular vortices
increases gradually with the increase in detuning. Moreover if we
compare the structure of molecular lattices (Figs. 7(b), 7(e),
7(h) and 2(e)) it is evident that the distance between molecular
vortices gradually decreases with increase in detuning and finally
completely superpose on each other and become indistinguishable.
Hence two molecular vortices which were well separated for 
$\epsilon_1$= -0.5 and 0 (Fig. 7(b) and Fig. 2(e))
approach towards each with further increase in detuning
(Fig. 7(e)) and finally merge with each other (Fig. 7(h)). Since the atomic vortex is in between the two molecular
vortices (Fig. 7(c)), with the increase in detuning two molecular
vortices approach towards the atomic vortex and completely overlap
with each other and also with the atomic vortex (Fig. 7(i)). Hence in the projection on the $x-y$ plane
three overlapped vortices look like a single vortex. To analyze
these two features of vortex lattices: (i) increase in the number
of atomic and molecular vortices and (ii) decrease in the
distance between atomic and molecular vortices leading to the decrease in distance between
two molecular vortices with the
increase in the detuning, corresponding energy components for
$\epsilon_1$ = -0.5, 0, 2 and 5 are given in Table- 3. From Table-
3, we find that the rotational energy $E_{rot}$ and the ratio of
atom-molecular coupling energy and atom-molecular interaction
energy $|E_c/ {E_{a,m}}|$ both increase with $\epsilon_1$. The
increase in the rotational energy gives rise to more spread of the
condensate towards the edge of the trap leading to the increase in
the number of visible vortices. Whereas increase in $|E_c/
{E_{a,m}}|$ with the increase in $\epsilon_1$ leads to squeezing
of the vortices i.e. decrease in relative distance between atomic and
molecular vortices leading to decrease in the relative distance between two molecular
vortices. Hence the atomic and molecular vortices which are
separated for $\epsilon_1$= -0.5 (Fig. 7(c)) and 0 (Fig. 2(f)),
tend to merge with increase in $\epsilon_1$ [Fig. 7(f)] and
finally they superpose on each other at $\epsilon_1$= 5 (Fig.
7(i)). In Figs. 3(b) and 3(e) we have shown the phase distribution
of $\psi_a$ and $\psi_m$ for $\epsilon_1$= 0 (for $\chi_1$= 50 and
$\Omega_1$=0.95). In Fig. 8, we have plotted the phase
distribution of $\psi_a$ and $\psi_m$ for $\epsilon_1$= -0.5 (Figs. 8(a) and
8(b)), for $\epsilon_1$= 2 (Figs. 8(c) and 8(d)) and for $\epsilon_1$= 5 
(Figs. 8(e) and 8(f)). In section 3.1
we have shown how the Feynman rule is satisfied for $\epsilon_1$= 0
by considering hidden vortices (counting singularities along the
central barrier region in phase distribution) with visible vortices (from
the density distribution) both for atomic and molecular vortex
latices. Here values of atomic
(molecular) $l_z$ for $\epsilon_1$= -0.5, 2 and 5 are 10 (19), 14 (24) and 16 (26), 
respectively. However from vortex lattices shown in
Figs. 7(a), 7(b), 7(d), 7(e), 7(g) and 7(h) for $\epsilon_1$ = -0.5, 2
and 5 the number of visible atomic and molecular vortices are
$N_{v,a}$ ($N_{v,m}$)= 6 (12), 14 (28) and 18 (36), respectively.
If $N_{h,a}$ and $N_{h,m}$
are the total number of phase singularities corresponding to the 
atomic and molecular hidden vortices then from Fig. 8, $N_{h,a}$= 14 and $N_{h,m}$= 26 for 
all the detunings  $\epsilon_1$= -0.5, 2 and 5, respectively. Hence $l_z$ (atomic) should 
be equal to ($N_{v,a}$+$N_{h,a}$)/2= (6+14)/2= 10 and $l_z$ (molecular) should be equal 
to  ($N_{v,m}$+$N_{h,m}$)/2= (12+26)/2= 19 for $\epsilon_1$= -0.5. This shows Feynman rule 
is satisfied for both the atomic and molecular vortices for $\epsilon_1$= -0.5. However for $\epsilon_1$= 2 and 5 (from Figs. 7(d), 7(e), 
8(c), 8(d), 7(g), 7(h), 8(e) and 8(f)) it is found that for atomic vortices ($N_{v,a}$+$N_{h,a}$)/2= 
(14+14)/2=14 and ($N_{v,a}$+$N_{h,a}$)/2=(18+14)/2=16, respectively whereas for molecular vortices 
($N_{v,m}$+$N_{h,m}$)/2= (28+26)/2= 27 and ($N_{v,m}$+$N_{h,m}$)/2= (36+26)/2= 31, respectively. 
Therefore we find that the Feynman's rule is well satisfied for atomic vortices for all the values of 
detuning considered here. However for molecular vortices although the Feynman's rule is satisfied for 
smaller values of detuning ($\epsilon_1$= -0.5 and 0), it deviates for larger values of detuning 
($\epsilon_1$= 2 to 5) and the deviation from Feynman rule for molecular vortex lattice increases with increase in detuning.
It is found from Figs. 7 that
with increase in detuning from zero, carbon-dioxide type structure
of atomic and molecular vortices is overwritten by overlapping of
molecular vortices with each other as well as with the atomic
vortices in the middle. The number of overlapping molecular
vortices increases with increase in detuning ($\epsilon_1$= 2 to 5)
leading to increase in the disagreement of Feynman rule for
molecular vortices. In the definition of average angular momentum
[Eq.(21)] it is shown that calculation of $l_z$ involves spatial
integration. Therefore when the molecular vortices superpose on
each other it becomes indistinguishable for integration over
space, effectively leading to a lower value of $l_z$. As a result
the discrepancy between value of $l_z$ and half of the total
number of vortices increases. However while counting the total
number of molecular vortices, if the two superposed molecular
vortices are counted as single molecular vortex effectively the
difference between $l_z$ and half of the total number of vortices
reduces to be one or zero. 

 This is a new feature of molecular vortex lattices, that the
dynamics of molecular vortices i.e. the distance between
two molecular vortices can be controlled by varying the
Raman detuning. This can be experimentally observed by in situ imaging \cite{Wilson} 
of molecular vortex lattices in atom molecular coupled BEC (which are coherently 
coupled by two-photon Raman photoassociation \cite{Donley}) by varying the 
Raman detuning. When the Raman detuning is varied
from lower to higher values the dynamics of molecular
vortices i.e. approach of two molecular vortices towards
each other and finally merging on each other to be
indistinguishable in the molecular vortex lattices can be
observed. Moreover in atom-molecular combined vortex lattice system 
one can observe that from $CO_2$ like structure of atomic with molecular vortices will 
start squeezing towards each other and finally three vortices will merge with the increase 
in Raman detuning.

\begin{figure}\begin{subfigure}[t]{0.4\textwidth}
\includegraphics[width=1.5in,height=1.8in,angle=0]{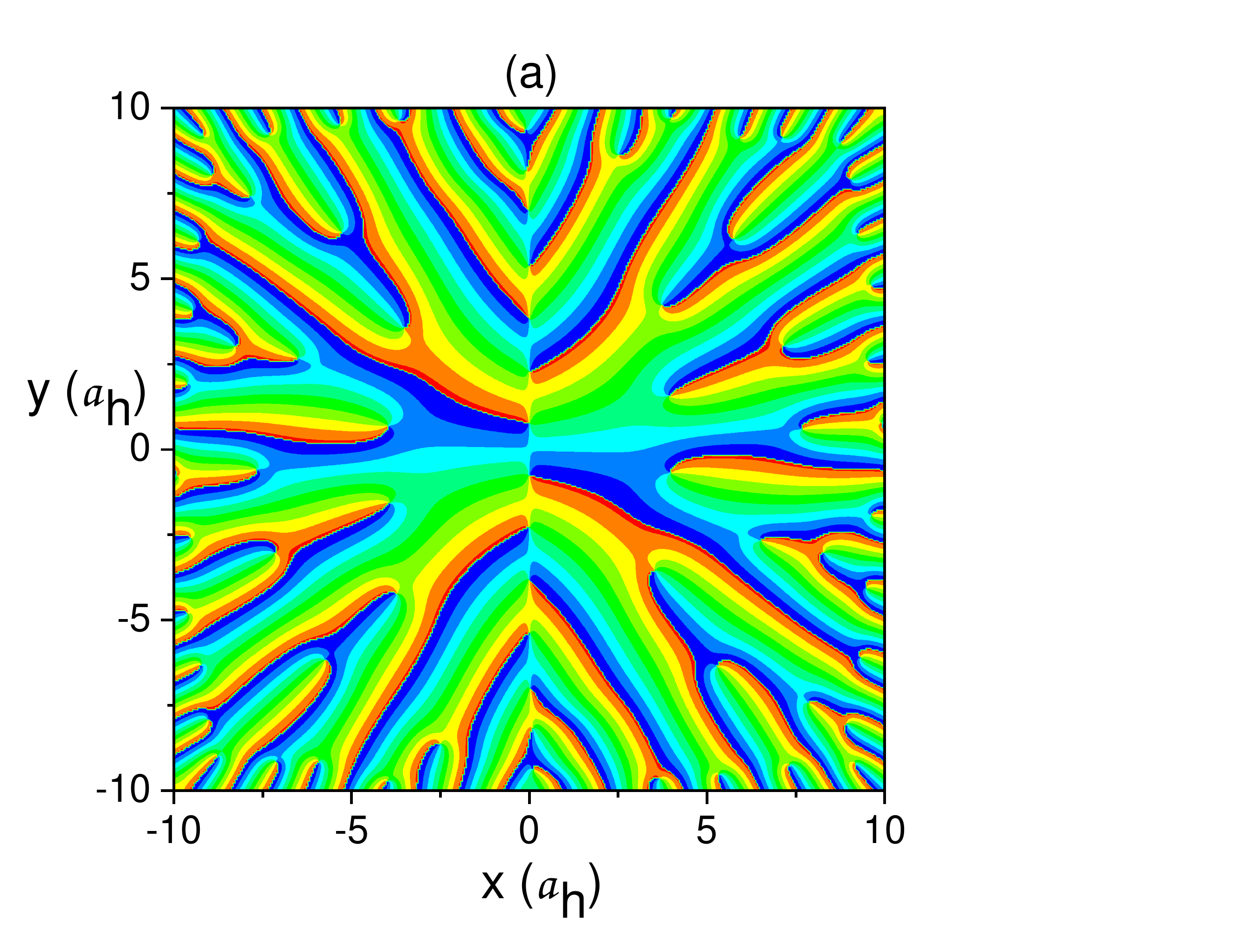}
\end{subfigure}\hfill
\begin{subfigure}[t]{0.4\textwidth}
\includegraphics[width=1.5in,height=1.8in,angle=0]{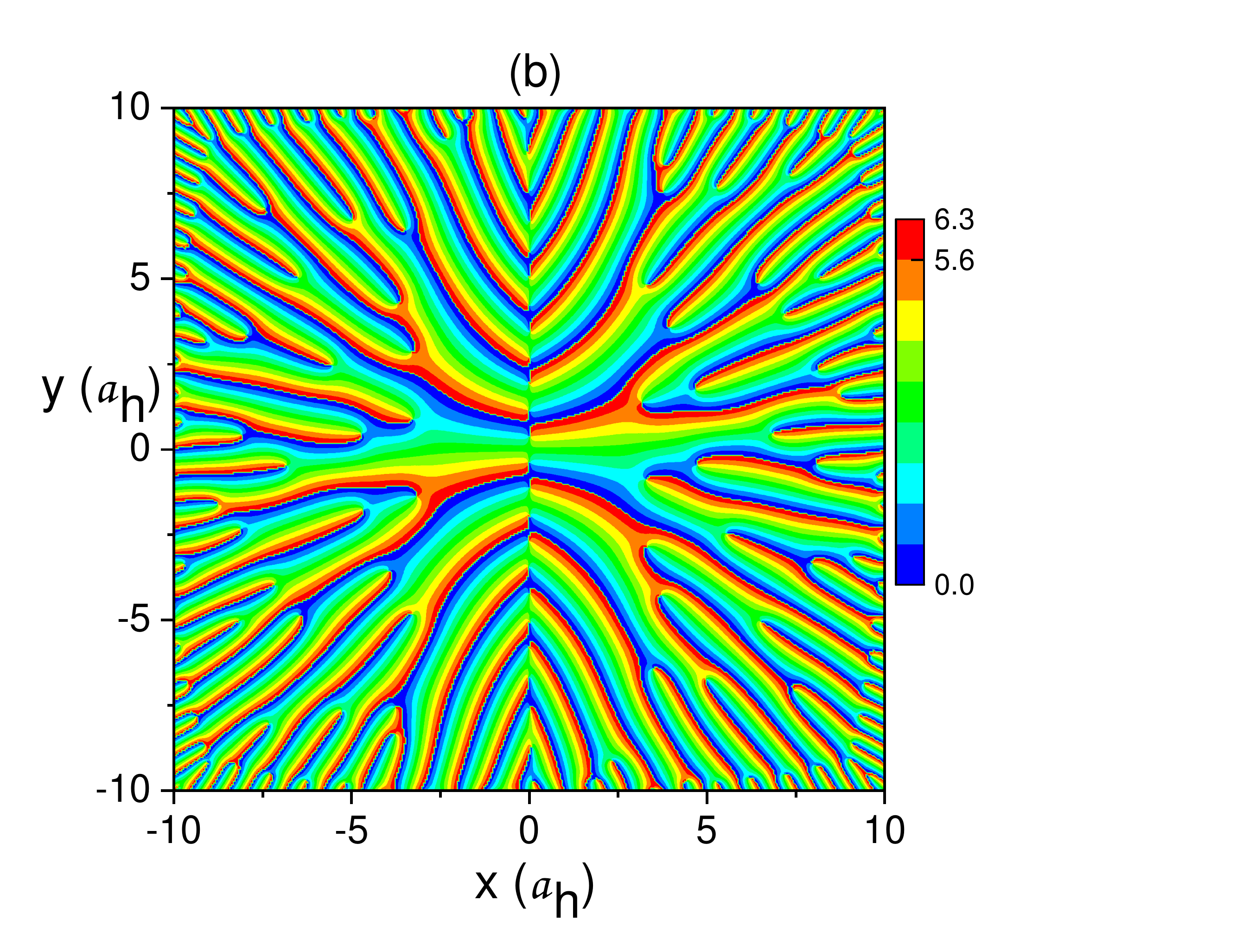}
\end{subfigure}

\begin{minipage}{5in}
\caption{(Color online) Phase profile of $\psi_a$ [Fig. (a)] and
$\psi_m$ [Fig. (b)] for rotation frequency $\Omega_1$= 0.9 with
$\chi_1$= 30 and $\epsilon_1$= 0 at $t_1$= 300. Red color
corresponds to higher values and blue color corresponds to lower
values. Darker color corresponds to lower phase. x and y are in
the units of $a_h$= $\sqrt{\hbar\over {2m\omega_\perp}}$.}
\end{minipage}
\end{figure}

\section{Conclusions}
In conclusion, we have studied in detail the atomic-molecular
vortex lattice formation of a rotating BEC in a double well trap
by numerically solving the two-dimensional coupled GP like
equations. We considered two photon Raman photoassociation for
atom to molecule conversion. We found that both the atomic and
molecular vortices those are hidden in density distribution could
be revealed in phase distribution and adding the number of hidden
vortices (along the barrier of the double well trap potential)
with the visible vortices in the trap the Feynman rule is
satisfied in general, however for large detuning a deviation from
the Feynman rule for molecular vortices is observed and this can
be attributed to the structural change in the vortex lattice i.e.
carbon-dioxide type structure is overwritten by overlapping of molecular 
vortices with atomic vortices in the middle and hence three overlapped 
vortices two molecular and one atomic appear to be a single vortex. 
The number of vortices in
molecular BEC is always twice than that of atomic BEC as the mass
of a molecule is twice of that of an atom and number of vortices
is proportional to the constituent masses. We explored the
dependence of vortex lattice structure on the system parameters
such as relative strength of atom-molecular coupling and
atom-molecular interaction, rotational frequency and the Raman
detuning. Our investigation reveals that the competition between
atom-molecular coupling strength and atom-molecular interaction
controls the spacing between atomic and molecular vortices and the
rotational energy controls the number of atomic and molecular
vortices in the vortex lattices. When coupling overpowers
atom-molecular scattering, distance between atomic and molecular
vortices decreases and tends to overlap with each other. Whereas
when the rotational energy increases it leads to increase in the
spread of the lattices towards the edge of the trap as well as the
number of vortices in the vortex lattices. The Raman detuning
parameter also controls the coupled vortex system by changing the
relative strength of atom-molecular coupling and atom-molecular
interaction as well as rotational energy and hence it controls
both the spacing between vortices and the number of vortices. To
explain all the features obtained in vortex lattices by varying
the system parameters such as relative strength of atom-molecular
coupling and the atom molecular interaction, rotational frequency
of the trap and the Raman detuning we have analyzed different
energy components as a function of these parameters. It is found
that although the energy components, angular momentum and number
of visible vortices of the coupled condensates largely depend on
the atom-molecule coupling, rotation frequency and Raman detuning,
the number of atomic and molecular hidden vortices those are
evident in the phase singularities along the central barrier
region of the trap remains nearly unaffected by the variation of
these parameters. Therefore this atomic-molecular vortex system in
a rotating double well trap offers some effective tools to control
the coupled vortex dynamics confining the system in the stability
domain.

\begin{figure}\begin{subfigure}[t]{0.3\textwidth}
\includegraphics[width=1.5in,height=1.8in,angle=0]{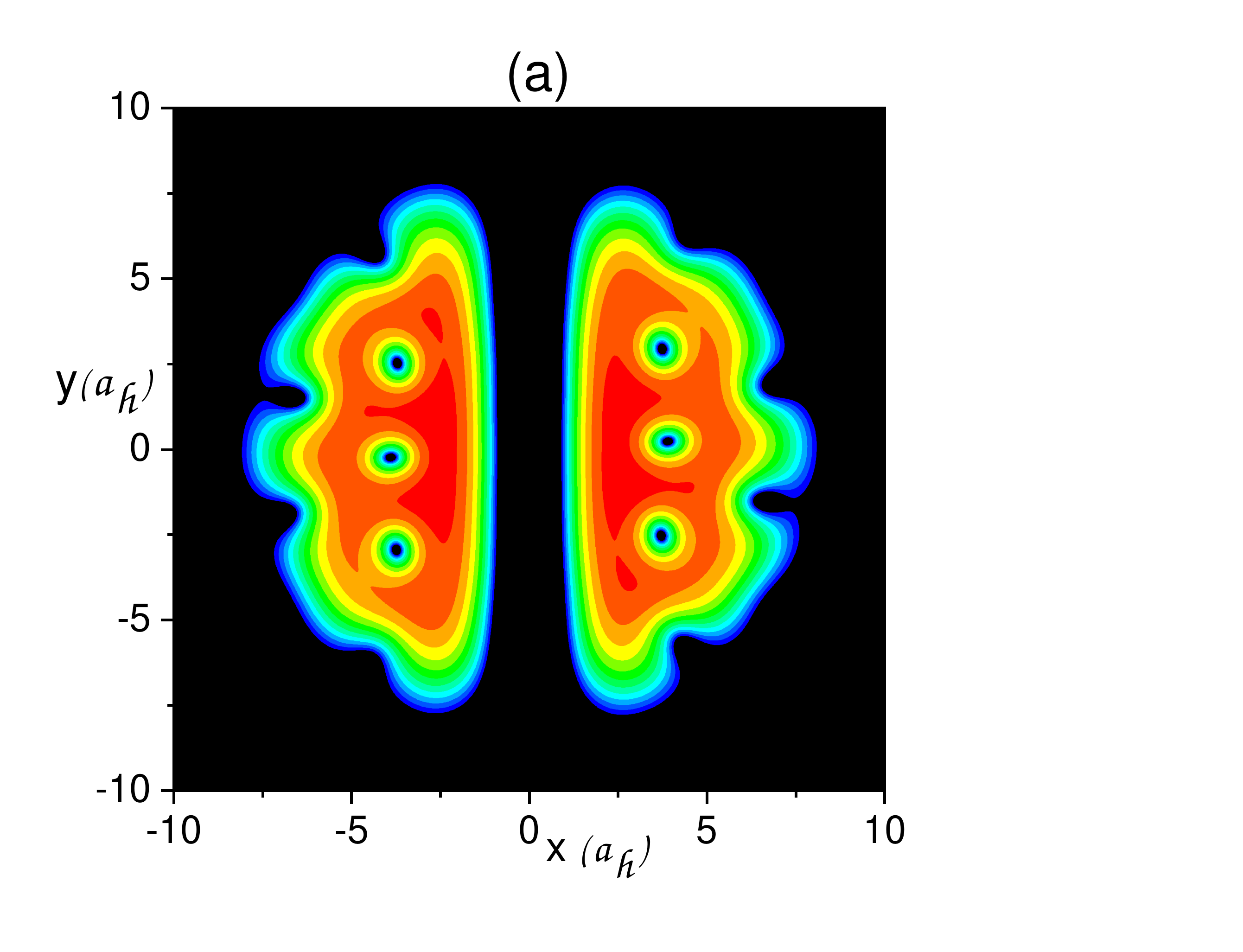}
\end{subfigure}\hfill
\begin{subfigure}[t]{0.3\textwidth}
\includegraphics[width=1.5in,height=1.8in,angle=0]{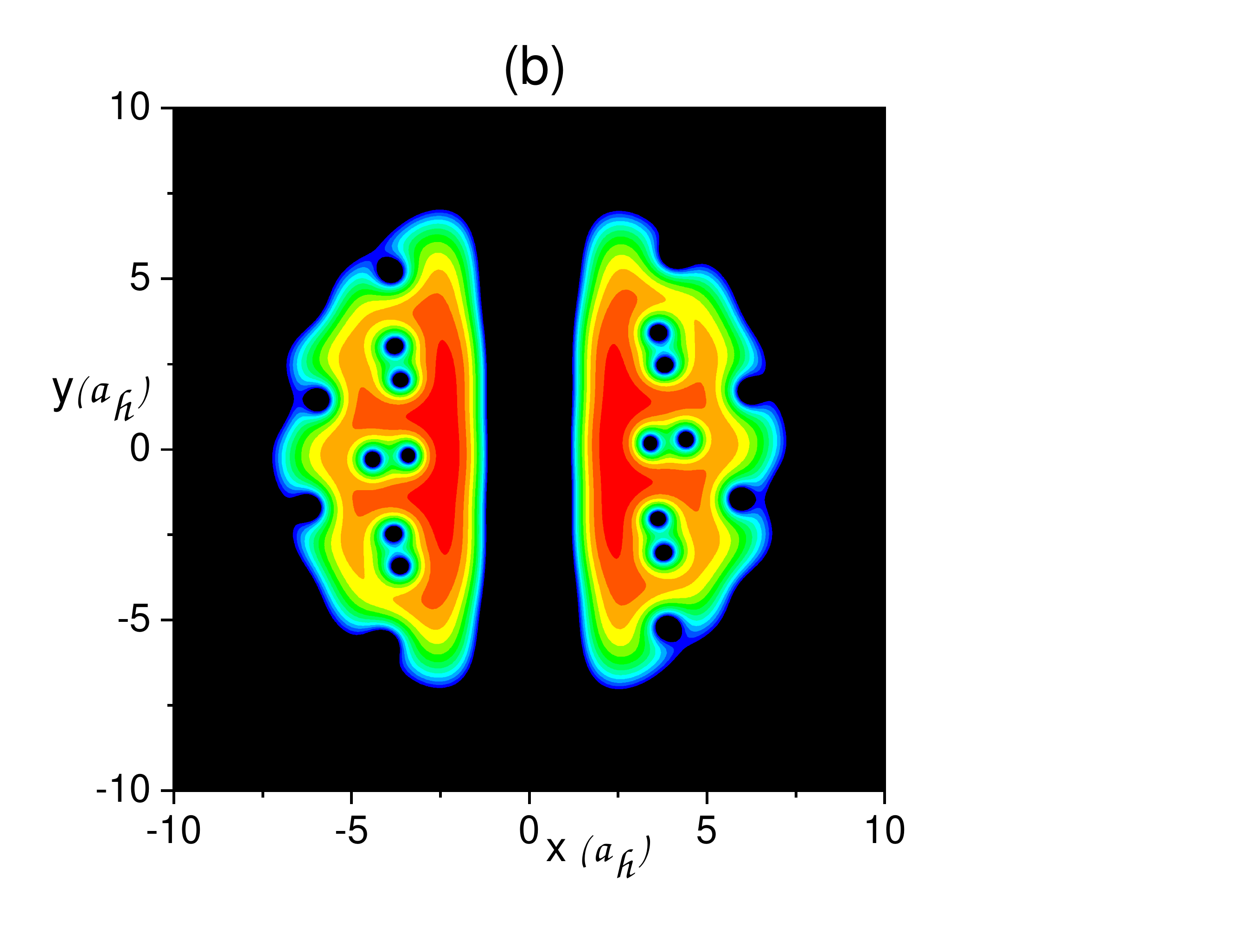}
\end{subfigure}\hfill
\begin{subfigure}[t]{0.3\textwidth}
\includegraphics[width=1.5in,height=1.8in,angle=0]{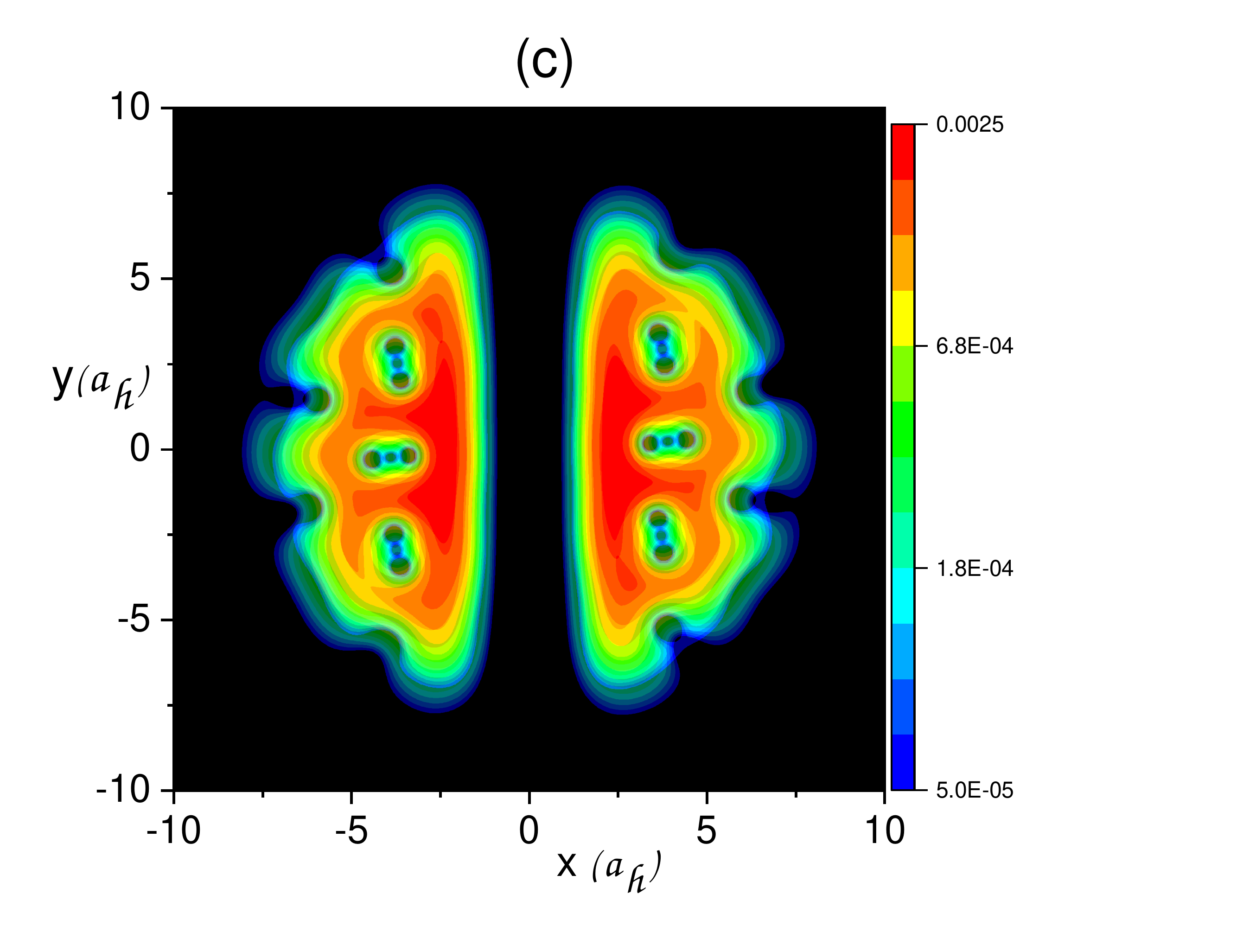}
\end{subfigure}

\begin{subfigure}[t]{0.3\textwidth}
\includegraphics[width=1.5in,height=1.8in,angle=0]{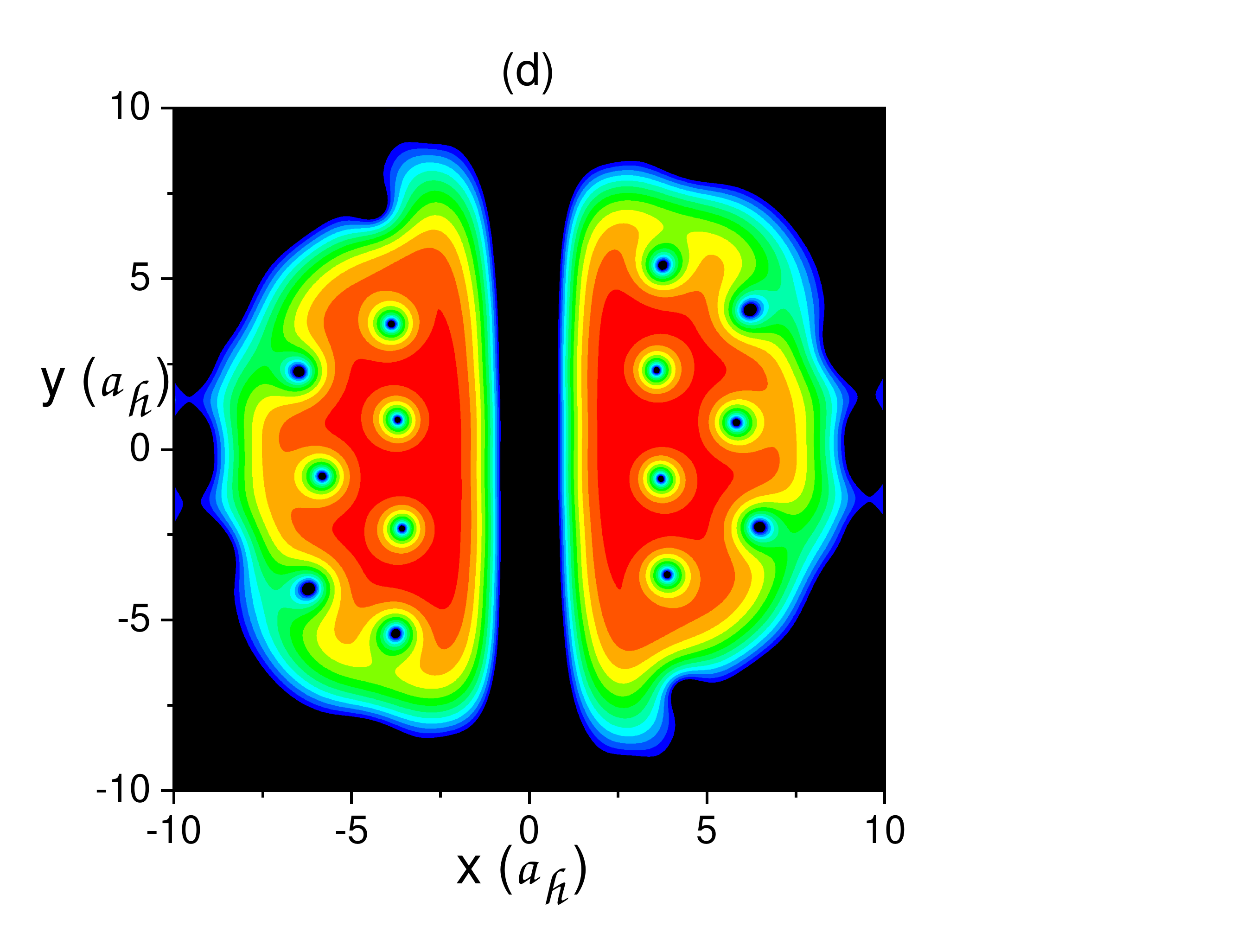}
\end{subfigure}\hfill
\begin{subfigure}[t]{0.3\textwidth}
\includegraphics[width=1.5in,height=1.8in,angle=0]{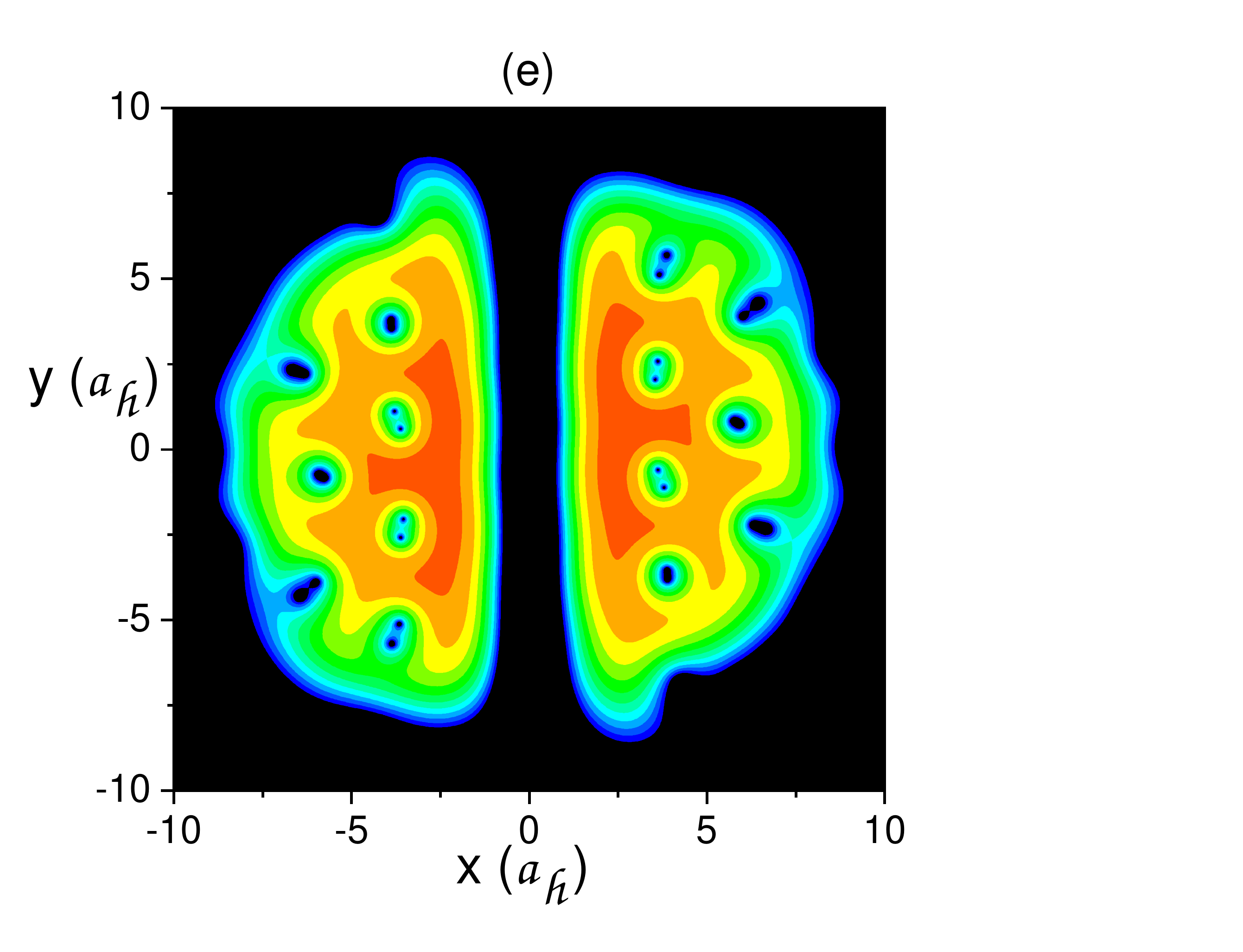}
\end{subfigure}\hfill
\begin{subfigure}[t]{0.3\textwidth}
\includegraphics[width=1.5in,height=1.8in,angle=0]{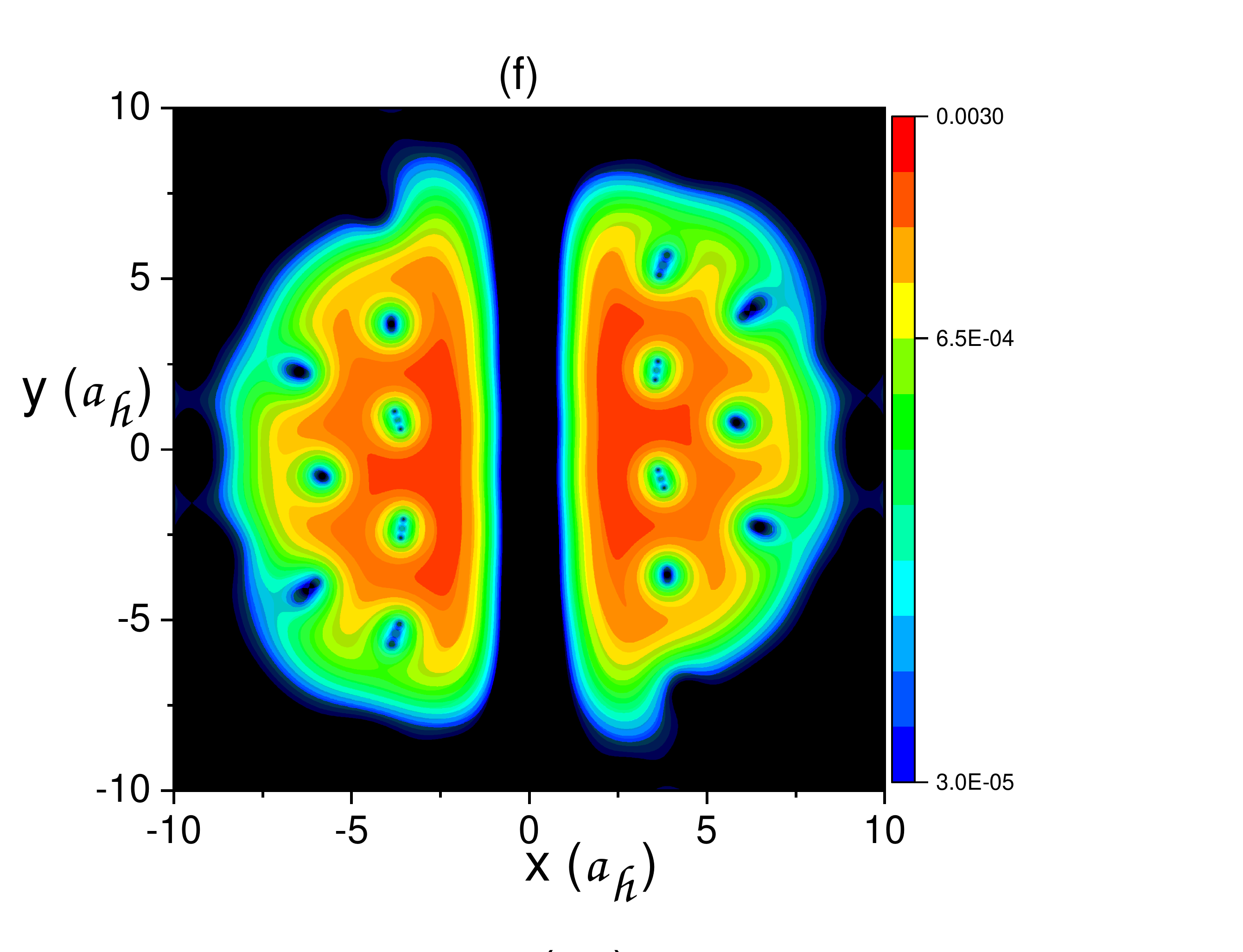}
\end{subfigure}

\begin{subfigure}[t]{0.3\textwidth}
\includegraphics[width=1.5in,height=1.8in,angle=0]{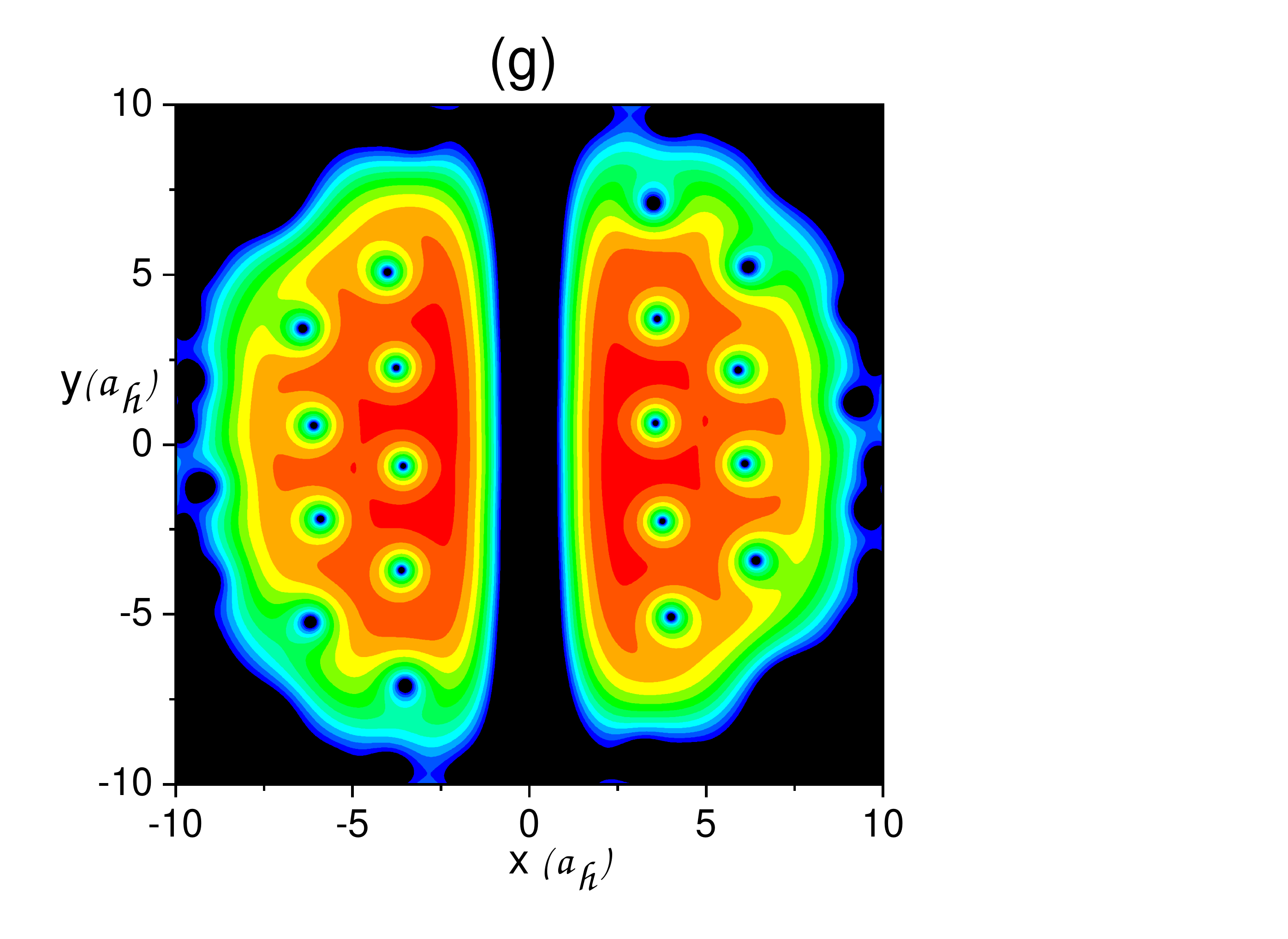}
\end{subfigure}\hfill
\begin{subfigure}[t]{0.3\textwidth}
\includegraphics[width=1.5in,height=1.8in,angle=0]{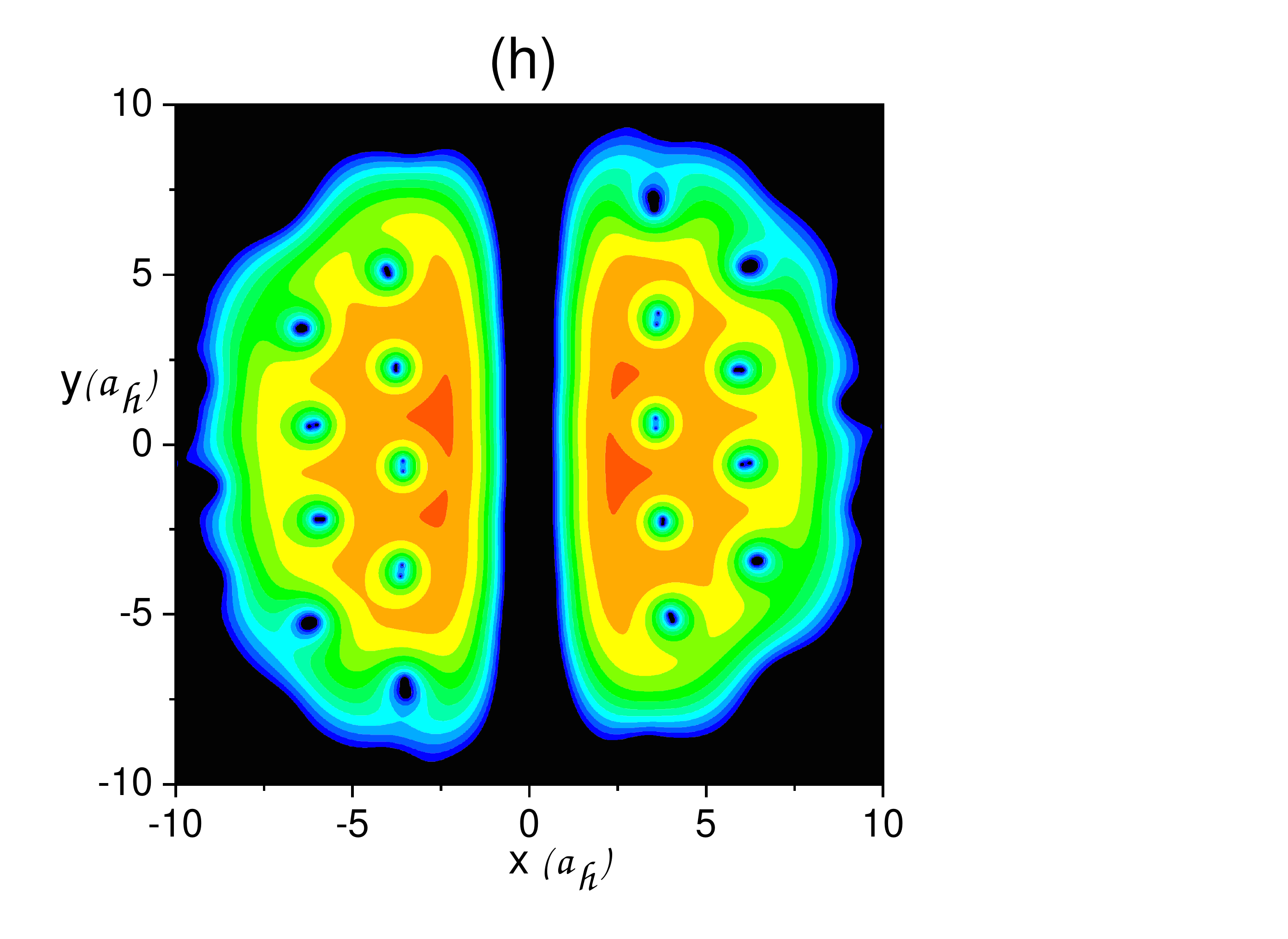}
\end{subfigure}\hfill
\begin{subfigure}[t]{0.3\textwidth}
\includegraphics[width=1.5in,height=1.8in,angle=0]{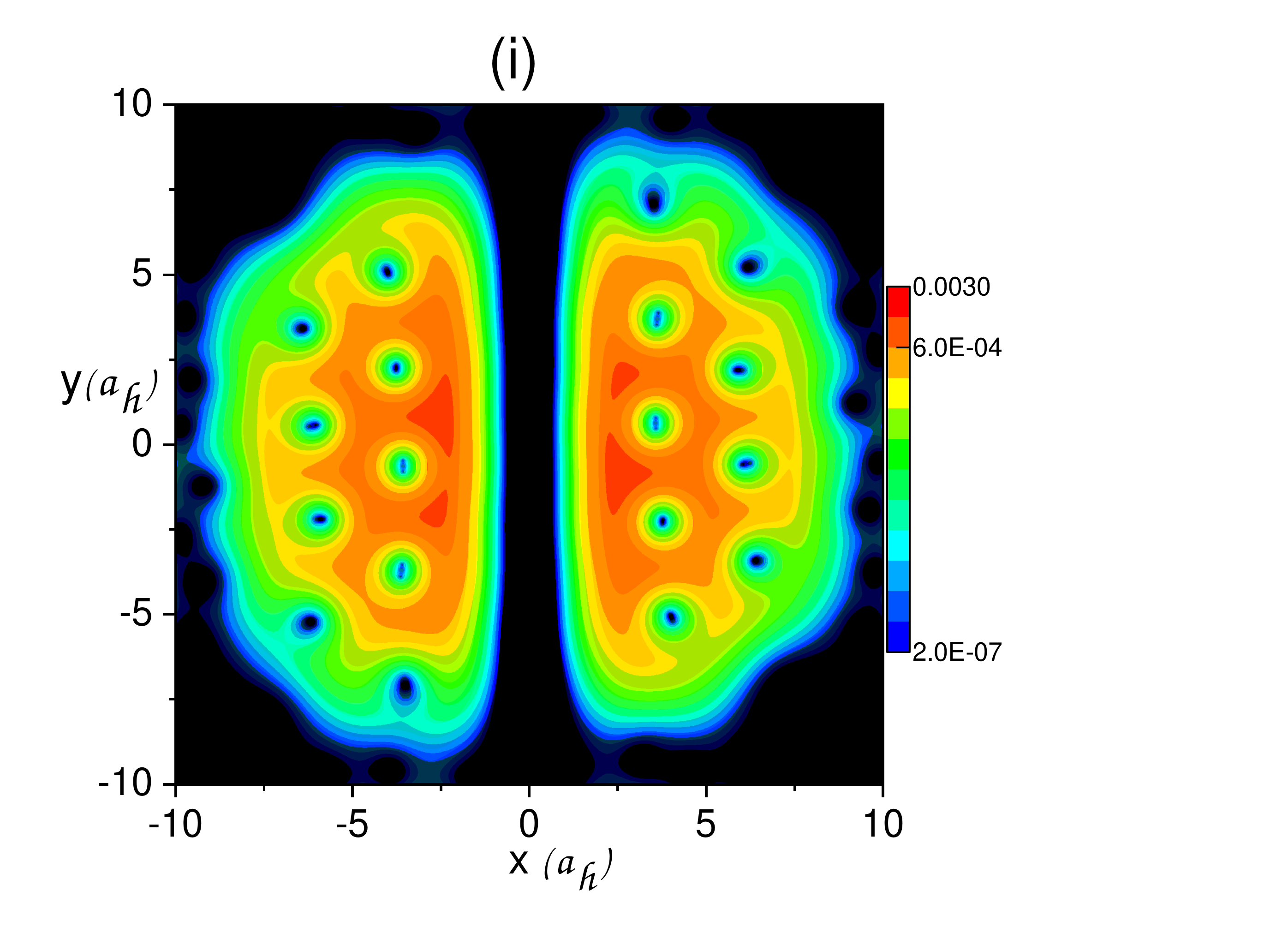}
\end{subfigure}

\begin{minipage}{5in}
\caption{(Color online) Density distributions for atomic [Figs. (a), (d), (g)], 
molecular  [Figs. (b), (e), (h)] and atomic with molecular  [Figs. (c), (f), (i)] 
vortex lattice configuration for different detuning parameters $\epsilon_1$= -0.5 
[Figs. (a), (b), (c)], 2 [Figs. (d), (e), (f)] and 5 [Figs. (g), (h), (i)] at $t_1$= 300. $\Omega_1$= 0.95 and
$\chi_1$= 50. Red color corresponds to higher
densities and blue color corresponds to lower densities. Darker
color corresponds to lower density. x and y are in the units of
$a_h$=$\sqrt{\hbar\over{2m\omega_\perp}}$.}
\end{minipage}
\end{figure}

\begin{figure}\begin{subfigure}[t]{0.25\textwidth}
\includegraphics[width=1.5in,height=1.8in,angle=0]{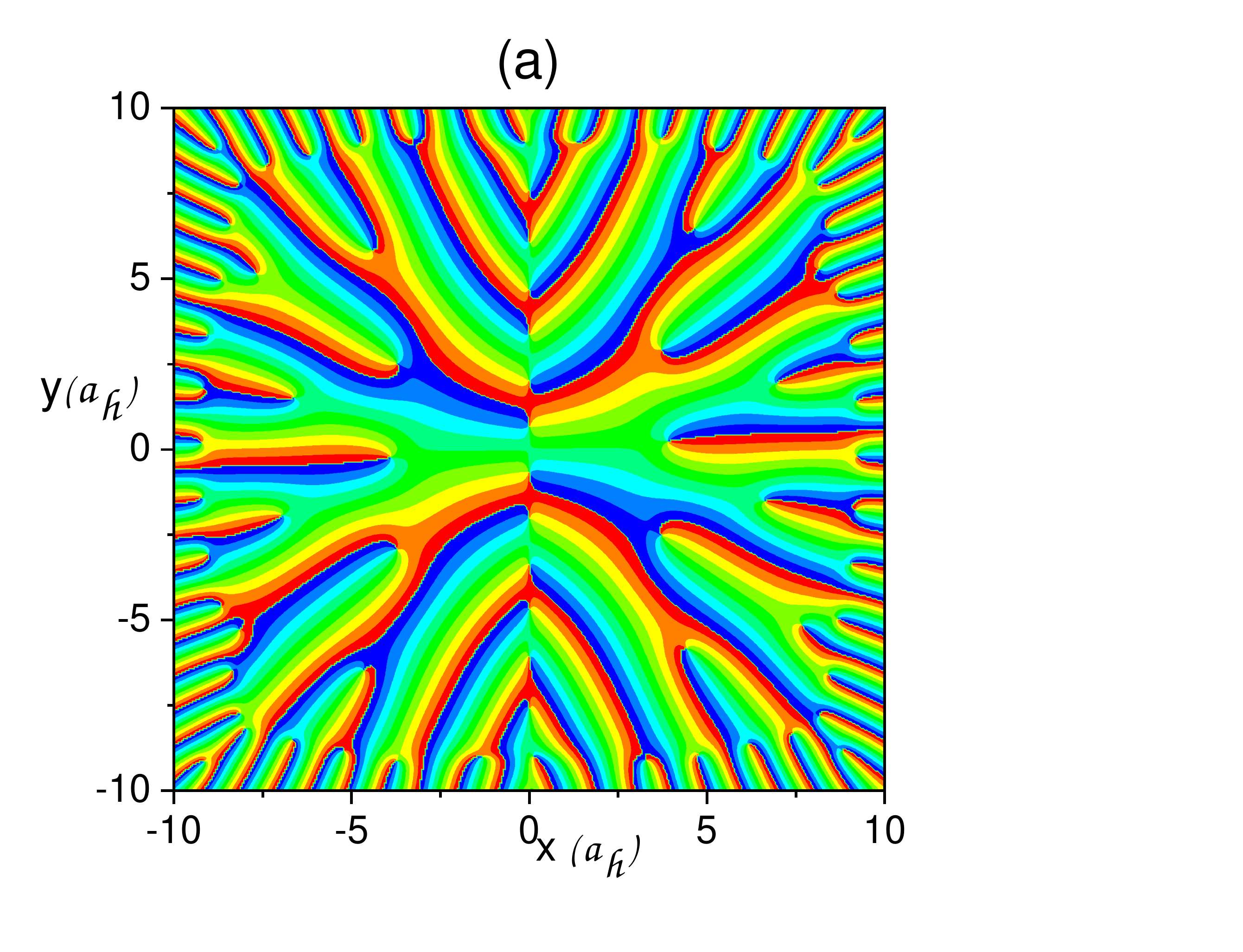}
\end{subfigure}\hfill
\begin{subfigure}[t]{0.25\textwidth}
\includegraphics[width=1.5in,height=1.8in,angle=0]{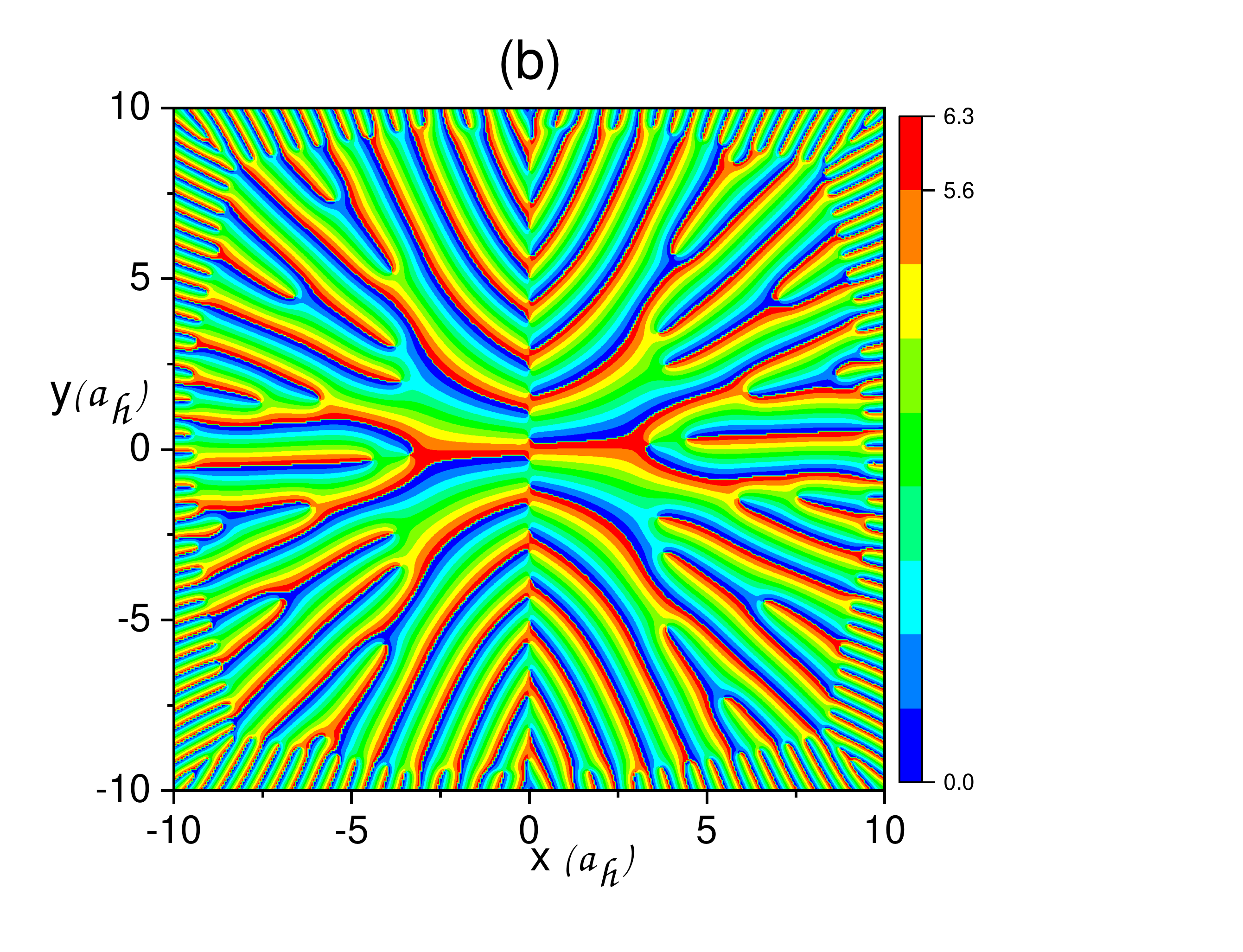}
\end{subfigure}\hfill
\begin{subfigure}[t]{0.25\textwidth}
\includegraphics[width=1.5in,height=1.8in,angle=0]{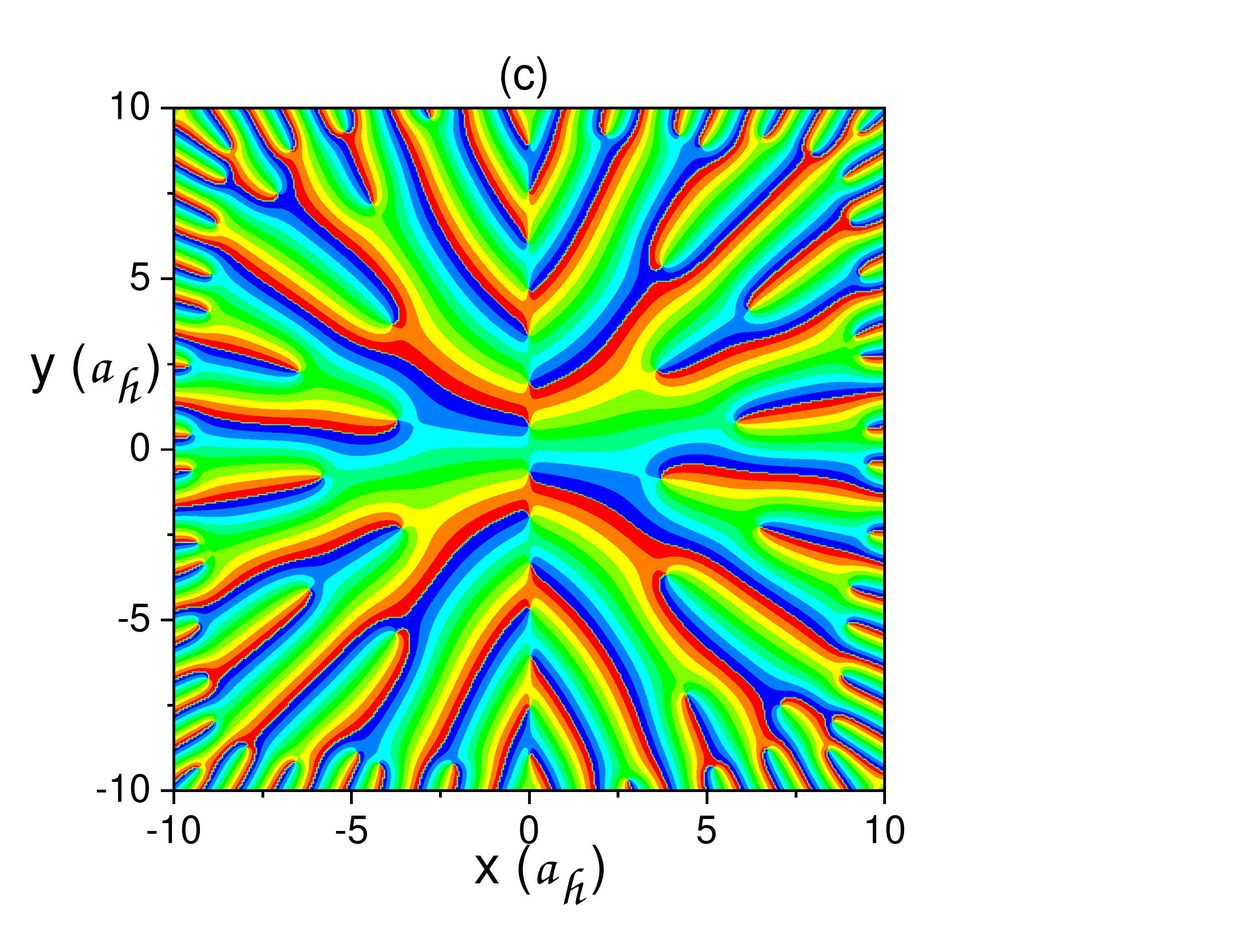}
\end{subfigure}

\begin{subfigure}[t]{0.25\textwidth}
\includegraphics[width=1.5in,height=1.8in,angle=0]{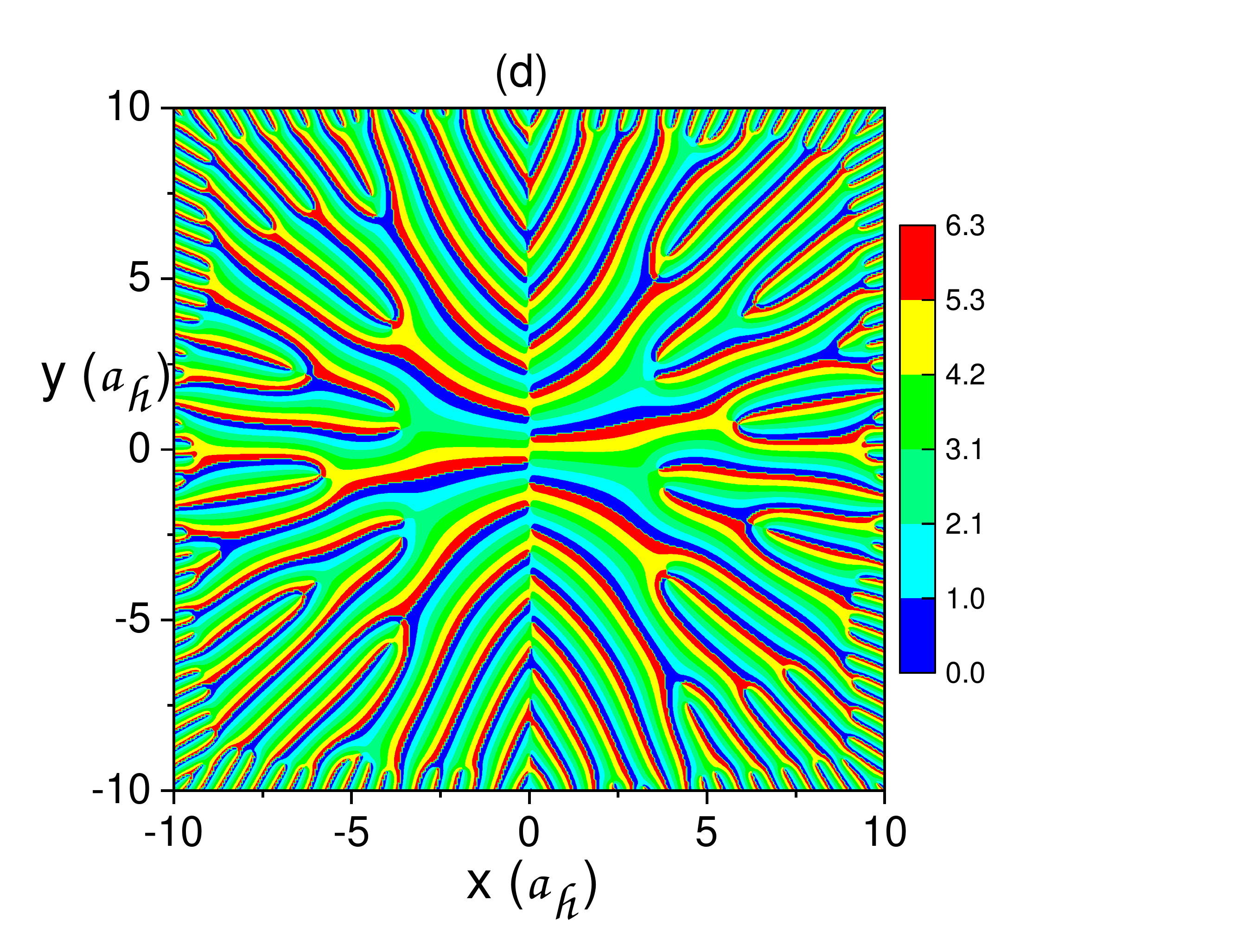}
\end{subfigure}\hfill
\begin{subfigure}[t]{0.25\textwidth}
\includegraphics[width=1.5in,height=1.8in,angle=0]{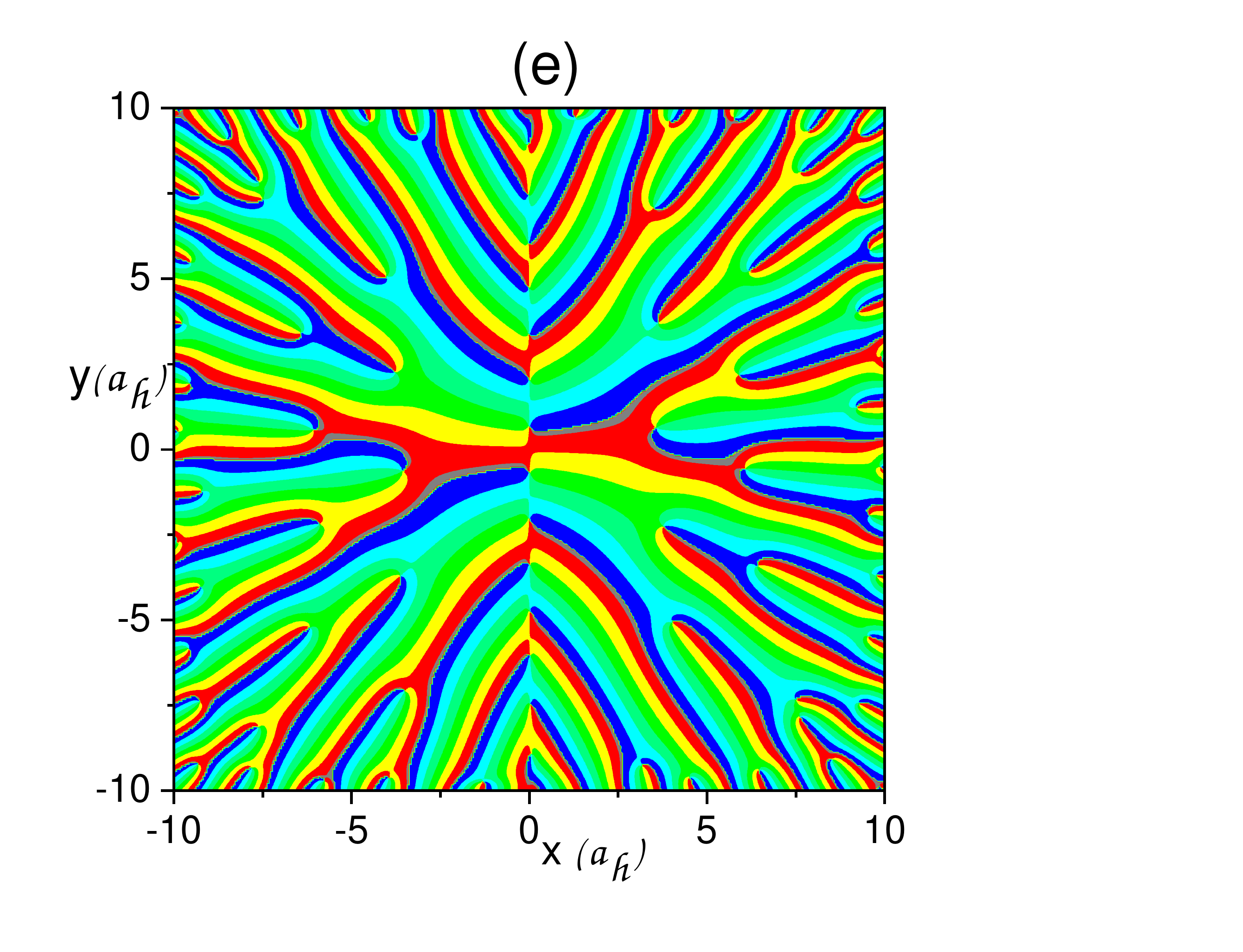}
\end{subfigure}\hfill
\begin{subfigure}[t]{0.25\textwidth}
\includegraphics[width=1.5in,height=1.8in,angle=0]{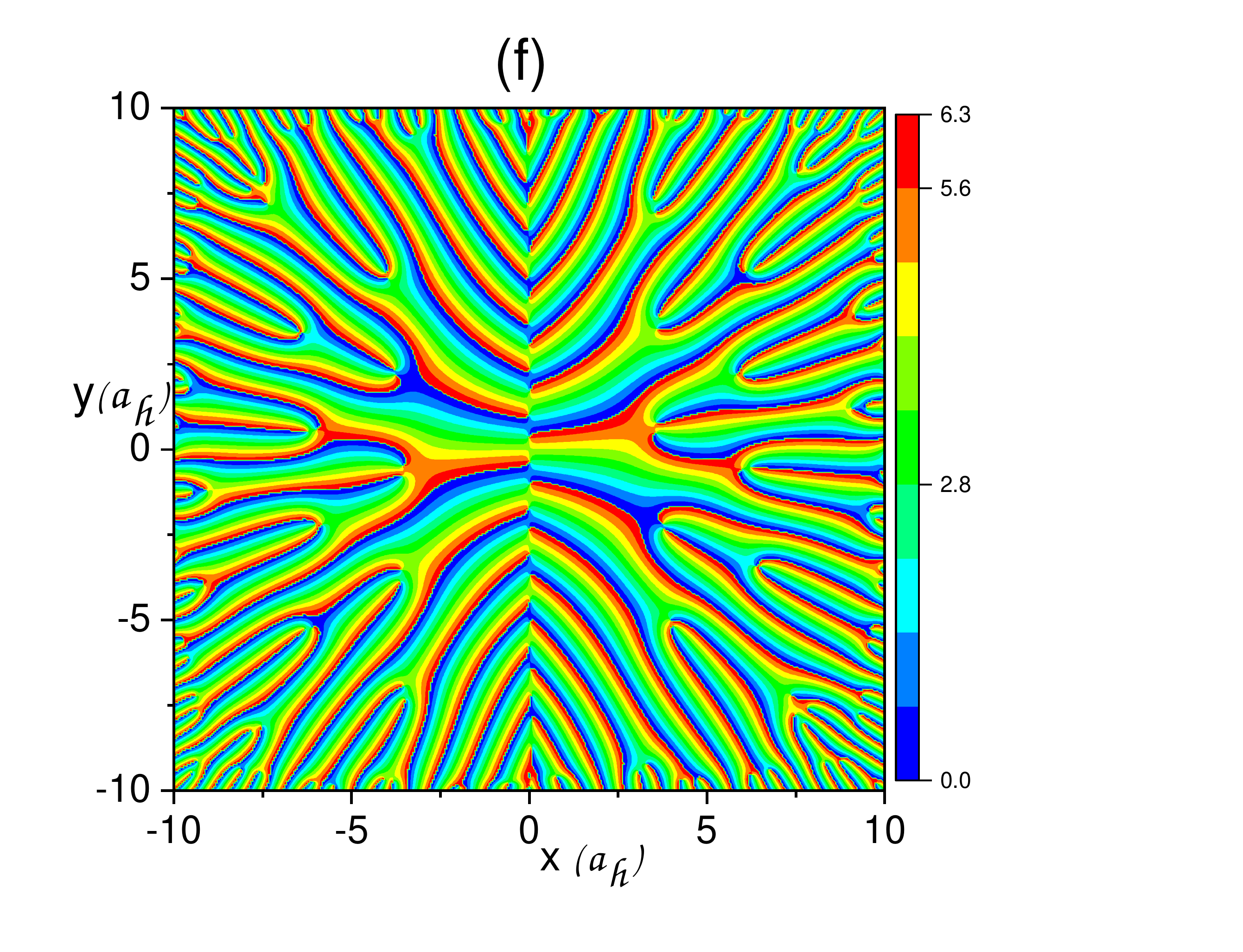}
\end{subfigure}

\begin{minipage}{5in}
\caption{(Color online) Phase profile of atoms $\psi_a$ for different
detuning parameters $\epsilon_1$= -0.5 [Fig. (a)], 2
[Fig. (c)], 5 [Fig. (e)]  and  phase profile of molecules $\psi_m$ for different
$\epsilon_1$= -0.5 [Fig. (b)], 2
[Fig. (d)], 5 [Fig. (f)] at $t_1$= 300. $\chi_1$= 50 and $\epsilon_1$= 0.
Phase varies from 0 to 2$\pi$. Red color corresponds to higher
values and blue color corresponds to lower values. Darker color
corresponds to lower phase. x and y are in the units of $a_h$=
$\sqrt{\hbar\over {2m\omega_\perp}}$}
\end{minipage}
\end{figure}

\section {Acknowledgement}

We thank Bimalendu Deb, School of Physics, Indian Association for
the Cultivation of Science for his interest in this work.

\end{document}